\RequirePackage{lineno} 
\documentclass[aps,prl,twocolumn,superscriptaddress,showkeys,showpacs,nofootinbib]{revtex4-2}

\usepackage[dvipsnames]{xcolor}
\usepackage[normalem]{ulem}
\usepackage{comment}
\usepackage{overpic}
\usepackage{multirow}
\usepackage{booktabs} 
\usepackage{array}

\graphicspath{{./figures/}}
\usepackage[colorlinks,linkcolor=blue,anchorcolor=blue,citecolor=blue,urlcolor=blue,breaklinks=true]{hyperref}
\newsavebox{\tablebox}
\usepackage{orcidlink}

\input{alias.tex}

\def\Lcp {\Lambda{}_{c}^{+}} 
\def\Lcm {\Lbar{}^{-}_{c}}

\def\LcToLamKp {{\Lambda_c^+\to\Lambda\Kp}}
\def\LcToLamPip {{\Lambda_c^+\to\Lambda\pip}}
\def\LcToSigKp {{\Lambda_c^+\to\Sigma^{0}\Kp}}
\def\LcToSigPip {{\Lambda_c^+\to\Sigma^{0}\pip}}
\def\LcToLamHp {{\Lambda_c^+\to\Lambda h^+}}
\def\LcToSigHp {{\Lambda_c^+\to\Sigma^{0} h^+}}
\def\LcmToLamKp {{\Lcm\to\Lbar\Km}}
\def\LcmToLamPip {{\Lcm\to\Lbar\pim}}
\def\LcmToSigKp {{\Lcm\to\Sigbar{}^{0}\Km}}
\def\LcmToSigPip {{\Lcm\to\Sigbar{}^{0}\pim}}

\begin{document}

\title{\boldmath Search for $\CP$ violation and measurement of branching fractions and decay asymmetry parameters for $\LcToLamHp$ and $\LcToSigHp$ ($h\!=\!K,\,\pi$)} 

\collaboration{The Belle Collaboration}

\begin{abstract}
We report a study of $\LcToLamHp$ and $\LcToSigHp$ ($h\!=\!K,\,\pi$) decays based on a data sample of ${980~\invfb}$ collected with the Belle detector at the KEKB energy-asymmetric $e^+e^-$ collider. 
The first results of direct $\CP$ asymmetry in two-body singly Cabibbo-suppressed (SCS) decays of charmed baryons are measured, $\Acp^{\rm{dir}}(\LcToLamKp)\!=\!+0.021\pm0.026\pm0.001$ and $\Acp^{\rm{dir}}(\LcToSigKp)\!=\!+0.025\pm0.054\pm0.004$.
We also make the most precise measurement of the decay asymmetry parameters ($\alpha$) for the four modes of interest and search for $\CP$ violation via the $\alpha$-induced $\CP$ asymmetry ($\Acp^{\alpha}$). 
We measure
$\Acp^{\alpha}(\LcToLamKp)\!=\!{-0.023\pm0.086\pm0.071}$ and 
$\Acp^{\alpha}(\LcToSigKp)\!=\!{+0.08\pm 0.35\pm 0.14}$, 
which are the first $\Acp^{\alpha}$ results for SCS decays of charmed baryons.
We search for $\Lambda$-hyperon $\CP$ violation
in $\Lcp\to(\Lambda,\,\Sigma^0)\pip$ and find $\Acp^{\alpha}(\Lambda\to p\pim)\!=\!{+0.013\pm0.007\pm0.011}$. 
This is the first time that hyperon $\CP$ violation has been measured via Cabibbo-favored charm decays.
No evidence of baryon $\CP$ violation is found. 
We also obtain the most precise branching fractions for two SCS $\Lcp$ decays, 
$\BR(\LcToLamKp)\!=\!(6.57\pm0.17\pm0.11\pm0.35)\times10^{-4}$ and 
$\BR(\LcToSigKp)\!=\!(3.58\pm0.19\pm0.06\pm0.19)\times10^{-4}$. The first uncertainties are statistical and the second systematic, while the third uncertainties come from the uncertainties on the world average branching fractions of $\Lcp\to(\Lambda,\,\Sigma^0)\pip$.

\end{abstract}


\maketitle

\section{Introduction}
Charge-parity~($\CP$) violation is one of the conditions necessary to explain the matter-antimatter asymmetry of the universe~\cite{Sakharov:1967dj}. 
The single complex phase in the Cabibbo-Kobayashi-Maskawa matrix provides the only source of $\CP$ violation~(CPV) in the standard model~(SM), but it is not large enough to explain the observed matter-antimatter asymmetry.
Baryogenesis, the process by which the baryon-antibaryon asymmetry of the universe developed, is directly related to baryon CPV~\cite{Sakharov:1988vdp,Shaposhnikov:1987tw}. 
To date, CPV has been observed in the open-flavored meson sector,
but not yet established in the baryon sector. 
Since CPV in charm decays is predicted in the SM to be at the level of $10^{-3}$ or smaller~\cite{Brod:2011re,Li:2012cfa,Cheng:2019ggx,Kagan:2020vri,Cheng:2021yrn},
an observation of CPV in charm decays much greater than $10^{-3}$ could indicate new physics beyond the SM~\cite{Delepine:2019cpp,Dery:2019ysp,Chala:2019fdb,Saur:2020rgd}. 

Singly Cabibbo-suppressed~(SCS) decays of charm hadrons 
provide an ideal laboratory for studying CPV as they are a unique window on the physics of decay-rate dynamics in the charm sector~\cite{Cheng:2021yrn,Saur:2020rgd}. 
The only observation of CPV in the charm sector was made by the LHCb collaboration 
in SCS charmed meson decays, $D^0\to h^+h^-$ ($h\!=\!K,\,\pi$ throughout this paper)~\cite{LHCb:2019hro}.
Measurements of the direct $\CP$ asymmetry~($A_{\CP}^{\rm dir}$), induced by the partial widths, in SCS charmed baryon decays are experimentally more challenging than in charmed meson decays and relatively unexplored.
Searches for direct CPV in SCS charmed baryon decays were made in $\Lcp\to{}ph^+h^-$~\cite{LHCb:2017hwf} and $\Xi_c^+\to{}p\Km\pip$~\cite{LHCb:2020zkk}. 
No direct CPV searches in two-body SCS decays of charmed baryons have been made.

In addition to $A_{\CP}^{\rm dir}$, the $\alpha$-induced $\CP$ asymmetry ($A_{\CP}^{\alpha}$) is an essential observable to search for CPV in baryon decays. 
Here $\alpha$ is the decay asymmetry parameter introduced 
to study the parity-violating and parity-conserving amplitudes in weak hyperon decays~\cite{Lee:1957qs}.
In a weak decay of $\Lcp$ into a spin $1/2$ baryon with positive parity and a pseudoscalar meson, 
$\alpha \!\equiv\! {2\cdot {\rm Re}(S^{*}P)/(|S|^2 + |P|^2)}$, 
where $S$ and $P$ denote the parity-violating $S$-wave and parity-conserving $P$-wave amplitudes, respectively. 
Since $\alpha$ is $\CP$-odd, the $\alpha$-induced $\CP$ asymmetry for $\Lcp$ decays is defined as $\Acp^{\alpha} \!\equiv\! {(\alpha_{\Lcp}+\alpha_{\Lcm})/(\alpha_{\Lcp}-\alpha_{\Lcm})}$.
In the case that $\Acp^{\rm dir}$ is zero, $\Acp^{\alpha}$ is given by the CPV in ${\rm Re}(S^{*}P)$.
Therefore, $\Acp^{\alpha}$ provides an observable complementary to $\Acp^{\rm dir}$.
To date, there is only one $\Acp^{\alpha}$ measurement for hadronic $\Lcp$ decays, $\Acp^{\alpha}({\Lcp\to\Lambda \pip})\!=\!-0.07 \pm 0.22$~\cite{FOCUS:2005vxq}. 
Using the precisely measured $\alpha_{\mp}$ in $\Lambda\to p\pim$ decays~\cite{BESIII:2022qax} and the high-statistics $\Lcp$ sample at Belle, 
we obtain the $\alpha_{\Lcp}$ and $\alpha_{\Lcm}$ values in $\Lcp\to(\Lambda,\,\Sigma^0)h^+$ decays, described in detail in Sec.~\ref{sec:method}, make the first measurements of $\Acp^{\alpha}$ in $\LcToLamKp$ and $\LcToSigHp$ decays, and measure $\Acp^{\alpha}$ with improved precision in $\LcToLamPip$. 

The $\Lambda$-hyperon $\CP$ asymmetry $\Acp^{\alpha}({\Lambda\to{}p\pim})$ can be extracted from the total $\alpha$-induced $\CP$ asymmetry ($\Acp^{\alpha}({\rm total}) \!\equiv\! {(\alpha_{\Lcp}\alpha_{-} - \alpha_{\Lcm}\alpha_{+})/(\alpha_{\Lcp}\alpha_{-} +\alpha_{\Lcm} \alpha_{+})}$) for Cabibbo-favored (CF) decays $\Lcp\to(\Lambda,\,\Sigma^0)\pip$ with $\alpha_{\Lcp}\!=\!-\alpha_{\Lcm}$ since no $\CP$ asymmetry is expected in the SM. 
CPV in hyperon decays is predicted to be at the level of $\mathcal{O}(10^{-4})$ or smaller in the SM~\cite{Donoghue:1985ww,Donoghue:1986hh,Tandean:2002vy,Salone:2022lpt} and can be enhanced to reach the level of $10^{-3}$ in some new physics models~\cite{Chang:1994wk,He:1999bv,Chen:2001cv,Tandean:2003fr,Salone:2022lpt}.
This analysis is a novel and complementary method, proposed in Ref.~\cite{Prof.Yu}, for $\Lambda$-hyperon CPV searches.

Since the $\Lcp$ was discovered, many efforts have been made to predict the branching fractions~(BF) and $\alpha$ parameters of its hadronic decays using phenomenological models such as current algebra~\cite{Uppal:1994pt}, pole model~\cite{Cheng:2018hwl,Zou:2019kzq} and SU(3)${}_{\rm F}$ symmetry~\cite{Geng:2019xbo,Wang:1979dx,Savage:1989qr,Sharma:1996sc,Lu:2016ogy}.
These predictions  
are nontrivial due to non-perturbative strong dynamics, which complicate the calculation of non-factorizable contributions
~\cite{Koniuk:1979vy,Cheng:2021qpd}.
Experimentally, studies of charmed baryon decays are more challenging than those of charmed mesons due to lower production rates. 
The current world averages $\BR(\LcToLamKp)\!=\!{(6.1\pm 1.2)\!\times\!10^{-4}}$ and $\BR(\LcToSigPip)\!=\!{(5.2\pm 0.8)\!\times\!10^{-4}}$~\cite{bib:PDG2022}, rely on measurements with partial datasets from Belle and BaBar~\cite{Belle:2001hyr,BaBar:2006eah}. 
We perform a measurement based on a dataset thirty times larger than previously used, superseding the result in Ref.~\cite{Belle:2001hyr}.

In this paper, we report $A_{\CP}^{\rm dir}$ and BF measurements for the SCS decays $\LcToLamKp$ and $\LcToSigKp$, using the CF decays $\LcToLamPip$ and $\LcToSigPip$ as reference modes. Inclusion of charge conjugate states is implicit, unless otherwise stated.
We also measure $\alpha$ and $\Acp^{\alpha}$ in these four decays and search for $\Lambda$-hyperon CPV in the CF $\Lcp$ decays. 

\section{Detector and data set}
This analysis is based on the full data set recorded by the Belle detector~\cite{Belle:2000cnh,Belle:2012iwr} 
operating at the KEKB~\cite{Kurokawa:2001nw,Abe:2013kxa}
asymmetric-energy $e^+e^-$ collider. 
This data sample corresponds to a total integrated luminosity of $980~\invfb$ collected at or near the $\Upsilon(nS)$ ($n\!=\!1,\,2,\,3,\,4,\,5$) resonances. 
The Belle detector is a large-solid-angle magnetic spectrometer consisting of a silicon vertex detector~(SVD), a central drift chamber~(CDC), an array of aerogel threshold Cherenkov counters~(ACC), a barrel-like arrangement of time-of-flight scintillation counters~(TOF), and an electromagnetic calorimeter~(ECL) consisting of CsI(Tl) crystals. These components are all located inside a superconducting solenoid coil that provides a 1.5~T magnetic field. The iron flux-return of the magnet is instrumented to detect $K^0_L$ mesons and to identify muons~(KLM). The detector is described in detail elsewhere~\cite{Belle:2000cnh,Belle:2012iwr}.

Monte Carlo~(MC) simulated events are generated with {\sc{evtgen}}~\cite{Lange:2001uf} and {\sc{pythia}}~\cite{Sjostrand:2000wi}, 
and are subsequently processed through a full detector simulation based on {\sc{geant3}}~\cite{Brun:1987ma}. 
Final-state radiation from charged particles is included at event generation using {\sc{photos}}~\cite{Barberio:1993qi}. 
Signal $\Lcp$ baryons are produced via the inclusive process $e^+e^-\to{}c\bar{c}\to\Lcp+{\rm anything}$ and $\LcToLamHp,\,\Sigma^0h^+$ decays, where $\Sigma^0\to\Lambda\gamma$ and $\Lambda\to{}p\pim$.

\section{Measurement methods\label{sec:method}}
The direct $\CP$ asymmetry, taking $\Lcp$ decays as an example, is defined as 
\begin{eqnarray}
\Acp^{\rm dir} = \frac{\Gamma(\Lcp\to f)-\Gamma(\Lcm\to\fbar)}{\Gamma(\Lcp\to f)+\Gamma(\Lcm\to \fbar)}\,, \label{eqn:Acpdir} 
\end{eqnarray}
where $\Gamma(\Lcp\to f)$ and $\Gamma(\Lcm\to\fbar)$ are the partial decay widths for the final state $f$ and its $\CP$-conjugate state $\fbar$.
The raw asymmetry in the decays of $\Lcp\to f$ and $\Lcm\to\fbar$ is defined with signal yields $N$ as follows:  
\begin{eqnarray}
A_{\rm raw} = \frac{N(\Lcp\to f)-N(\Lcm\to\fbar)}{N(\Lcp\to f) + N(\Lcm\to\fbar)}\,. \label{eqn:acp}
\end{eqnarray}
Several sources contribute to the raw asymmetry, which for $\LcToLamKp$ is given by  
\begin{eqnarray}
\hskip-15pt
A_{\rm raw} & = & \Acp^{\LcToLamKp} + \Acp^{\Lambda\to p\pim} + A_{\varepsilon}^{\Lambda}  + A_{\varepsilon}^{\Kp} + A_{\rm FB}^{\Lcp} ,
\end{eqnarray}
where all terms are small (at the level of $10^{-2}$ or smaller).
Here $\Acp^{\LcToLamKp}$ ($\Acp^{\Lambda\to{}p\pim}$) is the direct $\CP$ asymmetry associated with the $\Lcp$ ($\Lambda$) decay,
$A_{\varepsilon}^{\Lambda}$ ($A_{\varepsilon}^{\Kp}$) is the detection asymmetry resulting from differences in the reconstruction efficiency between $\Lambda$ ($K^+$) and its anti-particle $\Lbar$ ($K^-$), and
$A_{\rm FB}^{\Lcp}$ arises from the forward-backward asymmetry~(FBA) of $\Lcp$ production due to $\gamma$-$Z^0$ interference and higher-order QED effects in ${\epem\to c\cbar}$ collisions~\cite{Brown:1973ji,Cashmore:1985vp}.
The FBA is an odd function in $\cos\theta^*$, where $\theta^*$ is the $\Lcp$ production polar angle in the $e^+e^-$ center-of-mass frame, but due to asymmetric acceptance, small residual asymmetry remains after integrating over $\cos\theta^{*}$.

We weight $\Lambda_c^{\pm}$ candidates with factors $1\mp A_{\varepsilon}^{h^+}$ to remove the $\Kp$ or $\pip$ detection asymmetry from the raw asymmetry in $\Lcp\to(\Lambda,\Sigma^0)K^+$ or $\Lcp\to(\Lambda,\Sigma^0)\pi^+$. 
We use $A_{\rm raw}^{\rm corr}$ to indicate this corrected raw asymmetry. 
Here $A_{\varepsilon}^{h^+}$ depends on the cosine of the polar angle and transverse momentum of the $h^+$ tracks in the laboratory frame and was determined at Belle using $\Dz\to\Km\pip$ and $\Dsp\to\phi\pip$ events for $\mathcal{A}_{\varepsilon}^{\Kp}$~\cite{Belle:2012ygx} and 
$\Dp\to\Km\pip\pip$ and $\Dz\to\Km\pip\piz$ events for $\mathcal{A}_{\varepsilon}^{\pip}$~\cite{Belle:2012ygt}. 
The signal modes and corresponding reference modes have nearly the same $\Lambda$ kinematic distributions, including the $\Lambda$ decay length, 
the polar angle with respect to the direction opposite the positron beam and the momentum of the proton and pion in the laboratory reference frame. 
Asymmetries common between the signal and reference modes therefore cancel. 

The difference of the corrected raw asymmetries is 
\begin{eqnarray}
A_{\rm raw}^{\rm corr}(\LcToLamKp)-A_{\rm raw}^{\rm corr}(\LcToLamPip)  \nonumber \\
= \Acp^{\rm dir}(\LcToLamKp) - \Acp^{\rm dir}(\LcToLamPip)\,. \label{eqn:DeltaAcp}
\end{eqnarray}
The direct $\CP$ asymmetry for $\LcToLamPip$, a CF process with an amplitude that has only one weak phase, can be set to be zero. 
Thus, the measured asymmetry difference in Eq.~(\ref{eqn:DeltaAcp}) is equal to $\Acp^{\rm dir}$ for $\LcToLamKp$. 

The BFs of signal modes are measured relative to those of the reference modes using 
\begin{eqnarray}
\frac{\mathcal{B}_{\rm sig}}{\mathcal{B}_{\rm ref}} = \frac{N_{\rm sig}/\varepsilon_{\rm sig}}{ N_{\rm ref}/\varepsilon_{\rm ref} }\,, \label{eqn:BRratio}
\end{eqnarray}
where $N_{\rm sig}$ is the extracted signal yield and $\eff$ is the reconstruction efficiency.
The world average values $\BR(\LcToLamPip)\!=\!(1.30\pm0.07)\%$ and $\BR(\LcToSigPip)\!=\!(1.29\pm0.07)\%$~\cite{bib:PDG2022} are used for the reference modes. 
The common systematic uncertainties between the signal modes and reference modes, such as the inclusive $\Lcp$ yield produced from $e^+e^-\to c\cbar$ and the mass resolution of the $\Lambda$ and $\Sigma^0$, cancel in the ratio.

For $\LcToLamHp$ decays, the differential decay rate depends on $\alpha$ parameters and one helicity angle as
\begin{eqnarray}
\frac{dN}{d\cos\theta_{\Lambda}} \propto 1+\alpha_{\Lcp}\alpha_{-}\cos\theta_{\Lambda}\,, \label{eqn:alpha_LcToLamHp} 
\end{eqnarray}
where $\alpha_{\Lcp}$ is the decay asymmetry parameter of $\LcToLamHp$, and $\theta_{\Lambda}$ is the angle between the proton momentum and the direction opposite the $\Lcp$ momentum in the $\Lambda$ rest frame, 
as illustrated in 
the supplemental materials.
For $\LcToSigHp$ decays, considering $\alpha(\Sigma^0\to\gamma\Lambda)$ is zero due to parity conservation for an electromagnetic decay, the differential decay rate  
is given by 
\begin{eqnarray}
\frac{dN}{d\cos\theta_{\Sigma^0}d\cos\theta_{\Lambda}} \propto 1 - \alpha_{\Lcp}\alpha_{-}\cos\theta_{\Sigma^{0}}\cos\theta_{\Lambda}\,, \label{eqn:alpha_LcToSigHp}
\end{eqnarray}
where $\theta_{\Lambda}$ ($\theta_{\Sigma^0}$) is the angle between the proton ($\Lambda$) momentum and the direction opposite the $\Sigma^0$ ($\Lcp$) momentum in the $\Lambda$ ($\Sigma^0$) rest frame, as illustrated in 
the supplemental materials.

\section{Event selection and optimization}
The $h^+$ candidates from $\Lcp$ decays are selected as follows. 
Charged tracks satisfying $\mathcal{R}(K|\pi)\!=\!{\mathcal{L}_K/(\mathcal{L}_K+\mathcal{L}_{\pi})}\!>\!0.7$ are identified as kaons, while those satisfying $\mathcal{R}(K|\pi)\!<\!0.7$ are identified as pions. 
Here $\mathcal{L}_i$ ($i\!=\!\pi,\,K,\,p$) is the particle identification (PID) likelihood for a given particle hypothesis, which is calculated from the photon yield in the ACC, energy-loss measurements in the CDC, and time-of-flight information from the TOF~\cite{Nakano:2002jw}.
The highly proton-like tracks with $\mathcal{R}(p|K)\!>\!0.8$ and $\mathcal{R}(p|\pi)\!>\!0.8$ are rejected as $h^+$ candidates for signal modes and reference modes, respectively.
To suppress the background from $\Lambda_c^+$ semileptonic decays, tracks that are highly electron-like (${\mathcal{L}_e/(\mathcal{L}_{e}+\mathcal{L}_{\text{non-}e})}\!>\!0.95$) or muon-like (${\mathcal{L}_{\mu}/(\mathcal{L}_{\mu}+\mathcal{L}_{\pi}+\mathcal{L}_{K})}\!>\!0.95$) are rejected. 
The electron and muon likelihoods depend primarily on the information from the ECL and KLM, respectively~\cite{Hanagaki:2001fz,Abashian:2002bd}.
The signal efficiency after applying PID requirements is $83\%$ for signal modes and $96\%$ for reference modes. About $44\%$ and $9\%$ of total backgrounds are rejected for signal modes and reference modes, respectively. 
We require the $h^+$ candidates to have at least two hits in the SVD to improve their impact parameter resolution with respect to the interaction point.

The $\Lambda$ candidates are reconstructed from one $p$ and one $\pi$ candidate, which a fit requires to originate from a common vertex. 
We require $|M_{\Lambda}-m_{\Lambda}|\!<\!3$ MeV/$c^2$, corresponding to approximately 2.5 standard deviations of the $M_{\Lambda}$ resolution. 
Proton candidates are required to have $\mathcal{R}(p|K)\!>\!0.2$. 
To suppress the non-$\Lambda$ background, we calculate the significance of the $\Lambda$ decay length ($L/\sigma_L$), where 
$L$ is the projection of the $\Lambda$ displacement vector, relative to
the production vertex, onto its momentum direction. 
The corresponding uncertainty 
$\sigma_L$ is calculated by propagating uncertainties in 
the vertices and the $\Lambda$ momentum, including their correlations. 
We require $L/\sigma_L\!>\!4$ to suppress the non-$\Lambda$ background. 
The signal efficiency loss due to this requirement is 5\% for all decay modes and the background rejection rate is 22\% for$\LcToLamKp$, 35\% for $\LcToLamPip$, 19\% for $\LcToSigKp$ and 23\% for $\LcToSigPip$. 

Photon candidates are identified as energy clusters in the ECL that are not associated with any charged track.
The ratio of the energy deposited in the 3$\times$3 array of crystals centered on the crystal with the highest energy to the energy deposited in the corresponding 5$\times$5 array is required to be greater than 0.85.
Candidate $\Sigma^0\to\Lambda\gamma$ decays are formed by combining the $\Lambda$ candidate with a photon candidate that has an ECL cluster energy above 0.1 GeV. 
The $\Sigma^0$ candidate is required to have $|M(\Sigma^0)-m_{\Sigma^0}|\!<\!6$ MeV/$c^2$, corresponding to 1.5 standard deviations of the $M(\Sigma^0)$ resolution.

Candidate $\LcToLamHp$ and $\LcToSigHp$ decays are reconstructed by combining $\Lambda$ or $\Sigma^0$ candidate with a $h^+$ candidate. 
A fit constrains the $\Lambda$ and $h^+$ candidates to originate from a common vertex and the $\chi^2$ of the fit is required to be less than 9.
To suppress combinatorial backgrounds, the normalized momentum $x_p\!=\!p^{*}c/\sqrt{s/4-M^2(\Lcp)\cdot{c}^{4}}$ is required to be greater than 0.5, where $p^{*}$ is the $\Lcp$ momentum in $\epem$ center-of-mass frame and $\sqrt{s}$ is the center-of-mass energy. 

We improve the invariant mass resolution by calculating the corrected mass difference wherever the final state includes a hyperon. Taking $\LcToLamHp$ as an example, the corrected mass is $M(\Lcp)\!=\!M_{\Lcp}-M_{\Lambda}+m_{\Lambda}$ where $M_{X}$ is the invariant mass of reconstructed particle $X$ and $m_X$ represents its nominal mass~\cite{bib:PDG2022}.
The event selection criteria above are optimized with a figure-of-merit (FOM), which is defined as $S/\sqrt{S+B}$ where $S$ and $B$ are the expected signal and background yields in the signal region. The signal region is defined as $|M(\Lcp)-m_{\Lcp}|\!<\!15$~MeV/$c^2$, corresponding to 2.5 standard deviations in the $M(\Lcp)$ resolution.

After applying the optimized requirements, the $\Lcp$ candidate multiplicity is greater than one for $1\%$, $7\%$, $7\%$, and $11\%$ of events for $\LcToLamKp,\,\Lambda\pip,\,\Sigma^0\Kp$, and $\Sigma^0\pip$, respectively. 
For modes including a $\Sigma^0$, the multiplicity is predominantly from multiple photons. 
We perform a best candidate selection (BCS) for events with multiple candidates by retaining candidates with the smallest sum of $\chi^2$ from the vertex fits of the $\Lambda$ and $\Lcp$ candidates for $\LcToLamHp$ modes. For $\LcToSigHp$ modes, an additional term given by $(M(\Sigma^0)-m_{\Sigma^0})^2/\sigma_{M}^{2}$ where $\sigma_M\!=\!4$ MeV/$c^2$ is the $\Sigma^0$ mass resolution, is added. 
The BCS has a signal efficiency of 60\% for events with multiple candidates and does not introduce any peaking backgrounds. 

\section{Direct $\CP$ asymmetry}\label{sec:BR}
The signal probability density function (PDF) is described by a sum of three or four asymmetric Gaussian functions for SCS or CF modes, respectively. 
These Gaussian functions share a common mean parameter but have different width parameters. 
For modes that include a $\Sigma^0$, an additional component denoted broken-$\Sigma^0$ signal, which is the signal decay but with the $\gamma$ in $\Sigma^0\to\Lambda\gamma$ replaced by a random photon in the event, is added into the signal and its shape and ratio to the total signal are fixed according to the results of a fit to the MC sample. Such ratio is $16.2\%$ in $\LcToSigKp$ and $15.5\%$ in $\LcToSigPip$ and the shape is shown in the supplemental materials. 
The signal parameters are fixed to the fitted results of truth-matched signal, but with a common shift ($\delta_\mu$) for the mean parameter and a common scaling factor ($k_{\sigma}$) for all width parameters to account for discrepancies between the experimental data and simulated samples. 

The background PDF is constructed from a sum of empirical shapes based on truth-matched background events in simulation and a second-order polynomial function for $\LcToLamKp$ or a third-order polynomial for the other modes. 
For $\LcToLamKp$, the empirical backgrounds include $\LcToLamPip$ decays with the $\pip$ misidentified as a $\Kp$, a feed-down background from $\LcToSigKp$ with a missing $\gamma$, and a wide enhancement of $\LcToSigPip$ with a misidentified $\pip$ and a missing $\gamma$.
For $\LcToLamPip$, the empirical backgrounds include a feed-down background from $\LcToSigPip$, and a feed-down $\Xi_c$ background from $\Xi_c^{0,+}\to\Xi^{-,0}\pip$ where $\Xi^{-,0}\to\Lambda\pi^{-,0}$ with one missing pion.
For $\LcToSigKp$, the empirical backgrounds include a background from $\LcToSigPip$ with a misidentified $\pip$ and a feed-down background from $\Lcp\to\Xi^0\Kp$ where $\Xi^0\to\Lambda\piz,\,\piz\to\gamma\gamma$ with one missing photon. 
For $\LcToSigPip$, the empirical backgrounds include a reflection background from $\LcToLamPip$ where $\Lambda$ is combined with a random $\gamma$ to form fake $\Sigma^0$ candidate.
The yields of each component and the parameters of the polynomial functions are floated to account for discrepancies between the experimental data and simulated samples.

\begin{figure*}
  \begin{centering}%
  \begin{overpic}[width=0.48\textwidth]{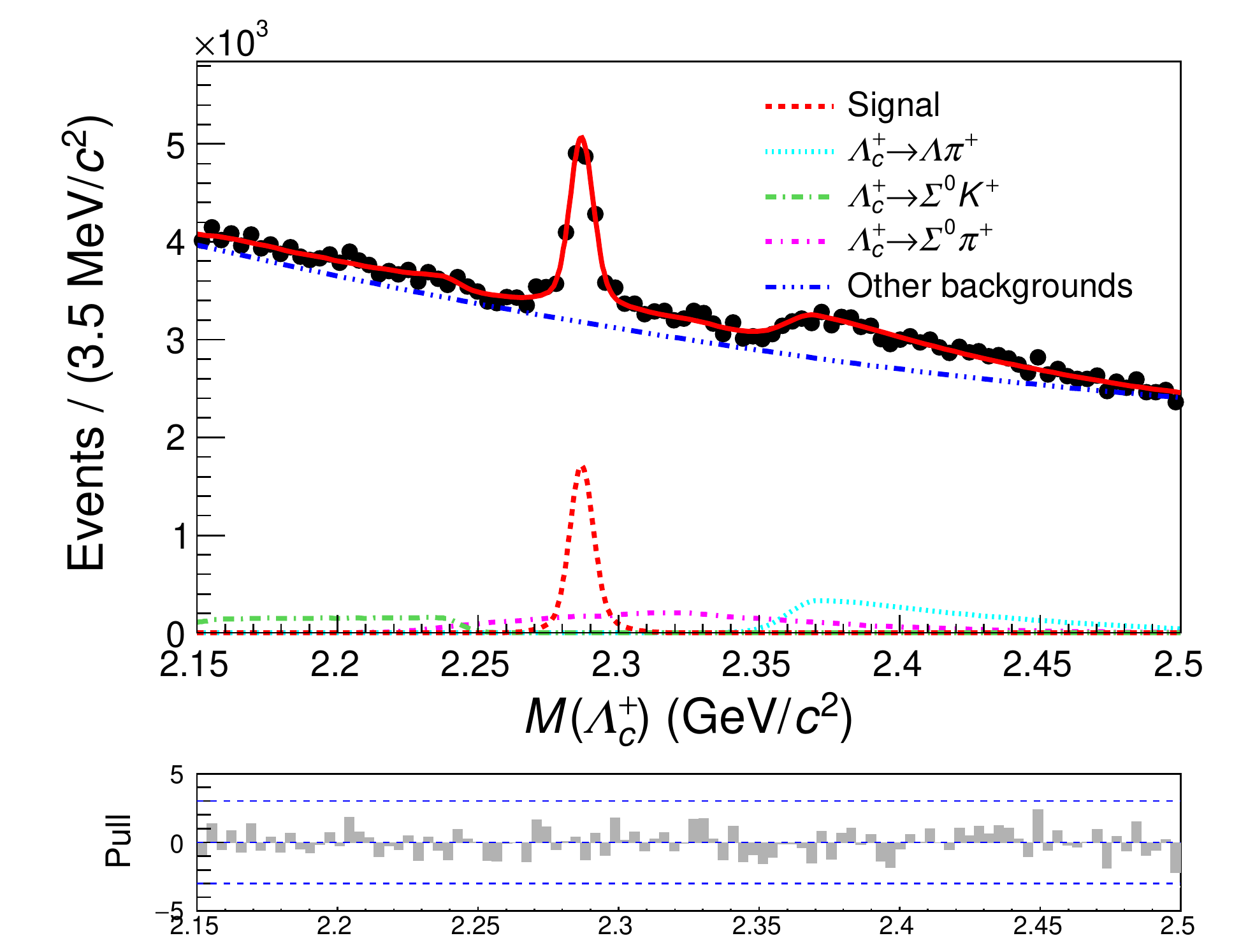}%
  \put(20,65){$\LcToLamKp$}
  \end{overpic}%
  \begin{overpic}[width=0.48\textwidth]{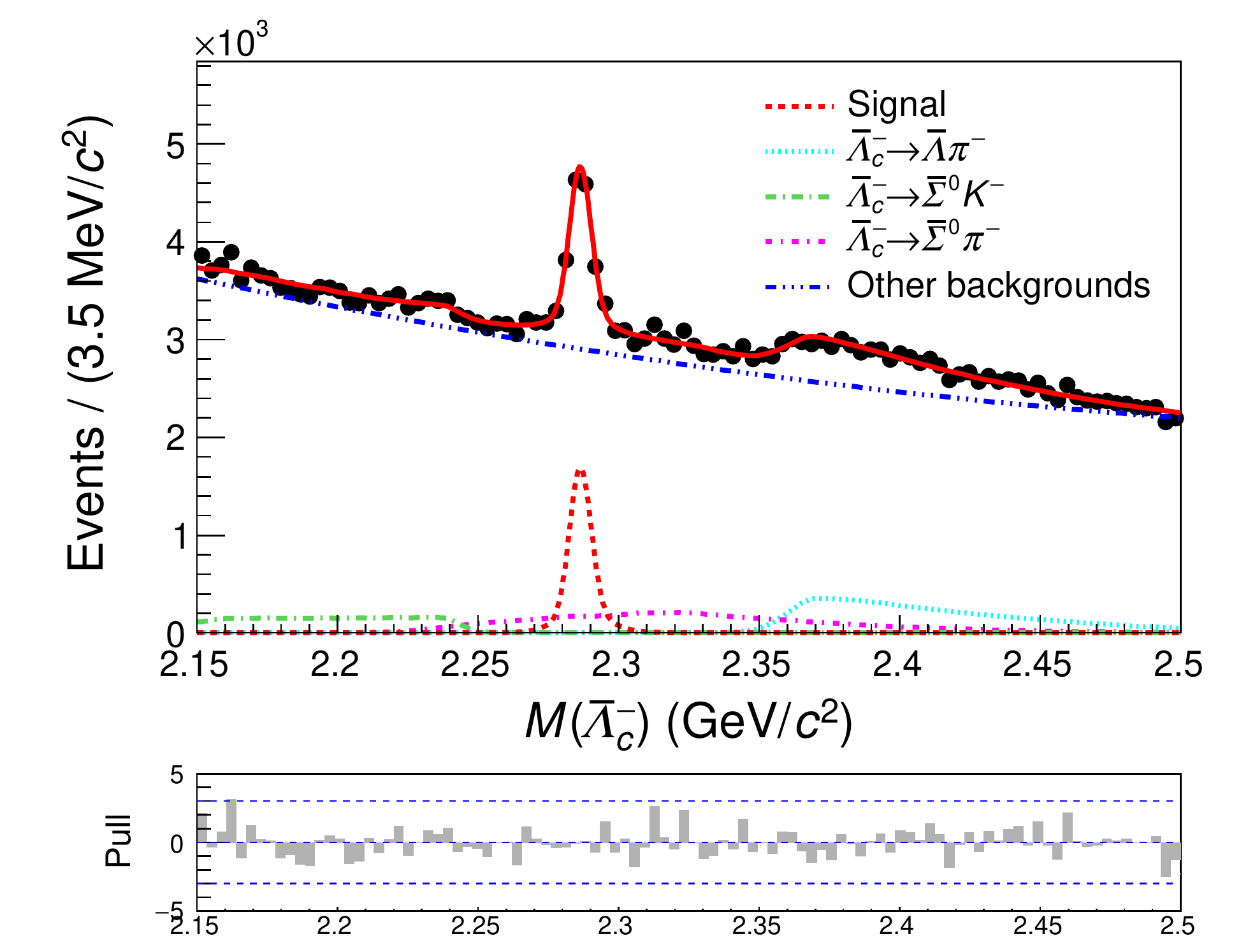}%
  \put(20,65){$\LcmToLamKp$}
  \end{overpic}\\
  \begin{overpic}[width=0.48\textwidth]{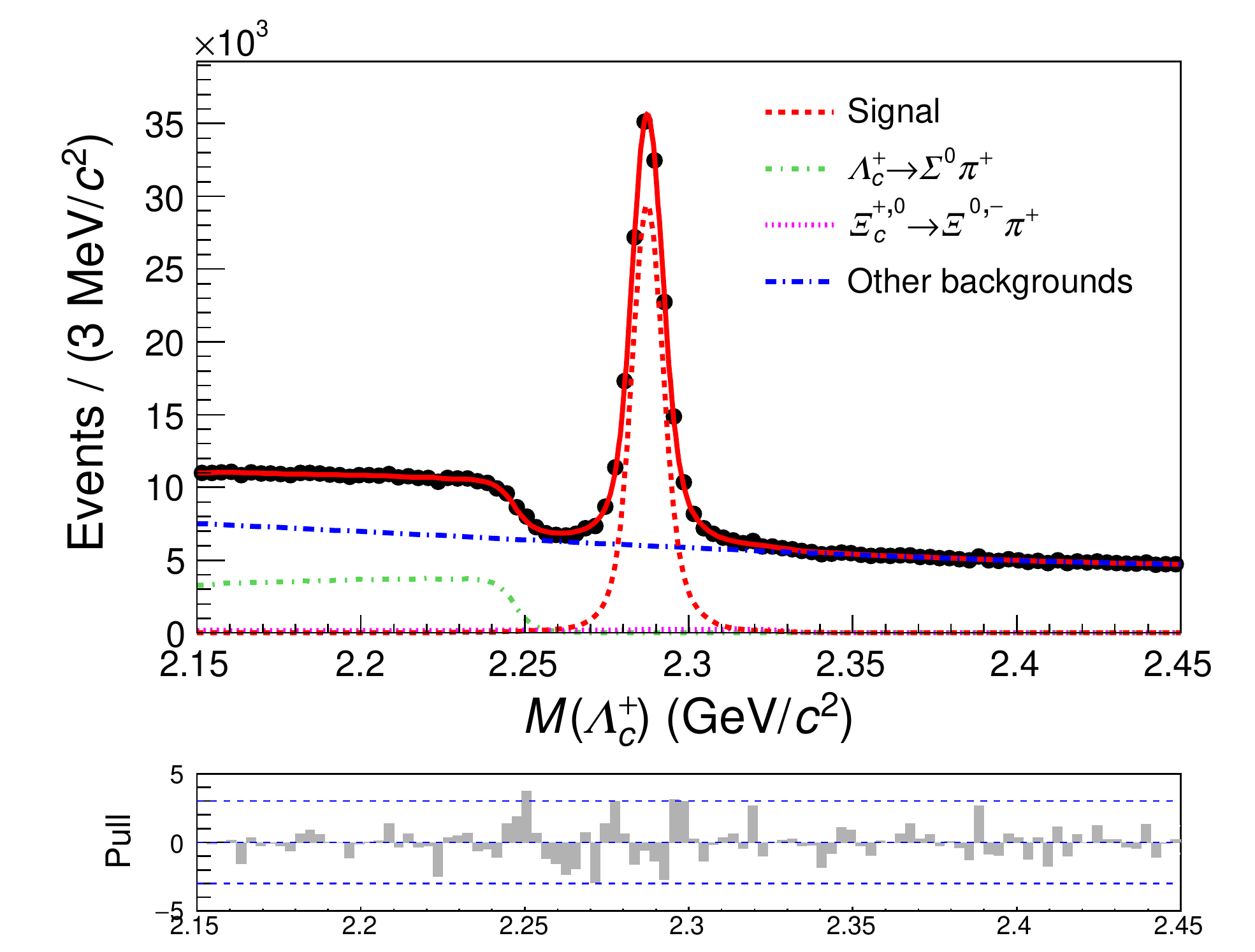}%
  \put(20,65){$\LcToLamPip$}
  \end{overpic}%
  \begin{overpic}[width=0.48\textwidth]{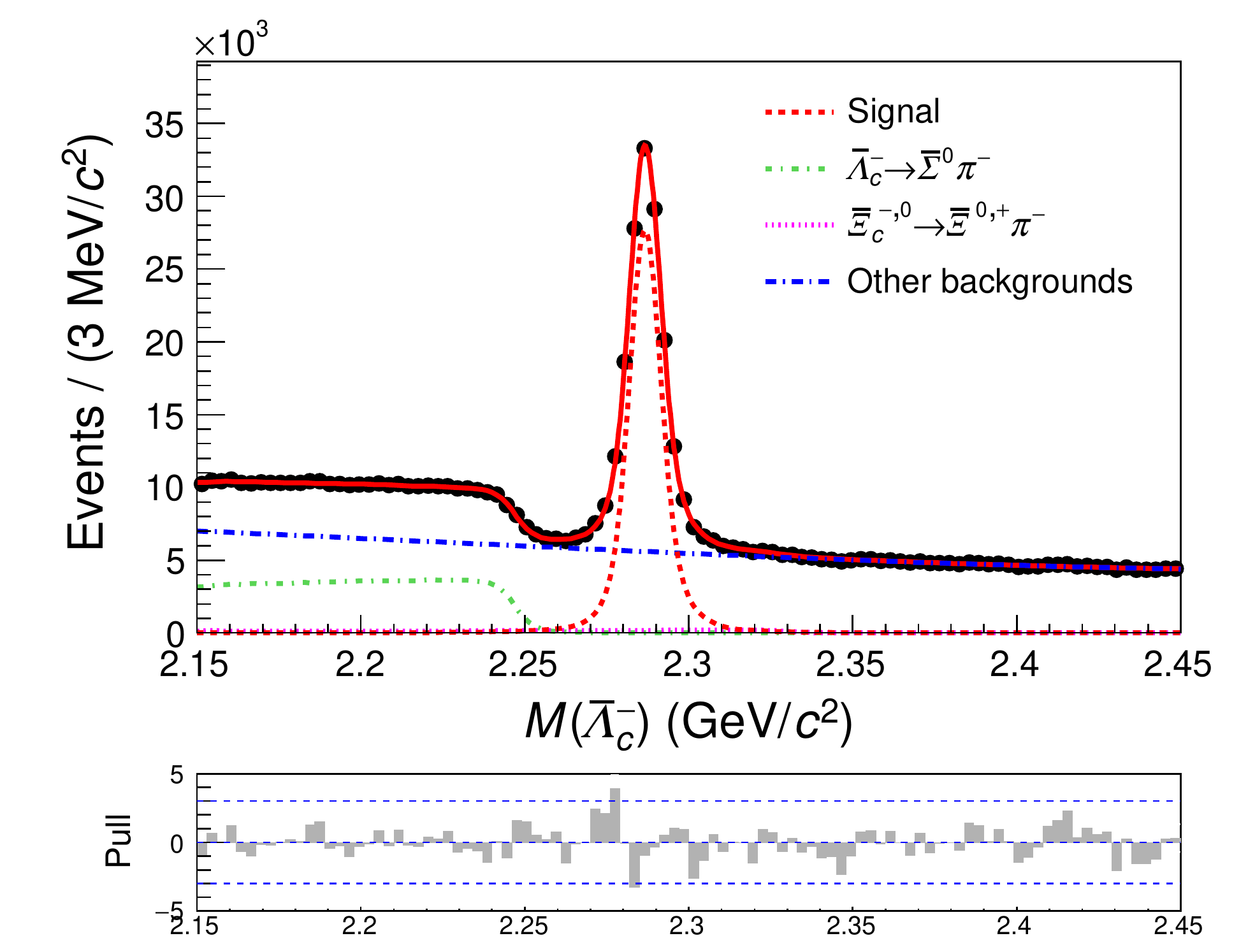}%
  \put(20,65){$\LcmToLamPip$}
  \end{overpic}
  \vskip-10pt
  \caption{\label{fig:CPasym_Final1}The simultaneous fit to $\Lcp$ (left) and $\Lcm$ (right) samples from real data for $\LcToLamKp$ (top) and $\LcToLamPip$ (bottom). The red curve is the total fitting result. The dashed lines show the components of signal and backgrounds (see text).}
  \end{centering}
\end{figure*}

\begin{figure*}
  \begin{centering}%
  \begin{overpic}[width=0.48\textwidth]{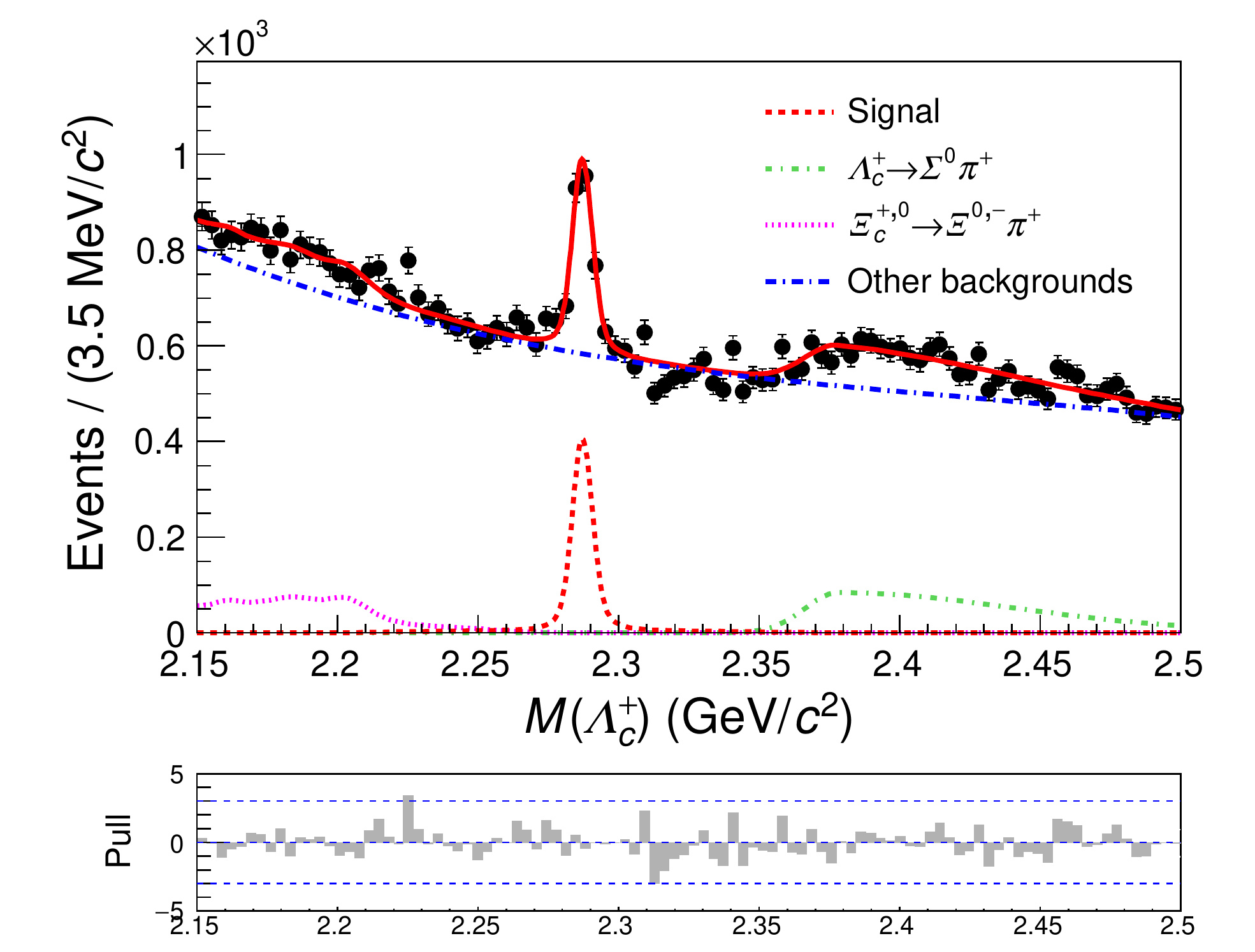}%
  \put(20,65){$\LcToSigKp$}
  \end{overpic}%
  \begin{overpic}[width=0.48\textwidth]{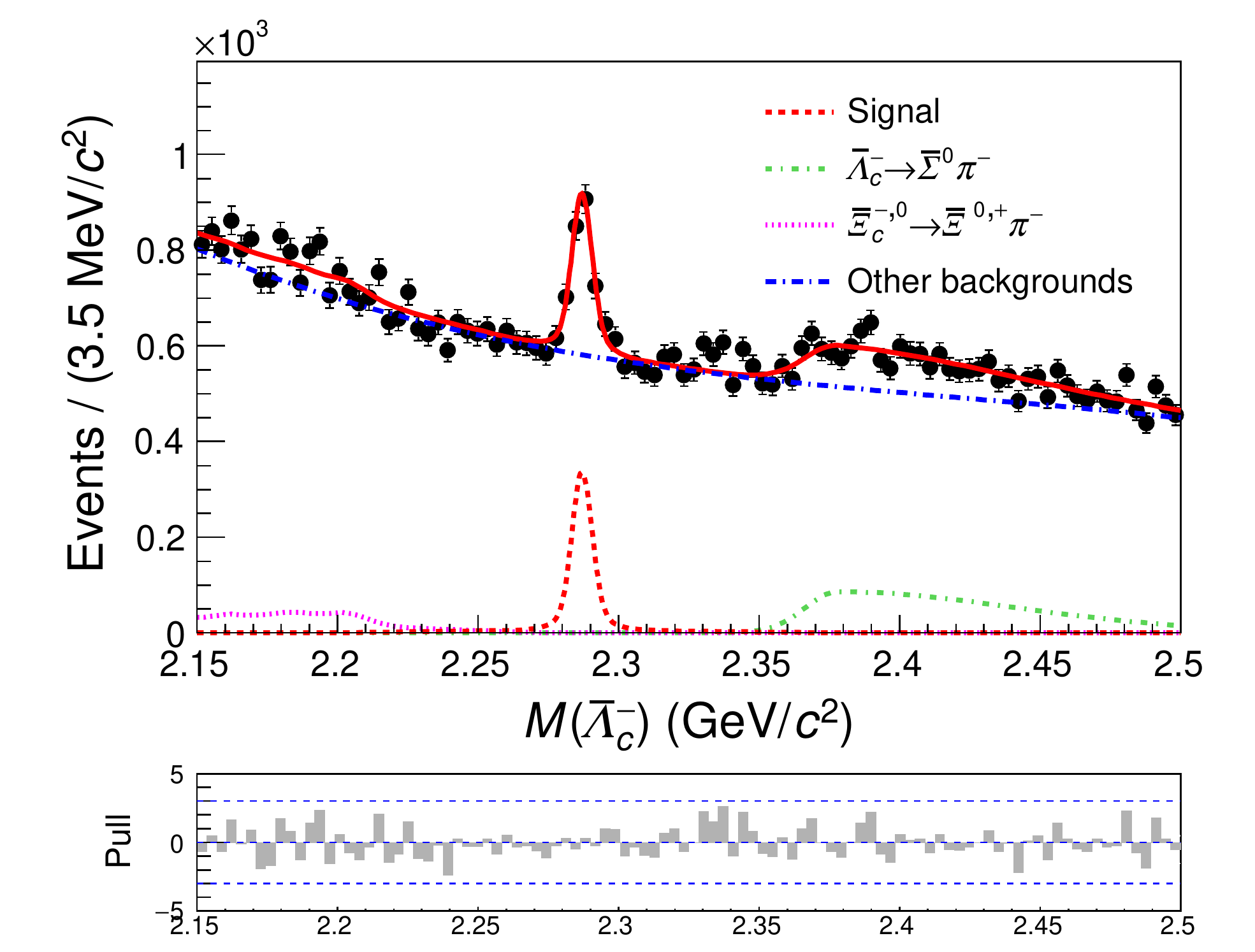}%
  \put(20,65){$\LcmToSigKp$}
  \end{overpic}\\
  \begin{overpic}[width=0.48\textwidth]{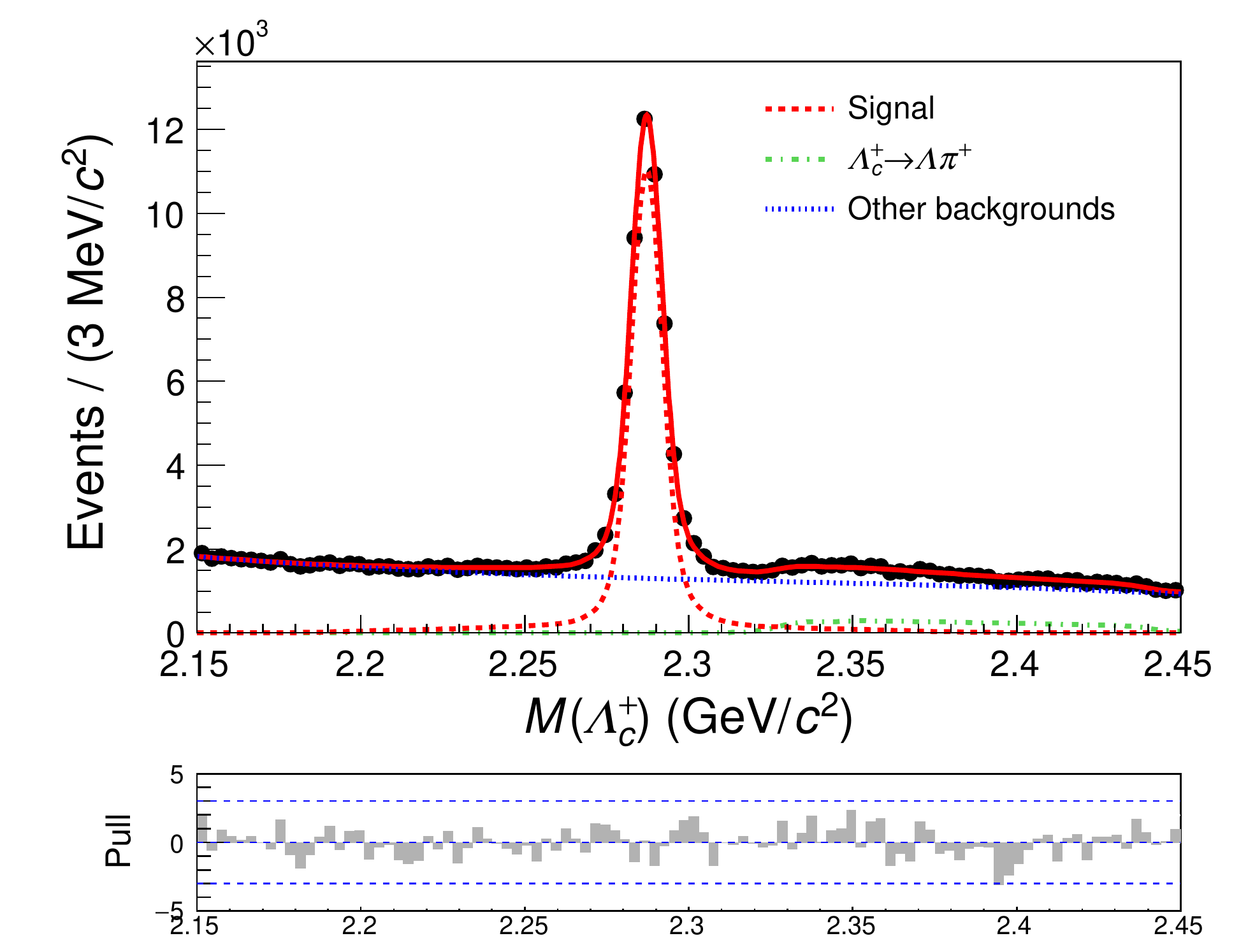}%
   \put(20,65){$\LcToSigPip$}
 \end{overpic}%
  \begin{overpic}[width=0.48\textwidth]{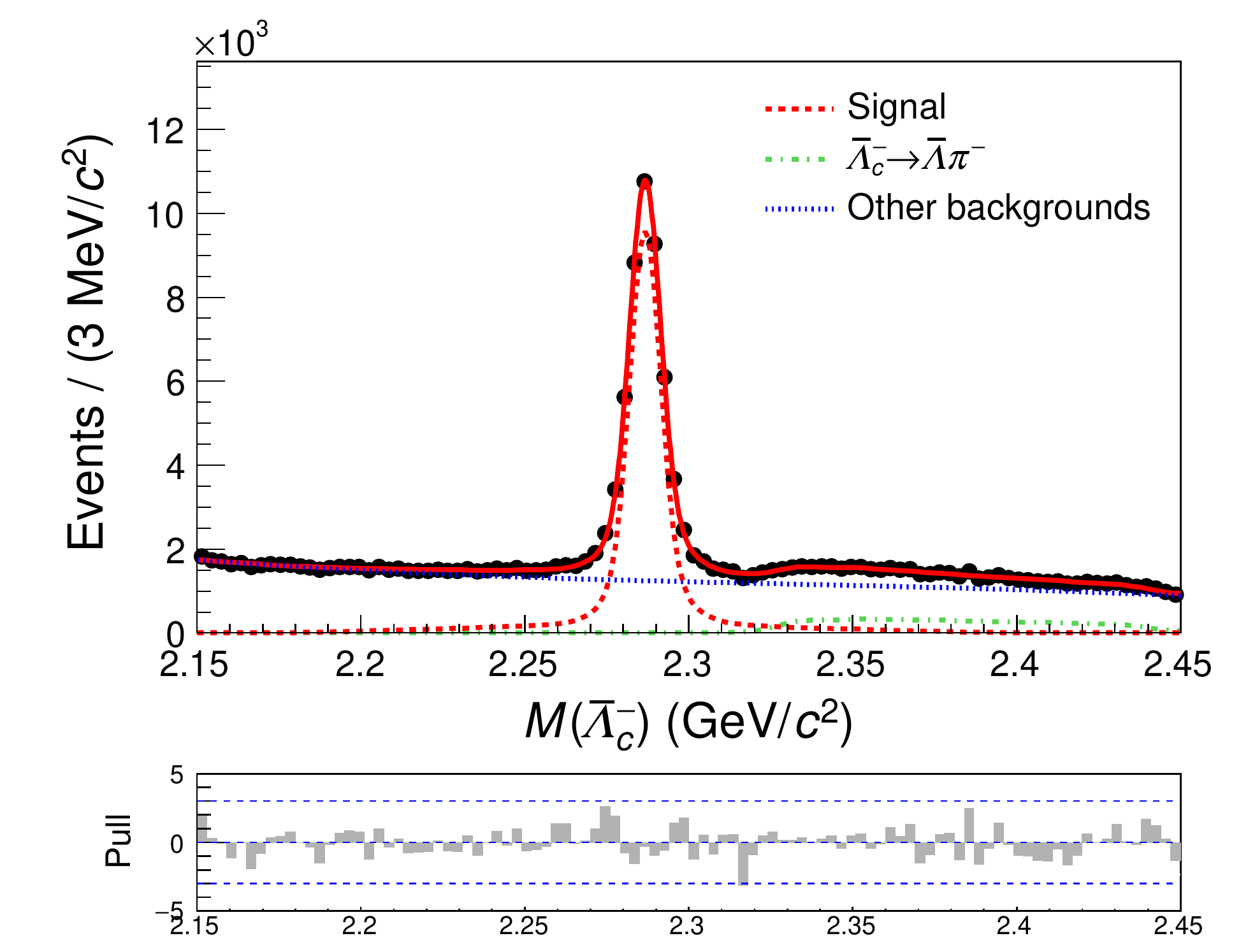}%
  \put(20,65){$\LcmToSigPip$}
  \end{overpic}    
  \vskip-10pt
  \caption{\label{fig:CPasym_Final2}The simultaneous fit to $\Lcp$ (left) and $\Lcm$ (right) samples from real data for $\LcToSigKp$ (top) and $\LcToSigPip$ (bottom). The red curve is the total fitting result. The dashed lines show the components of signal and backgrounds (see text).}
  \end{centering}
\end{figure*}

We perform an unbinned extended maximum likelihood fit on the $M(\Lambda_c^{\pm})$ distributions of the weighted $\Lcp$ and $\Lcm$ samples simultaneously to measure the corrected raw asymmetries. 
In the fit, the mass resolution of $\Lcp$ and $\Lcm$ are allowed to differ.
The fractions of broken-$\Sigma^0$ signal are fixed to those for the $\Lcp$ and $\Lcm$ MC samples, separately.
The fit projections are shown in Fig.~\ref{fig:CPasym_Final1} for $\LcToLamHp$ and in Fig.~\ref{fig:CPasym_Final2} for $\LcToSigHp$, along with the distribution of pull values, 
defined as $(N_{\rm data}-N_{\rm fit})/\sigma_{\rm data}$ where $\sigma_{\rm data}$ is the uncertainty on $N_{\rm data}$. 
The fitted $A_{\rm raw}^{\rm corr}$ values with statistical uncertainties)~\footnote{The difference in $A_{\rm raw}^{\rm corr}$ between the two CF modes is mainly due to the efficiency asymmetry of $\Sigma^0\to\Lambda\gamma$ reconstruction (due to the extra fake photons from anti-proton annihilation in the ECL) according to a MC study, but this asymmetry cancels in the $\Acp^{\rm dir}$ measurement.} are 
\begin{eqnarray}
A_{\rm raw}^{\rm corr}(\LcToLamKp) & = & (+3.66 \pm 2.59)\%\,,    \\
A_{\rm raw}^{\rm corr}(\LcToLamPip) & = & (+1.55 \pm 0.30)\%\,,    \\ 
A_{\rm raw}^{\rm corr}(\LcToSigKp) & = & (+7.71 \pm 5.35)\%\,,    \\
A_{\rm raw}^{\rm corr}(\LcToSigPip) & = & (+5.23 \pm 0.40)\%\,.  
\end{eqnarray}
Using Eq.~(\ref{eqn:DeltaAcp}), we measure the $\CP$ asymmetries: 
\begin{eqnarray}
\Acp^{\rm dir}(\LcToLamKp) & = & (+2.1 \pm 2.6 \pm 0.1)\% \,,   \label{eqn:Adir1}\\
\Acp^{\rm dir}(\LcToSigKp)  & = & (+2.5 \pm 5.4 \pm 0.4)\% \,,  \label{eqn:Adir2}  
\end{eqnarray}
where the first uncertainties are statistical and the second are systematic, which are discussed in detail below.
No evidence of charm $\CP$ violation is found. 
This is the first direct $\CP$ asymmetry measurement for SCS two-body decays of charmed baryons. 

For the measurements of $A_{\CP}^{\rm dir}$ described here, as well as the BF, $\alpha$, and $A_{\CP}^{\alpha}$ measurements described later, 
we validated our fitting procedure using simulated samples, along with ``toy" MC samples in which events were generated by sampling from the PDFs that were fit to the data.
In all cases, the fit results were consistent with the input values used to generate events and with correct fit uncertainties.

\section{Branching fraction}
To measure the BF, we perform a fit to the $M(\Lcp)$ distribution for the combined $\Lcp$ and $\Lcm$ sample. The fitted signal yields are listed in Table~\ref{tab:BR}, along with the reconstruction efficiency ratio for the SCS modes relative to the CF modes. 
The efficiency is determined based on signal MC events, which are produced with a special angular distribution using our measured $\alpha$ values. 
An event-by-event correction (typically 0.3\% and 2.8\%) is applied to account for discrepancies in the $K^+$ and $\pi^+$ PID efficiencies between data and simulation. 
These correction factors depend on the momentum and polar angle of tracks and are determined using a sample of $D^{*+}\to[D^0\to\Km\pip]\pip$ decays.
Additional details are given in the supplemental materials.

\begin{table}
\begin{centering}
\caption{\label{tab:BR}The fitted yield ($N_{\rm sig}$), efficiency ($\eff$) ratio, and ratio of branching fractions ($\BR$) for signal modes $\Lcp\to(\Lambda,\,\Sigma^0)K^+$ relative to reference modes $\Lcp\to(\Lambda,\,\Sigma^0)\pi^+$, compared with the world average values (W.A.)~\cite{bib:PDG2022}.}
\begin{lrbox}{\tablebox}
\begin{tabular}{ccccc} \hline  
Channel  		& $N_{\rm sig}$		& $\eff_{\rm sig}/\eff_{\rm ref}$ 	& $\BR_{\rm sig}/\BR_{\rm ref}$	 (\%) & W.A.(\%) \\ \hline   
$\LcToLamKp$	& $\,\,\,11175 \pm 296$	& \multirow{2}{*}{$0.836$} 	&  \multirow{2}{*}{$5.05\pm 0.13\pm 0.09$} & \multirow{2}{*}{$4.7\pm 0.9$}		\\
$\LcToLamPip$	& $264470 \pm 787$	&		& 	&  \\ \hline 
$\LcToSigKp$ 	& $\,\,\,\,\,\,2436 \pm 132$	& \multirow{2}{*}{$0.835$}	&  \multirow{2}{*}{$2.78\pm 0.15\pm 0.05$} & \multirow{2}{*}{$4.0\pm 0.6$}	\\
$\LcToSigPip$	& $105018 \pm 475$	& 			&  & 			\\
\hline  
\end{tabular}
  \end{lrbox}
  \scalebox{0.95}{\usebox{\tablebox}}
\end{centering}  
\end{table} 

Using the fitted yields and efficiency ratios, we calculate the BF ratios according to Eq.~(\ref{eqn:BRratio}) as
\begin{eqnarray}
\frac{\mathcal{B}(\LcToLamKp)}{\mathcal{B}(\LcToLamPip)} & = & (5.05\pm 0.13\pm 0.09)\%\,,  \label{eqn:BR1}\\
\frac{\mathcal{B}(\LcToSigKp)}{\mathcal{B}(\LcToSigPip)} & = & (2.78\pm 0.15\pm 0.05)\%\,,  \label{eqn:BR2} 
\end{eqnarray}
where the first uncertainties are statistical and the second are systematic. Systematic uncertainties are described in detail in Sec.~\ref{sec:sys}. These results are consistent with the recent results from BESIII, $\frac{\mathcal{B}(\LcToLamKp)}{\mathcal{B}(\LcToLamPip)}\!=\!{(4.78\pm0.39)\%}$~\cite{BESIII:2022tnm} within $0.6\sigma$ and $\frac{\mathcal{B}(\LcToSigKp)}{\mathcal{B}(\LcToSigPip)}\!=\!{(3.61\pm0.73)\%}$~\cite{BESIII:2022wxj} within $1.1\sigma$, but with precision improved by threefold and fivefold, respectively.

Multiplying the BF results in Eqs.(\ref{eqn:BR1},\,\ref{eqn:BR2}) by the world average values for the BF of the appropriate reference mode, 
$\mathcal{B}(\LcToLamPip)\!=\!(1.30\pm0.07)\%$ and $\mathcal{B}(\LcToSigPip)\!=\!(1.29\pm0.07)\%$~\cite{bib:PDG2022}, 
we measure the absolute branching fraction for the SCS decays,
\begin{eqnarray}
\BR(\LcToLamKp) = \hskip115pt \quad \nonumber \\
                    \hskip40pt (6.57\pm 0.17\pm 0.11\pm 0.35)\times 10^{-4},  \\
\BR(\LcToSigKp) =  \hskip110pt \quad\nonumber \\
                    \hskip40pt (3.58\pm 0.19\pm 0.06\pm 0.19)\times 10^{-4},   
\end{eqnarray}
where the first uncertainties are statistical, the second are systematic, and the third are from the uncertainties on the BFs for the reference modes. 
These results are consistent with current world average values~\cite{bib:PDG2022}, 
$\BR(\LcToLamKp)\!=\!{(6.1\pm 1.2)\!\times\!10^{-4}}$ within $1\sigma$ and $\BR(\LcToSigKp)\!=\!{(5.2\pm 0.8)\!\times\!10^{-4}}$ within $2\sigma$, 
but with significantly improved precision.

\section{Decay asymmetry parameter $\alpha$}
To extract the $\alpha$ parameter, the $\cos\theta_{\Lambda}$ distributions of $\LcToLamHp$ modes are divided into 10 bins of uniform width. 
The $\cos\theta_{\Sigma^0}$ versus $\cos\theta_{\Lambda}$ distributions for $\LcToSigHp$ modes are similarly divided into 5$\times$5 bins for $\LcToSigKp$ and 6$\times$6 bins for $\LcToSigPip$, since the latter mode has much greater statistics.
To extract the per-bin yield, we fit the $M(\Lcp)$ distribution with signal parameters and background polynomial parameters fixed according to the fit to the full sample integrated over helicity angles.
In the $\LcToSigHp$ modes, the ratio of broken-$\Sigma^0$ signal to total signal depending on the $\cos\theta_{\Sigma^0}$ bin is fixed to the truth-matched results in simulation. 
In the $\LcToSigPip$ mode, the shape of the reflection background $\LcToLamPip$ is found to depend on the $\cos\theta_{\Sigma^0}$ bins and its shape in each bin is fixed to the results from a fit to simulation.

\begin{figure}
  \begin{centering}
   \hskip-10pt 
  \begin{overpic}[width=0.25\textwidth]{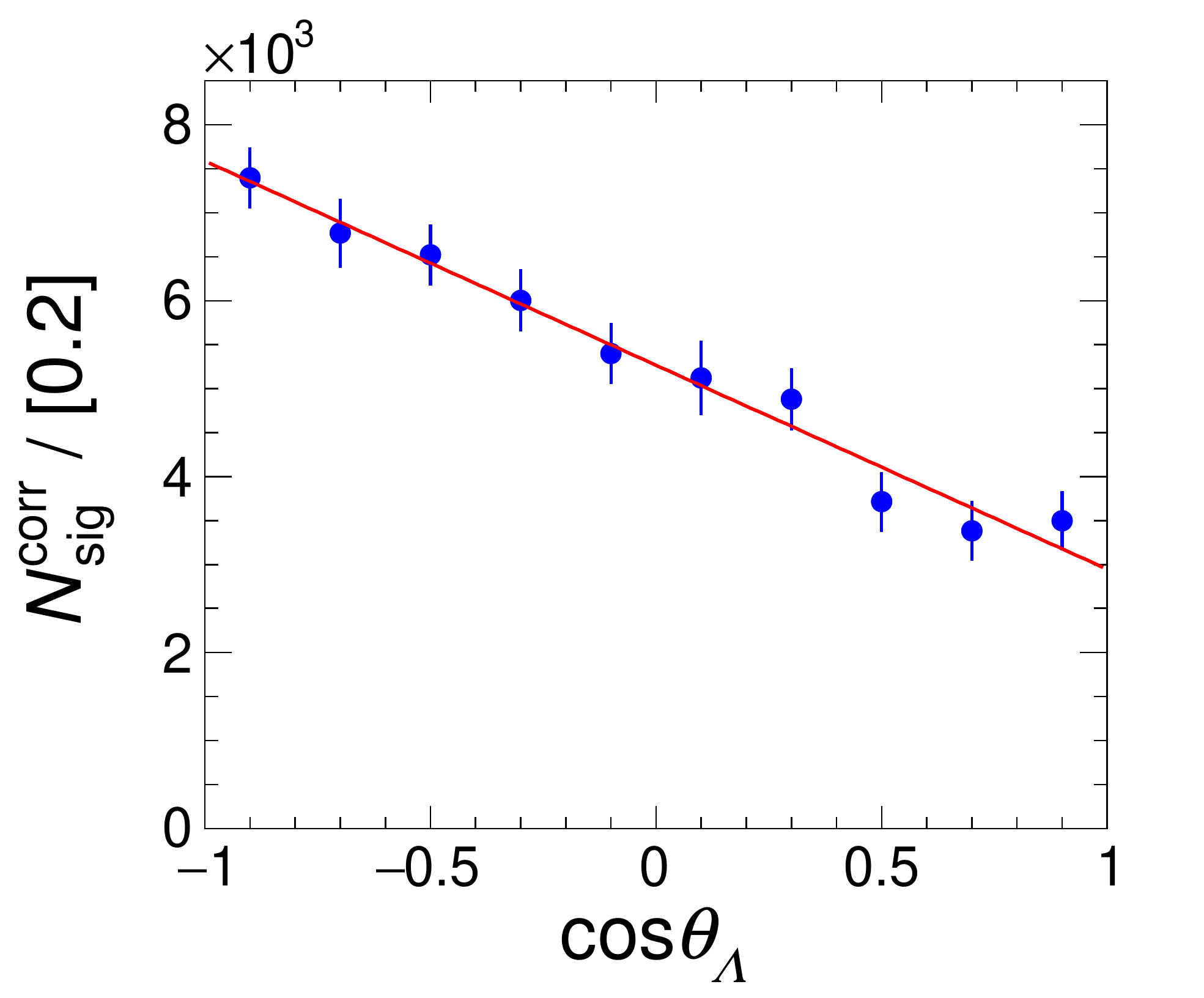}%
  \put(52,65){\small{$\LcToLamKp$}}%
  \end{overpic}%
  \begin{overpic}[width=0.25\textwidth]{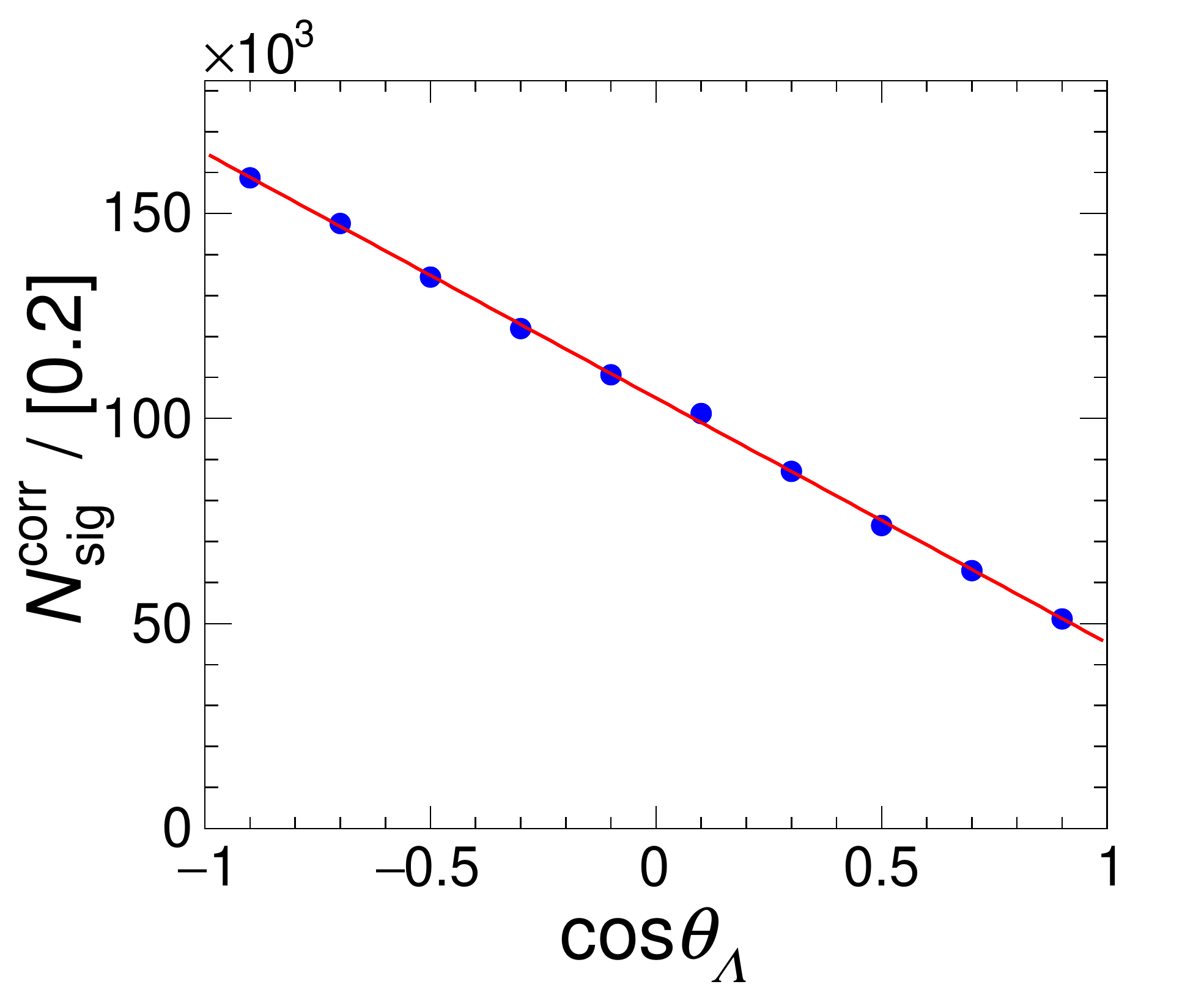}%
  \put(52,65){\small{$\LcToLamPip$}}%
  \end{overpic}%
  \vskip-10pt 
  \caption{\label{fig:AlphaFinal_LcToLamHp}The $\cos\theta_{\Lambda}$ distributions of $\LcToLamKp$ and $\LcToLamPip$ and their conjugated decays after efficiency corrections. The red curves show the fitted results with the $\chi^2$ divided by the number of degree of freedom, $\chi^2/9=0.43$ and 1.05, respectively.}
  \end{centering}
 \end{figure}

\begin{figure*}[!hbtp]
  \begin{centering}
  \hskip-10pt
  \begin{overpic}[width=0.25\textwidth]{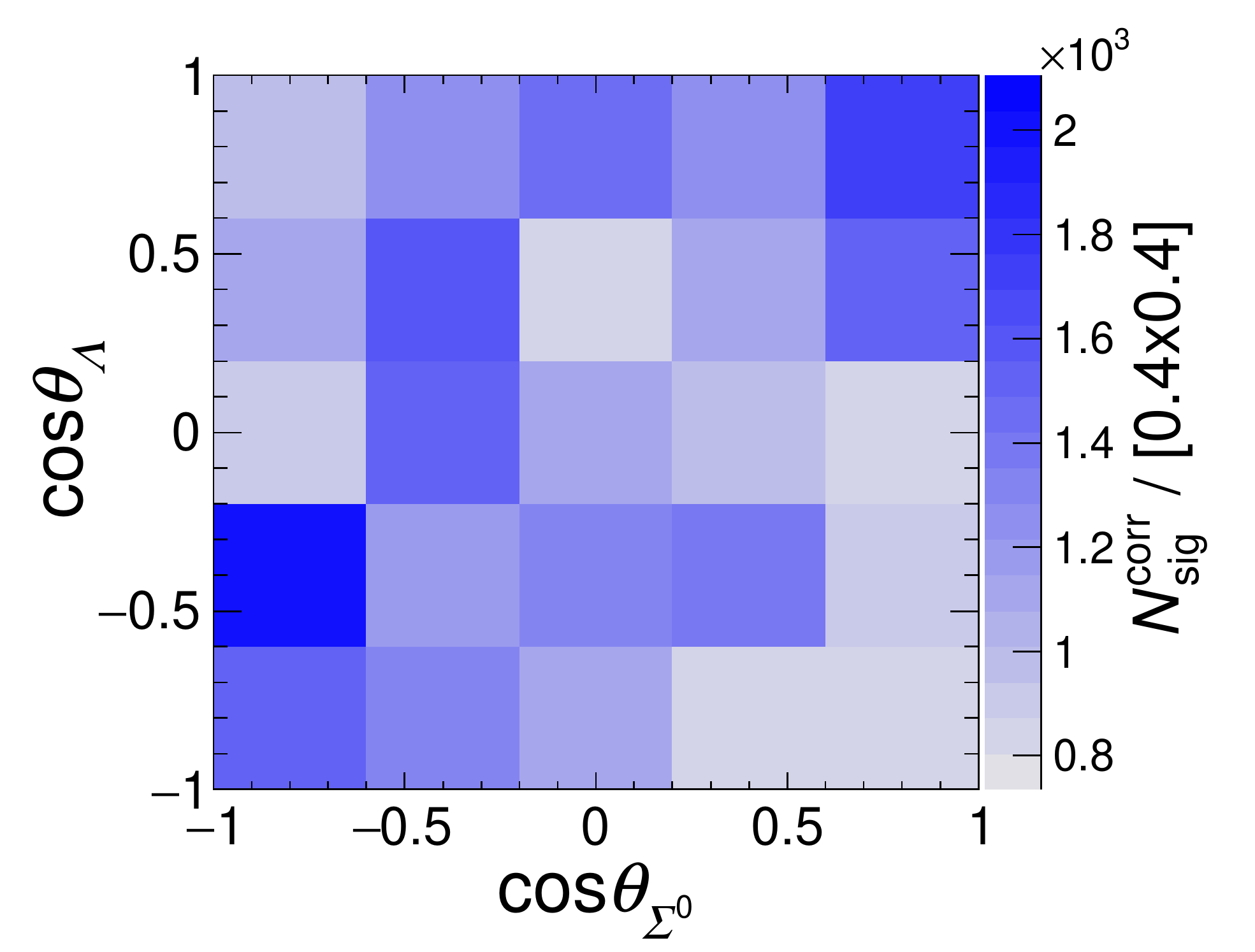}%
  \put(25,73){\small{$\LcToSigKp$}}%
  \end{overpic}%
  \begin{overpic}[width=0.25\textwidth]{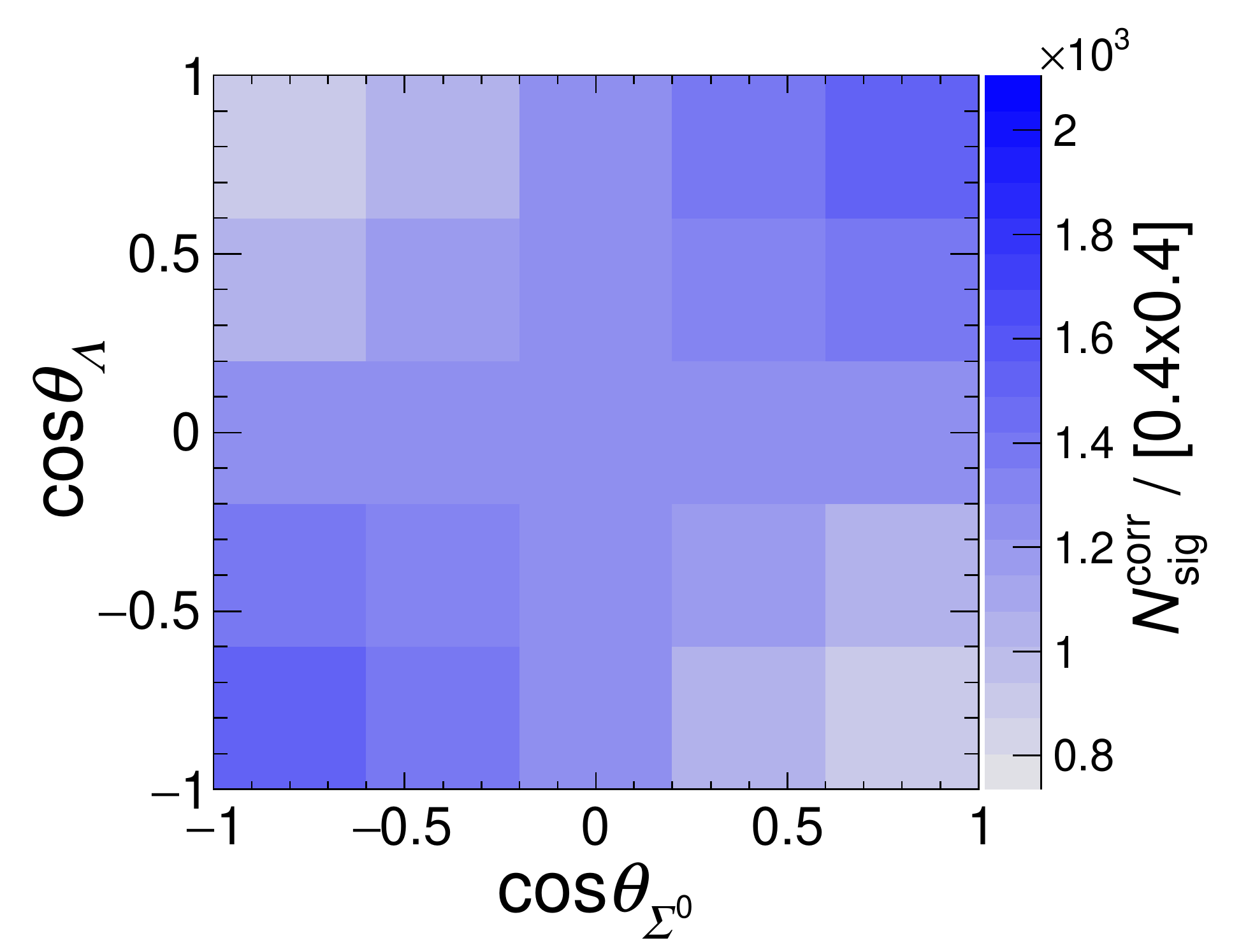}%
  \put(25,73){\small{Fit result}}%
  \end{overpic}
  \begin{overpic}[width=0.25\textwidth]{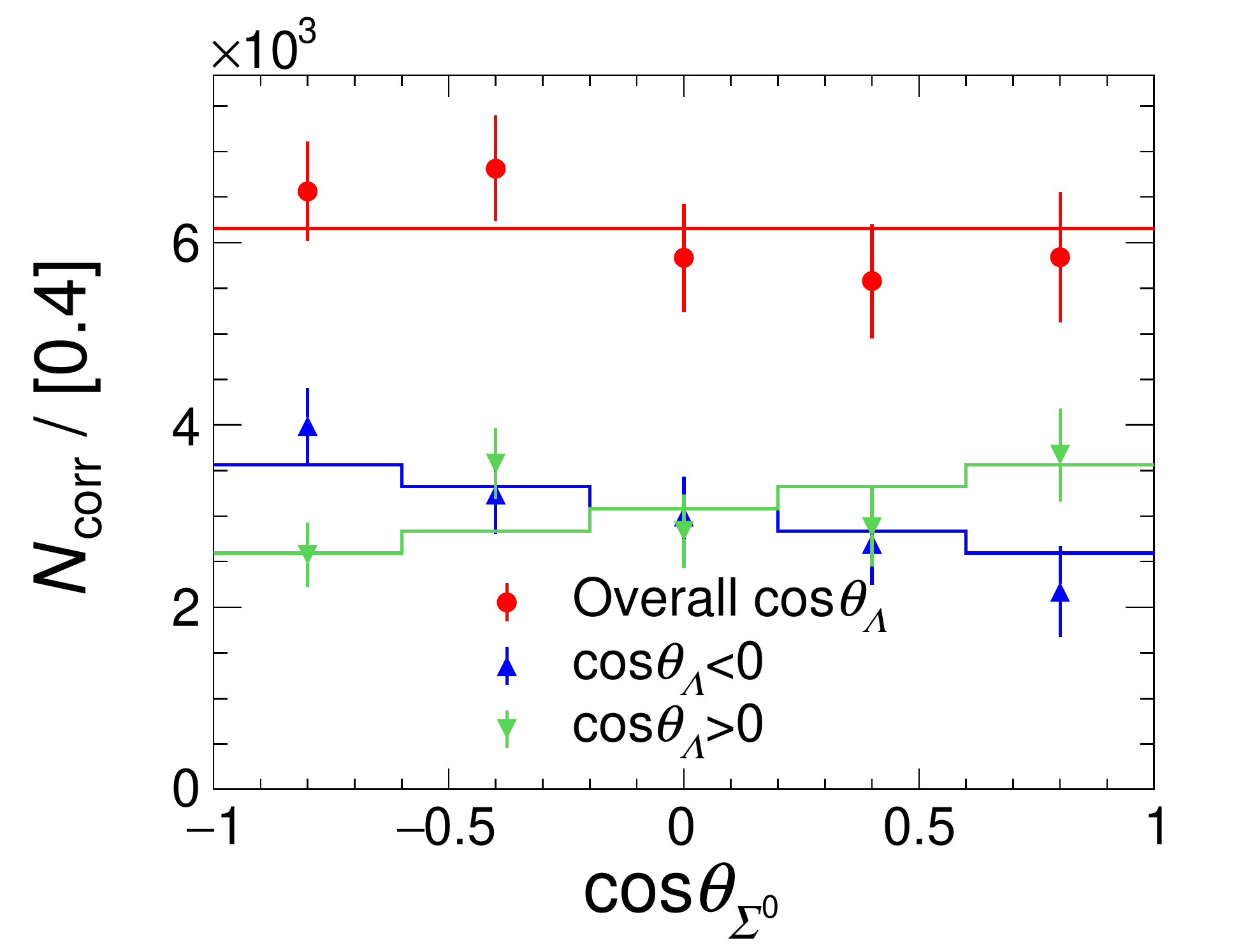}%
  \end{overpic}%
  \begin{overpic}[width=0.25\textwidth]{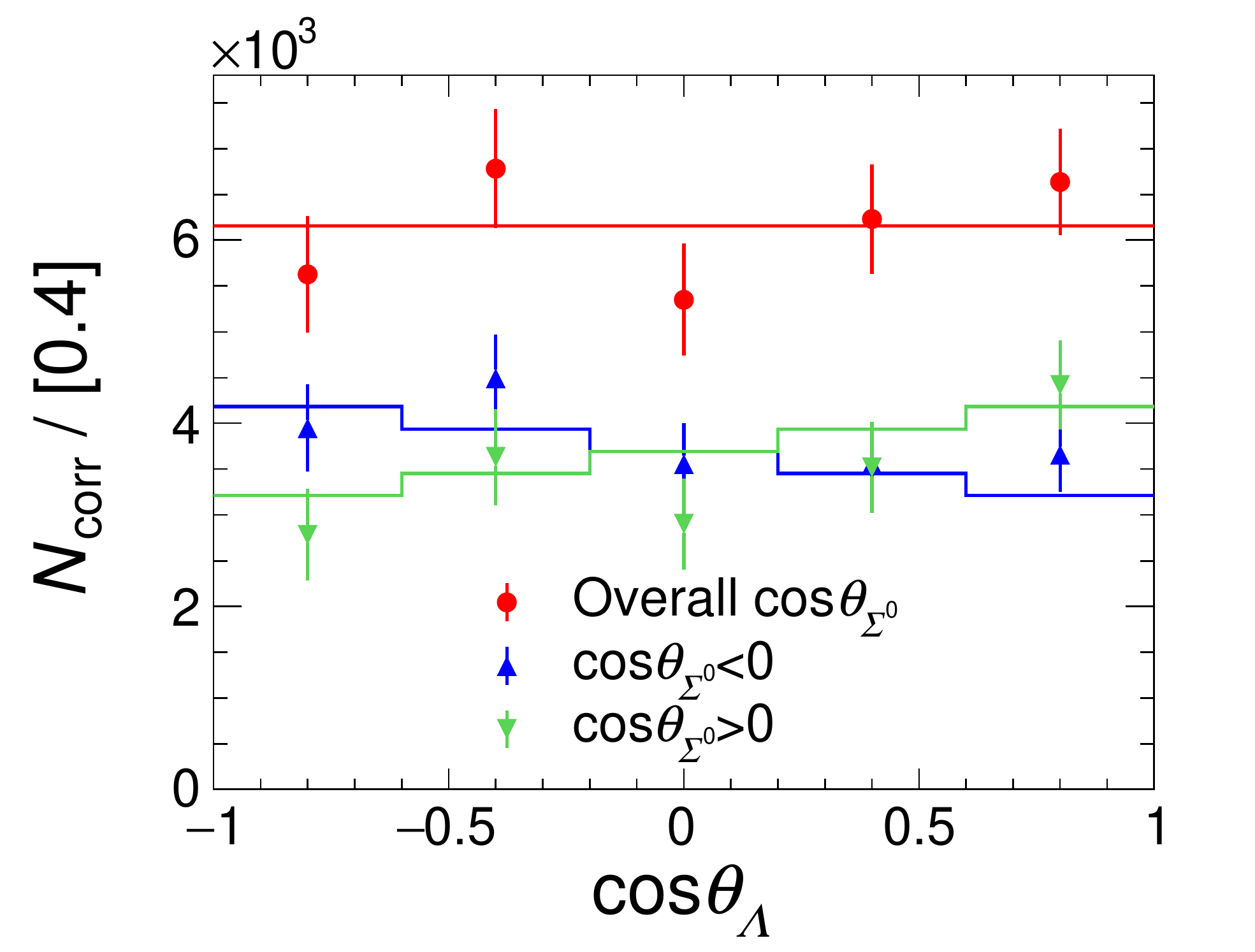}%
  \end{overpic}\\
  \vskip8pt   
  \hskip-10pt
  \begin{overpic}[width=0.25\textwidth]{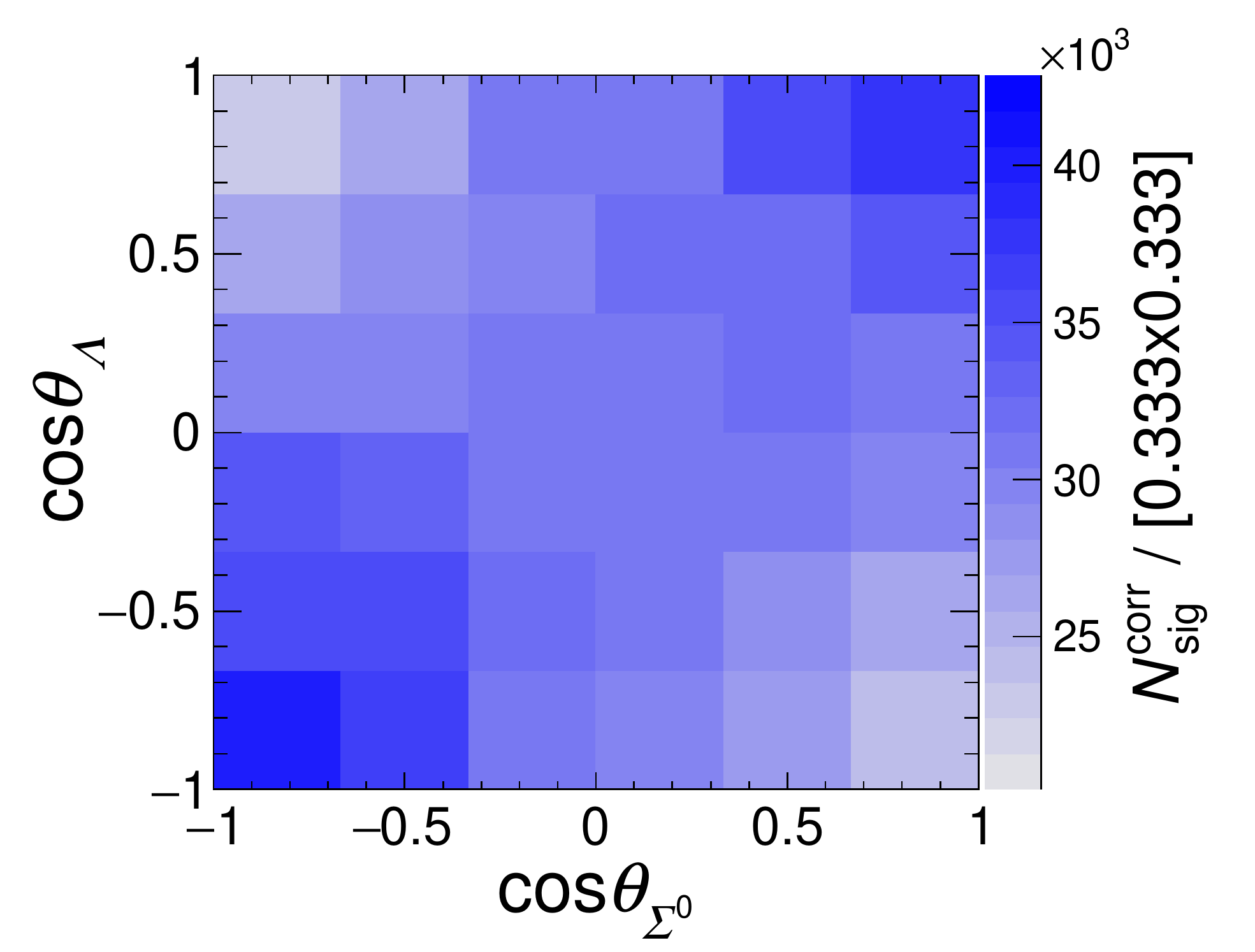}%
  \put(25,73){\small{$\LcToSigPip$}}%
  \end{overpic}%
  \begin{overpic}[width=0.25\textwidth]{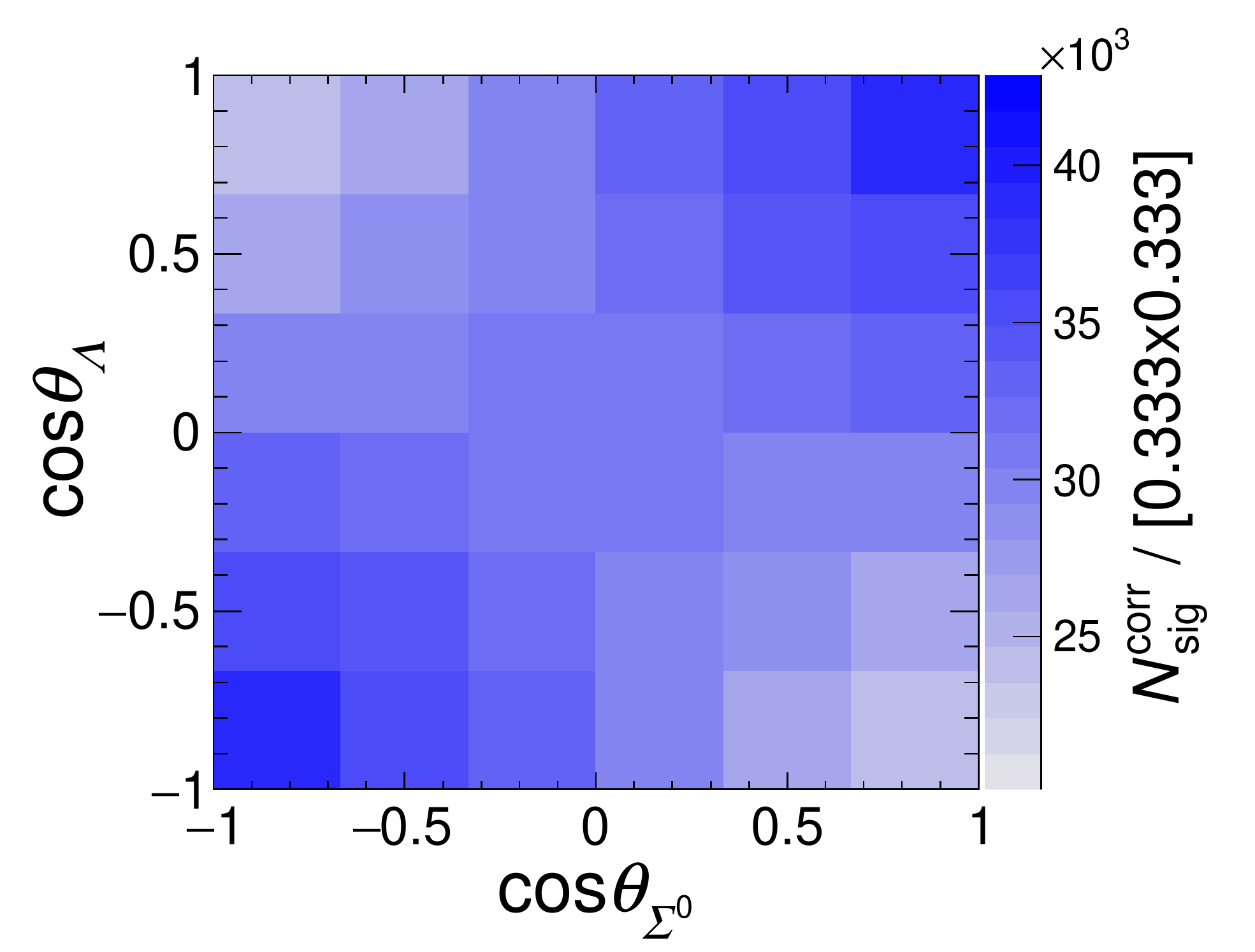}%
  \put(25,73){\small{Fit result}}%
  \end{overpic}%
  \begin{overpic}[width=0.25\textwidth]{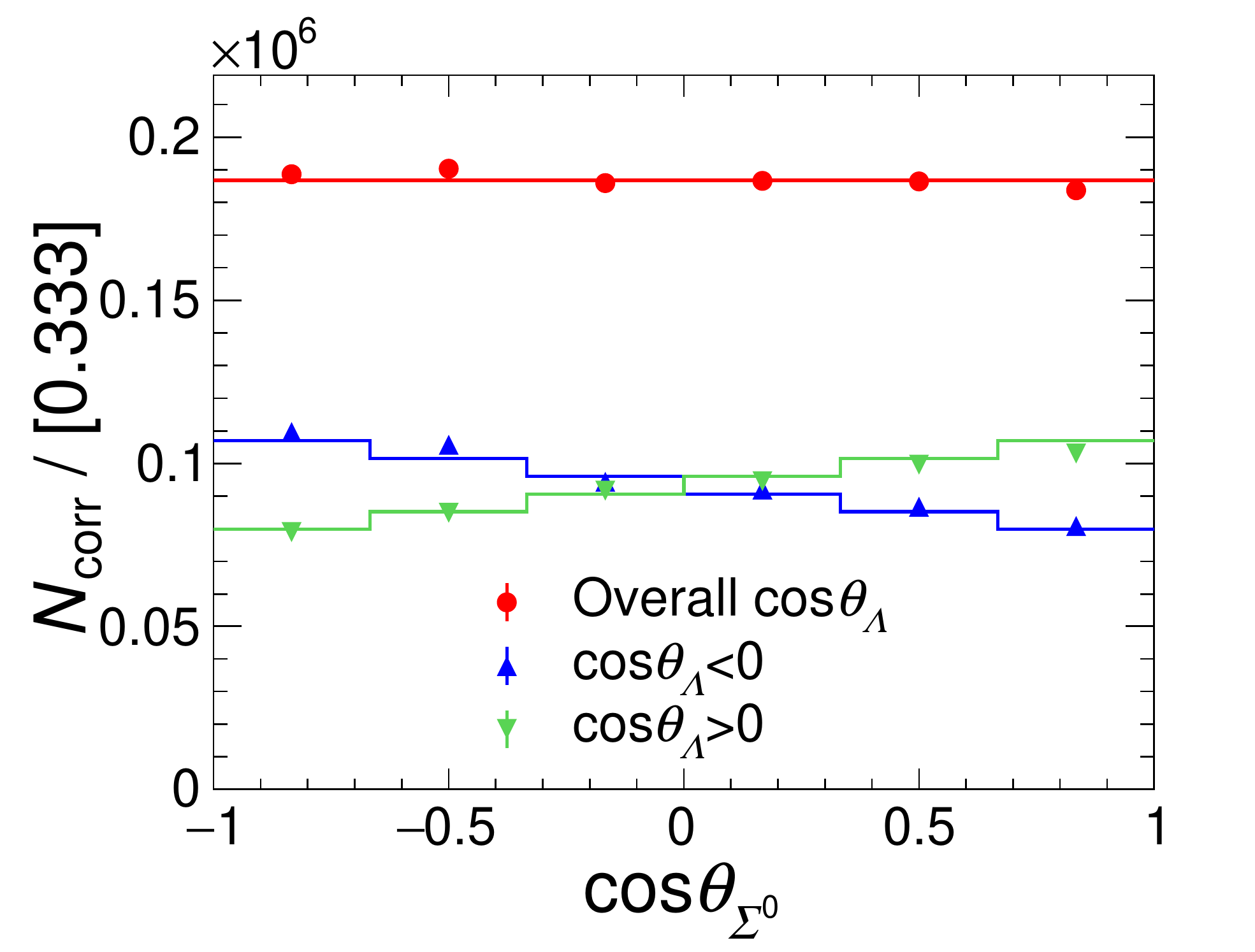}%
  \end{overpic}%
  \begin{overpic}[width=0.25\textwidth]{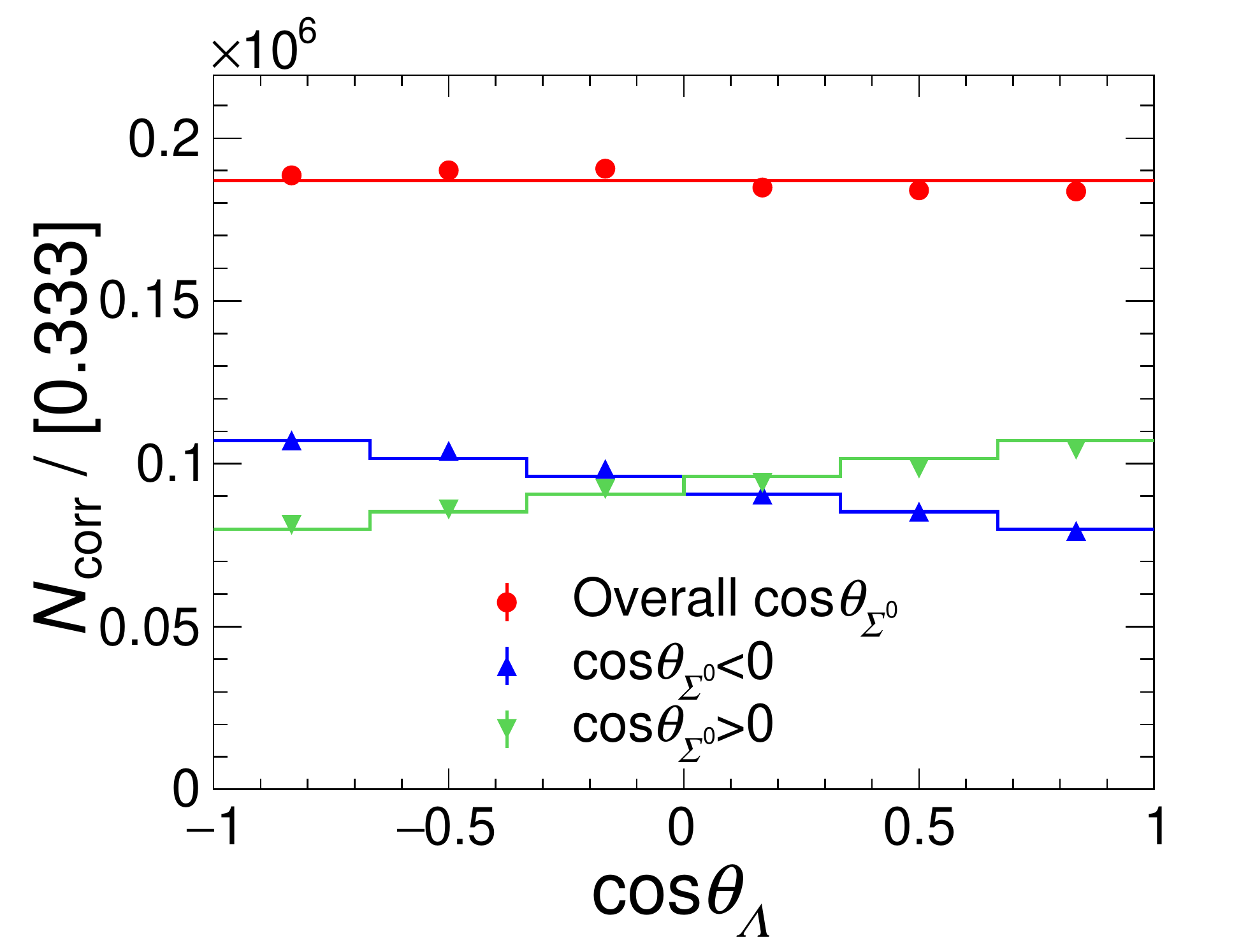}%
  \end{overpic}%
  \vskip-10pt 
  \caption{\label{fig:AlphaFinal_LcToSigHp}The first column shows the $[\cos\theta_{\Sigma^0},\,\cos\theta_{\Lambda}]$ distributions of $\LcToSigKp$ and $\LcToSigPip$ and their conjugated decays after efficiency correction; the second column shows the fitted results of the first column with the $\chi^2$ divided by the number of degree of freedom $\chi^2/24=0.87$ for $\LcToSigKp$ and $\chi^2/35=1.45$ for $\LcToSigPip$. The third column shows the projections of the $\cos\theta_{\Sigma^0}$ distributions (point with error) and the fit results (histograms) in overall (red) or negative (blue) or positive (green) $\cos\theta_{\Lambda}$ region; vice versa in fourth column. The absolute slopes of all projections in slices equal half of the fitted slope mentioned in text.}
  \end{centering}
 \end{figure*}

The fitted signal yields are corrected bin-by-bin with the signal efficiencies,
which are determined based on signal MC events produced with our measured angular distribution. 
Here the efficiency correction has effectively included the resolution of helicity angles because the efficiencies are calculated by the ratios between the reconstructed signals in $i$-th bin of the cosine of reconstructed helicity angles and the generated signals in $i$-th bin of the cosine of helicity angles.
These distributions are fitted according to Eqs.~(\ref{eqn:alpha_LcToLamHp},\,\ref{eqn:alpha_LcToSigHp}) 
and the fit results are shown in Fig.~\ref{fig:AlphaFinal_LcToLamHp} for $\LcToLamHp$ and Fig.~\ref{fig:AlphaFinal_LcToSigHp} for $\LcToSigHp$.
The fitted slope factors ($\alpha_{\Lcp}\alpha_{\Lambda}$) are 
\begin{eqnarray}  
\alpha_{\Lcp}^{\rm avg}(\LcToLamKp)\cdot\alpha_{\Lambda}^{\rm avg} & = & -0.441\pm 0.037\,, \quad \\ 
\alpha_{\Lcp}^{\rm avg}(\LcToLamPip)\cdot\alpha_{\Lambda}^{\rm avg} & = & -0.570\pm 0.004\,,  \\ 
\alpha_{\Lcp}^{\rm avg}(\LcToSigKp)\cdot\alpha_{\Lambda}^{\rm avg} & = & -0.41\phantom{0}\pm 0.14\phantom{0}\,,  \\ 
\alpha_{\Lcp}^{\rm avg}(\LcToSigPip)\cdot\alpha_{\Lambda}^{\rm avg} & = & -0.349\pm 0.012\,,  
\end{eqnarray}
where only statistical uncertainties are given. The superscript `avg' denotes the averaged $\alpha$ value for the combined $\Lcp$ ($\Lambda$) and $\Lcm$ ($\Lbar$) decays.
Dividing these results by the most precise
$\alpha_{\Lambda}^{\rm avg}\!=\!0.7542\pm 0.0026$ 
from BESIII~\cite{BESIII:2022qax} gives 
the final decay asymmetry parameters $\alpha_{\Lcp}^{\rm avg}$ for the combined $\Lcp$ and $\Lcm$ sample,
\begin{eqnarray}  
\alpha_{\Lcp}^{\rm avg}(\LcToLamKp) & = & -0.585\pm 0.049 \pm 0.018\,,\quad   \\
\alpha_{\Lcp}^{\rm avg}(\LcToLamPip) & = & -0.755\pm 0.005 \pm 0.003\,,  \\
\alpha_{\Lcp}^{\rm avg}(\LcToSigKp) & = & -0.54\phantom{0} \pm 0.18\phantom{0}\pm 0.09\,,   \\
\alpha_{\Lcp}^{\rm avg}(\LcToSigPip) & = & -0.463\pm 0.016 \pm 0.008\,,   
\end{eqnarray} 
where the first uncertainties are statistical and the second are systematic, which are described in detail in Sec.~\ref{sec:sys}. 
The measured values of $\alpha_{\Lcp}^{\rm avg}$ for the $\LcToLamKp$ and $\LcToSigKp$ modes are the first $\alpha$ results for SCS decays of charmed baryons.
The measured values of $\alpha_{\Lcp}^{\rm avg}$ for the $\LcToLamPip$ and $\LcToSigPip$ modes are consistent with the current world average values: $\alpha_{\Lcp}^{\rm avg}({\LcToLamPip})\!=\!-0.84\pm0.09$~\cite{bib:PDG2022} and $\alpha_{\Lcp}^{\rm avg}({\LcToSigPip})\!=\!-0.73\pm0.18$~\cite{BESIII:2019odb}, but with significantly improved precision: the uncertainty is improved from 11\% to 1\% for $\alpha_{\Lcp}^{\rm avg}(\LcToLamPip)$ and from 18\% to 4\% for $\alpha_{\Lcp}^{\rm avg}(\LcToSigPip)$.   

\section{$\alpha$-induced $\CP$ asymmetry}
We separate the $\Lcp$ and $\Lcm$ samples and measure $\alpha_{\Lcp}$ and $\alpha_{\Lcm}$ with the same method described above. 
The signal shape parameters for individual bins of helicity angles are fixed to the fitted results in the full sample integrated over helicity angles for $\Lcp$ and $\Lcm$ separately.  
The helicity angle distributions for the $\Lcp$ and $\Lcm$ samples are fitted separately, and the fitted slope factors, $\alpha_{\Lcp}\alpha_{-}$ and $\alpha_{\Lcm}\alpha_{+}$, are listed in Table~\ref{tab:alpha_AcpAlpha}. 
Additional details are given in the supplemental materials. 

Using the precise results 
$\alpha_{-}(\Lambda\to p\pim)\!=\!0.7519\pm 0.0043$ and 
$\alpha_{+}(\Lbar\to \overline{p}\pip)\!=\!-0.7559\pm 0.0047$ 
measured by BESIII~\cite{BESIII:2022qax}, we extract four $\alpha$-induced $\CP$ asymmetries as listed in Table~\ref{tab:alpha_AcpAlpha}, where $\Acp^{\alpha}$ for $\LcToLamKp$, $\LcToSigKp$, and $\LcToSigPip$ are measured for the first time. 
The measured $\Acp^{\alpha}$ for $\LcToLamPip$ is consistent with previous results, but with much better precision.

\begin{table*}[!htpb]
\begin{centering}
\caption{\label{tab:alpha_AcpAlpha}The fitted slopes $\alpha_{\Lambda_c^{\pm}}\alpha_{\mp}$ for $\Lcp$ and $\Lcm$ samples, and decay asymmetry parameters $\alpha_{\Lcp}$ and $\alpha_{\Lcm}$ for individual $\Lcp$ and $\Lcm$ samples using the most precise
$\alpha_{\mp}$ from BESIII recently~\cite{BESIII:2022qax}, and the corresponding $\alpha$-induced $\CP$ asymmetry $\Acp^{\alpha}$, comparing with current world averages (W.A.)~\cite{bib:PDG2022}.}
\begin{lrbox}{\tablebox}
\renewcommand\arraystretch{1.2}
\setlength{\tabcolsep}{1mm}{
\begin{tabular}{ccccccc}%
\hline 
Channel  		  & $\alpha_{\Lambda_c^{+}}\alpha_{-}$  & $\alpha_{\Lambda_c^{-}}\alpha_{+}$
														&   $\alpha_{\Lcp}$ 			  	& $\alpha_{\Lcm}$  			&  $\Acp^{\alpha}$   	& W.A. $\Acp^{\alpha}$ \\ \hline  
$\LcToLamKp$   & $-0.425\pm 0.053$ & $-0.448\pm 0.053$	& $-0.566\pm 0.071\pm 0.028$ 	& $0.592\pm 0.070\pm 0.079$ 	& $-0.023\pm 0.086\pm 0.071$		& -- \\
$\LcToLamPip$  & $-0.590\pm 0.006$  & $-0.570\pm 0.006$	& $-0.784\pm 0.008\pm 0.006$ 	& $0.754\pm 0.008\pm 0.018$ 	& $+0.020\pm 0.007\pm 0.014$	& $-0.07\pm0.22$ \\
$\LcToSigKp$    & $-0.43\phantom{0}\pm0.18\phantom{0}$ & $-0.37\phantom{0}\pm0.21\phantom{0}$	& $-0.58\phantom{0}\pm0.24\phantom{0}\pm 0.09\phantom{0}$ 	& $0.49\phantom{0}\pm0.28\phantom{0}\pm 0.14\phantom{0}$ & $+0.08\phantom{0}\pm0.35\phantom{0}\pm 0.14\phantom{0}$	& --	\\
$\LcToSigPip$   & $-0.340\pm 0.016$ & $-0.358\pm 0.017$	& $-0.452\pm 0.022\pm 0.023$ 	& $0.473\pm 0.023\pm 0.035$ 	& $-0.023\pm 0.034\pm 0.030$		&  -- \\
\hline 
\end{tabular}
}
  \end{lrbox}
  \scalebox{0.93}{\usebox{\tablebox}}
\end{centering}
\end{table*}

We search for hyperon CPV in $\Lambda\to{}p\pim$ in CF modes. Using the fitted slopes $\alpha_{\Lcp}\alpha_{-}$ and $\alpha_{\Lcm}\alpha_{+}$ for $\LcToLamPip$ and $\LcToSigPip$ as listed in Table~\ref{tab:alpha_AcpAlpha},
the $\alpha$-induced $\CP$ asymmetry of $\Lambda\to{}p\pim$ is measured to be
${+0.0169\pm 0.0073 \pm 0.0120 }$ in $\LcToLamPip$ and 
${-0.026\pm 0.034\pm 0.030 }$ in $\LcToSigPip$.
Finally, their average value is calculated to be  
\begin{eqnarray}
\Acp^{\alpha}(\Lambda\to p\pim) = +0.013\pm 0.007 \pm 0.011\,.  
\end{eqnarray}
This is the first measurement of hyperon CPV searches in CF charm decays.
No evidence of $\Lambda$-hyperon CPV is found.

\section{Systematic uncertainties}\label{sec:sys}
Most of the systematic uncertainties for the direct $\CP$ asymmetry cancel since they affect both $\Lcp$ and $\Lcm$ decays. 
The remaining systematic uncertainties are listed in Table~\ref{tab:sysBR}. 
The uncertainty due to each charged track asymmetry map is evaluated by varying the asymmetry value bin-by-bin by its uncertainty ($\pm1\sigma$) and repeating the measurement of the $\Acp^{\rm dir}$. 
The resulting deviations from the nominal $\Acp^{\rm dir}$ value are added in quadrature for positive and negative shifts, separately, and assigned as a systematic uncertainty.
We sample the parameters of the signal PDF, which are fixed in the nominal fit, from a multivariate Gaussian distribution that accounts for their uncertainties and correlations and re-fit for the signal yield.
The procedure is repeated 1000 times, and the root-mean-square of the distribution of fitted yields is taken as the systematic uncertainty due to the fixed parameters. 
To allow for the different background shapes for the $\Lcp$ and $\Lcm$ candidates, the background parameters are allowed to differ. 
The difference in the fitted results relative to the nominal results are assigned as a systematic uncertainty. 
We consider the possible fit bias with a linearity test for $\Acp^{\rm dir}$ with toy MC samples which are generated with five $A_{\rm raw}^{\rm corr}$ values per channel. A linear fit is applied to the measured $A_{\rm raw}^{\rm corr}$ distribution versus the generated values. 
The fitted slopes consistent with one indicate no fit bias. The relative shift between the fitted linear function and the nominal value is taken as a systematic uncertainty.
The remaining asymmetry due to the reconstruction of the $\Lambda$ or its children is considered as follows. The reference modes are weighted based on the ratio of the shapes for the momentum and polar angle of the $\Lambda$ between the signal and reference modes, causing the distributions in the reference modes to be the same as those for the signal modes. After this weighting, $A_{\rm raw}^{\rm corr}$ is remeasured and $A_{\CP}^{\rm dir}$ is calculated. The changes on $A_{\CP}^{\rm dir}$ are $-0.02\%$ for $\LcToLamKp$ and $-0.04\%$ for $\LcToSigKp$, which are assigned as systematic uncertainty.
The total systematic uncertainty is determined from the sum of all contributions in quadrature to be 
${}^{+1.2}_{-0.7}\!\times\!10^{-3}$ for $\Acp^{\rm dir}(\LcToLamKp)$ and ${}^{+3.0}_{-4.3}\!\times\!10^{-3}$ for $\Acp^{\rm dir}(\LcToSigKp)$. 
And considering the statistical uncertainties of $\Acp^{\rm dir}$ results are larger than $1\%$, we assign $0.1\%$ and $0.4\%$ as the final systematic uncertainties of $\Acp^{\rm dir}(\LcToLamKp)$ and $\Acp^{\rm dir}(\LcToSigKp)$, respectively, which are greatly smaller than the corresponding statistical uncertainties 2.6\% and 5.4\%. 

\begin{table}[!htpb]
\begin{centering}
\caption{\label{tab:sysAcp}The absolute systematic uncertainties (in units of $10^{-3}$) for $\CP$ asymmetry $\Acp^{\rm dir}$.}
\renewcommand\arraystretch{1.2}
\begin{tabular}{ccc} \hline 
Sources  					&  $\Acp^{\rm dir}(\LcToLamKp)$         & $\Acp^{\rm dir}(\LcToSigKp)$  \\ \hline  
$A_{\varepsilon}^{\Kp}$ map &  ${}^{+0.8}_{-0.2}$   &   $\pm0.4$     \\
$A_{\varepsilon}^{\pip}$ map&  $\pm0.4$     		&  ${}^{+0.5}_{-2.5}$  \\
Signal shape 				&   $\pm0.5$   		    & $\pm1.4$   \\ 
Background shape 			&  $-0.2$     			& $-3.1$    \\ 
Fit bias 					& $+0.6$				& $+2.6$ \\  
$\Lambda$ asymmetry         & $-0.2$               & $-0.4$ \\ 
\hline  
Total 		                &  $^{+1.2}_{-0.7}$ 	& $^{+3.0}_{-4.3}$  \\ 
\hline 
\end{tabular}
\end{centering}
\end{table}

For the measurement of BF ratio, most systematic uncertainties cancel since they affect both the signal and reference modes. 
The remaining systematic uncertainties are listed in Table~\ref{tab:sysBR}. 
Using the $D^{*+}\to[\Dz\to\Km\pip]\pip$ control sample, the PID uncertainties are estimated to be 0.9\% for $\LcToLamKp$, 0.8\% for $\LcToLamPip$, 0.9\% for $\LcToSigKp$, and 0.8\% for $\LcToSigPip$. 
Since the kaon and pion PID efficiency use the same control sample, 
we assign 1.7\% as the systematic uncertainty for both BF ratios.
The systematic uncertainties associated with the fixed parameters in the signal-yield fit is determined according to the same method as for $\Acp^{\rm dir}$ to be 0.2\% and 0.4\% for the $\Lambda$- and $\Sigma^0$-involved modes, respectively.
In modes that include a $\Sigma^0$, the broken-$\Sigma^0$ signal has a fixed ratio to signal based on MC simulation. The $M(\Lcp)$ distributions of the MC sample and experimental data in $M(\Sigma^0)$ sideband region have nearly same shapes, which suggests that the MC simulation is reliable for this broken-$\Sigma^0$ signal. 
We vary its ratio in the $M(\Lcp)$ fit by $\pm10\%$ and the larger deviation, $0.1\%$, is assigned as a conservative estimate. 
We consider the effects of the $\Xi_c$ background shape in the $\LcToSigPip$ mode by parameterizing it separately from the other backgrounds. The difference in the fitted signal yield is $0.1\%$, assigned as a systematic uncertainty.
Since the multiplicity of events for modes that include a $\Lambda$ is small, we remove events with multiple candidates and repeat the measurement. 
For modes that include a $\Sigma^0$, an alternative BCS method is applied to select the candidate with highest momentum $\gamma$ from the $\Sigma^0$ decay. 
The resulting changes in the branching fraction measurement are assigned as systematic uncertainties.
The $\alpha$ value used in signal MC production is varied by its uncertainty and the resulting change in the efficiency is assigned as a systematic uncertainty. A systematic uncertainty due to limited MC statistics is also considered.  
The total systematic uncertainty is determined by adding the uncertainties from all sources in quadrature, as given in Table~\ref{tab:sysBR}. 

\begin{table}
\begin{centering}
\caption{\label{tab:sysBR}Relative systematic uncertainties (in units of \%) for branching fractions.}
\setlength{\tabcolsep}{2mm}{
\begin{tabular}{ccc} 
\hline 
Sources  &   $\frac{\mathcal{B}(\LcToLamKp)}{\mathcal{B}(\LcToLamPip)}$  
                                & $\frac{\mathcal{B}(\LcToSigKp)}{\mathcal{B}(\LcToSigPip)}$  \\ \hline  
PID efficiency correction   	
				    &   1.7    	&  	1.7 	\\
Signal shape     	&   0.2   	&   0.4	\\
Background shape 	&   --   	& 	0.1	 \\  
BCS effect		    &   0.1	    &	0.4	\\
Efficiency ratio   	&   0.2    	&  	0.4 	\\  \hline 
Total      		    &   1.7  	&	1.8	 \\   
\hline 
\end{tabular}}
\end{centering}
\end{table}

\begin{table*}[!htpb]
\begin{centering}
\caption{\label{tab:sysAlphaAcp}Absolute systematic uncertainties (in units of $10^{-2}$) for decay asymmetry parameters and the $\alpha$-induced $\CP$ asymmetries: $\alpha_{\rm avg}$/$\alpha_{\Lcp}$/$\alpha_{\Lcm}$/$A_{\CP}^{\alpha}$ in each decay mode (the fifth items in $\LcToLamPip/\Sigma^0\pip$ are for $\Acp^{\alpha}(\Lambda)$).}
\renewcommand\arraystretch{1.2}
\setlength{\tabcolsep}{4mm}{
\begin{tabular}{lcccc} \hline 
Sources  			&  $\LcToLamKp$ 	   &  $\LcToLamPip$  	& $\LcToSigKp$   		& $\LcToSigPip$  \\ \hline  
$\cos\theta$ bins 	&  0.8/1.5/1.5/1.4     &  0.0/0.1/0.2/0.2/0.16  &  7.4/6.9/\phantom{0}9.9/\phantom{0}8.4  	&  0.7/1.5/3.3/1.9/1.9 \\
Efficiency curve 	&  0.2/0.5/0.1/0.4	   &  0.2/0.3/0.1/0.3/0.29	&  1.4/1.1/\phantom{0}1.8/\phantom{0}0.9	&  0.1/0.5/0.4/0.9/0.9 \\
Fit bias  			&  1.6/2.3/7.7/6.9	   &  0.2/0.3/1.7/1.2/1.15	&  4.1/5.7/10.3/11.7  		&  0.4/1.7/0.9/2.1/2.1	 \\
$\alpha_{\mp}(\Lambda\to p\pim)$    
			    &0.2/0.3/0.4/0.6   	   &                               0.2/0.4/0.5/0.6/--\phantom{.}\phantom{0}\phantom{0} &  0.2/0.3/\phantom{0}0.3/\phantom{0}0.6	&  0.1/0.2/0.3/0.6/--\phantom{.}\phantom{0}   \\ \hline 
Total 	        &   1.8/2.8/7.9/7.1   &  0.3/0.6/1.8/1.4/1.20   &  8.6/9.0/14.4/14.4  	&  0.8/2.3/3.5/3.0/3.0		\\  
\hline 
\end{tabular}}
\end{centering}
\end{table*}

For the $\alpha$ and $\Acp^{\alpha}$ measurements, we consider the systematic uncertainty due to the number of helicity angle bins, the efficiency curve, the fit bias, and the quoted uncertainty on $\alpha_{\mp}$. 
We change the number of helicity angle bins from 10 to 8 or 12 for $\LcToLamHp$, from 5$\times$5 to 4$\times$4 or 6$\times$6 for $\LcToSigKp$, and from 6$\times$6 to 5$\times$5 or 7$\times$7 for $\LcToSigPip$.
The $\alpha$ value used in signal MC production is varied by its uncertainty. The resulting changes in $\alpha$ or $\Acp^{\alpha}$ are assigned as systematic uncertainties. 
Additional signal MC samples are produced with different hypotheses for the $\Lcp$ polarization (${\bf P}=\pm0.4$), which may affect the efficiency as a function of helicity angle~\cite{CLEO:1990unw}. The maximum difference in the measured $\alpha$ relative to the unpolarized hypothesis, 0.001, is taken as a systematic uncertainty.
We consider the possible fit bias for $\alpha$ and $\Acp^{\alpha}$ with a linearity test, in which we replace the signal events in the MC sample with events produced with a special angular distribution using five $\alpha$ values. 
A linear fit is applied to the measured $\alpha$ distribution versus the generated values. The fitted slopes consistent with one indicate no fit bias. 
The relative shift between the fitted linear function and the nominal value is taken as a systematic uncertainty. 
The quoted uncertainties of $\alpha_{\Lambda}^{\rm avg}$ and $\alpha_{\mp}$ (with their correlation coefficient $\rho(\alpha_-,\,\alpha_+)=0.850$~\cite{BESIII:2022qax} considered) of $\Lambda\to p\pim$ decays are assigned as systematic uncertainties.
The effect arising from the spin precession of $\Lambda$-hyperons with average momentum 2~GeV/$c$ in the magnetic field of the detector (1.5~T) has been considered. 
The resulting systematic uncertainty is estimated at level of $\mathcal{O}(10^{-4})$~\cite{Li:2019knv}, which is negligible given the precision of this result.
The total systematic uncertainties for $\alpha_{\rm avg}$/$\alpha_{\Lcp}$/$\alpha_{\Lcm}$/$A_{\CP}^{\alpha}$/$\Acp^{\alpha}(\Lambda)$ are taken as the sum in quadrature of all contributions, as listed in Table~\ref{tab:sysAlphaAcp}.

\section{Summary}
In conclusion, based on the 980 $\invfb$ data set collected with the Belle detector, 
we make the first measurement of the direct $\CP$ asymmetry in SCS two-body decays of charmed baryons,
$\Acp^{\rm dir}(\LcToLamKp) \!=\! +0.021 \pm 0.026 \pm 0.001$ and $\Acp^{\rm dir}(\LcToSigKp) \!=\! +0.025 \pm 0.054 \pm 0.004$. 
The relative branching fractions are measured to be, 
$\BR(\LcToLamKp)/\BR(\LcToLamPip) \!=\! (5.05\pm 0.13 \pm 0.09)\%$ and 
$\BR(\LcToSigKp)/\BR(\LcToSigPip) \!=\! (2.78\pm 0.15 \pm 0.05)\%$, 
which supersede previous Belle measurements~\cite{Belle:2001hyr}.
Using the world average values for the branching fractions for $\Lcp\to(\Lambda,\,\Sigma^0)\pip$, 
we obtain 
$\BR(\LcToLamKp) \!=\! {(6.57\pm 0.17 \pm 0.11 \pm 0.35 )\times 10^{-4}}$ and 
$\BR(\LcToSigKp) \!=\! {(3.58\pm 0.19 \pm 0.06 \pm 0.19 )\times 10^{-4}}$.
These results are the most precise to date and significantly improve the precision of the world average values~\cite{bib:PDG2022}.

We obtain the averaged decay asymmetry parameters 
$\alpha_{\Lcp}^{\rm avg}({\Lcp\to\Lambda\Kp}) \!=\! -0.585\pm 0.049 \pm 0.018$ and 
$\alpha_{\Lcp}^{\rm avg}({\Lcp\to\Sigma^0\Kp}) \!=\! -0.54 \pm 0.18  \pm 0.09$ for the first time. 
We obtain 
$\alpha_{\Lcp}^{\rm avg}({\Lcp\to\Lambda\pip}) \!=\! -0.755\pm 0.005 \pm 0.003$ and 
$\alpha_{\Lcp}^{\rm avg}({\Lcp\to\Sigma^0\pip}) \!=\! -0.463\pm 0.016 \pm 0.008$, 
which are consistent with previous measurements~\cite{bib:PDG2022} but with significantly improved precision.
We also determine the $\alpha$-parameter for $\Lcp$ and $\Lcm$ individually and search for CPV via the $\alpha$-induced $\CP$ asymmetry, as listed in Table~\ref{tab:alpha_AcpAlpha}. These results include the first measurements of $\Acp^{\alpha}$ for SCS decays of charmed baryons, 
$\Acp^{\alpha}(\LcToLamKp)\!=\!-0.023\pm0.086\pm0.071$ and 
$\Acp^{\alpha}(\LcToSigKp)\!=\!+0.08\pm 0.35 \pm 0.14$.
We search for $\Lambda$ hyperon CPV via the $\alpha$-induced $\CP$ asymmetry in $\LcToLamPip$ and $\LcToSigPip$ decays, and determine $\Acp^{\alpha}({\Lambda\to p\pim}) \!=\! +0.013\pm 0.007\pm 0.011$ by combining the two modes.
No evidence of baryon CPV is found. 
The method used in our $\Acp^{\alpha}({\Lambda\to p\pim})$ measurement can be applied to other hyperons, such as $\Acp^{\alpha}({\Xi^{0,-}\to\Lambda\pi^{0,-}})$ in ${\Lcp\to\Xi^0\Kp}$ and ${\Xi_c^{+,0}\to\Xi^{0,-}\pi^+}$. 
Our measurement is a milestone for hyperon CPV searches in charm CF decays and this method is promising for precise measurements of various hyperon CPV at Belle II and LHCb.

\section*{Conflict of interest}
The authors declare that they have no conflict of interest.

\section*{Acknowledgments}
This work, based on data collected using the Belle detector, which was
operated until June 2010, was supported by 
the Ministry of Education, Culture, Sports, Science, and
Technology (MEXT) of Japan, the Japan Society for the 
Promotion of Science (JSPS), and the Tau-Lepton Physics 
Research Center of Nagoya University; 
the Australian Research Council including grants
DP180102629, 
DP170102389, 
DP170102204, 
DE220100462, 
DP150103061, 
FT130100303; 
Austrian Federal Ministry of Education, Science and Research (FWF) and
FWF Austrian Science Fund No.~P~31361-N36;
the National Natural Science Foundation of China under Contracts
No.~11675166,  
No.~11705209;  
No.~11805064;  
No.~11975076;  
No.~12135005;  
No.~12175041;  
No.~12161141008; 
Key Research Program of Frontier Sciences, Chinese Academy of Sciences (CAS), Grant No.~QYZDJ-SSW-SLH011; 
the Shanghai Science and Technology Committee (STCSM) under Grant No.~19ZR1403000; 
the Ministry of Education, Youth and Sports of the Czech
Republic under Contract No.~LTT17020;
the Czech Science Foundation Grant No. 22-18469S;
Horizon 2020 ERC Advanced Grant No.~884719 and ERC Starting Grant No.~947006 ``InterLeptons'' (European Union);
the Carl Zeiss Foundation, the Deutsche Forschungsgemeinschaft, the
Excellence Cluster Universe, and the VolkswagenStiftung;
the Department of Atomic Energy (Project Identification No. RTI 4002) and the Department of Science and Technology of India; 
the Istituto Nazionale di Fisica Nucleare of Italy; 
National Research Foundation (NRF) of Korea Grant
Nos.~2016R1\-D1A1B\-02012900, 2018R1\-A2B\-3003643,
2018R1\-A6A1A\-06024970, RS\-2022\-00197659,
2019R1\-I1A3A\-01058933, 2021R1\-A6A1A\-03043957,
2021R1\-F1A\-1060423, 2021R1\-F1A\-1064008, 2022R1\-A2C\-1003993;
Radiation Science Research Institute, Foreign Large-size Research Facility Application Supporting project, the Global Science Experimental Data Hub Center of the Korea Institute of Science and Technology Information and KREONET/GLORIAD;
the Polish Ministry of Science and Higher Education and 
the National Science Center;
the Ministry of Science and Higher Education of the Russian Federation, Agreement 14.W03.31.0026, 
and the HSE University Basic Research Program, Moscow; 
University of Tabuk research grants
S-1440-0321, S-0256-1438, and S-0280-1439 (Saudi Arabia);
the Slovenian Research Agency Grant Nos. J1-9124 and P1-0135;
Ikerbasque, Basque Foundation for Science, Spain;
the Swiss National Science Foundation; 
the Ministry of Education and the Ministry of Science and Technology of Taiwan;
and the United States Department of Energy and the National Science Foundation.
These acknowledgements are not to be interpreted as an endorsement of any
statement made by any of our institutes, funding agencies, governments, or
their representatives.
We thank the KEKB group for the excellent operation of the
accelerator; the KEK cryogenics group for the efficient
operation of the solenoid; and the KEK computer group and the Pacific Northwest National
Laboratory (PNNL) Environmental Molecular Sciences Laboratory (EMSL)
computing group for strong computing support; and the National
Institute of Informatics, and Science Information NETwork 6 (SINET6) for
valuable network support.
We warmly thank Fu-Sheng~Yu, Di~Wang, Chao-Qiang Geng, Chia-Wei Liu, and Xian-Wei Kang for valuable and helpful discussions.


\bibliographystyle{SciBull.bst}
\bibliography{reference.bib}


\section*{The Belle Experiment}
Belle is a first-generation ``asymmetric B-factory'' experiment at the
KEKB asymmetric-energy $e^+e^-$ collider in Tsukuba, Japan. 
Belle was designed to discover $C\!P$ violation in $B$ meson decays, 
as well as to cover a wide range of physics targets. 
Belle collected over 1~ab$^{-1}$ of collision data between 1999 and 2010, 
including 772 million $B\overline{B}$ pairs produced at the energy 
corresponding to the $\Upsilon(4S)$ resonance. 
Belle has also collected large samples of data from 
$\tau$-lepton, charm-, and light-hadron decays, 
and data at the energies of the $\Upsilon(1,2,3,5S)$ resonances. 
In 2001, Belle observed a large $C\!P$ asymmetry in $B$ meson decays, 
providing strong experimental evidence for the theoretical predictions 
from Kobayashi and Maskawa. 
In 2003, Belle discovered an unexpected new charmonium-like state, 
$X(3872)$, heralding a new era of hadron physics. 
After more than a decade since the end of data-taking, 
Belle is still actively analyzing data and reporting new results.

\section*{The Belle Collaboration} 
L. K. Li\,\orcidlink{0000-0002-7366-1307}, W. Shan\,\orcidlink{0000-0003-2811-2218}, K. Kinoshita\,\orcidlink{0000-0001-7175-4182}, I. Adachi\,\orcidlink{0000-0003-2287-0173}, H. Aihara\,\orcidlink{0000-0002-1907-5964}, D. M. Asner\,\orcidlink{0000-0002-1586-5790}, H. Atmacan\,\orcidlink{0000-0003-2435-501X}, T. Aushev\,\orcidlink{0000-0002-6347-7055}, V. Babu\,\orcidlink{0000-0003-0419-6912}, S. Bahinipati\,\orcidlink{0000-0002-3744-5332}
% ...

\section*{Appendix A. Supplementary material}
\begin{figure*}[!htbp]
  \begin{centering}%
  \begin{overpic}[width=0.46\textwidth]{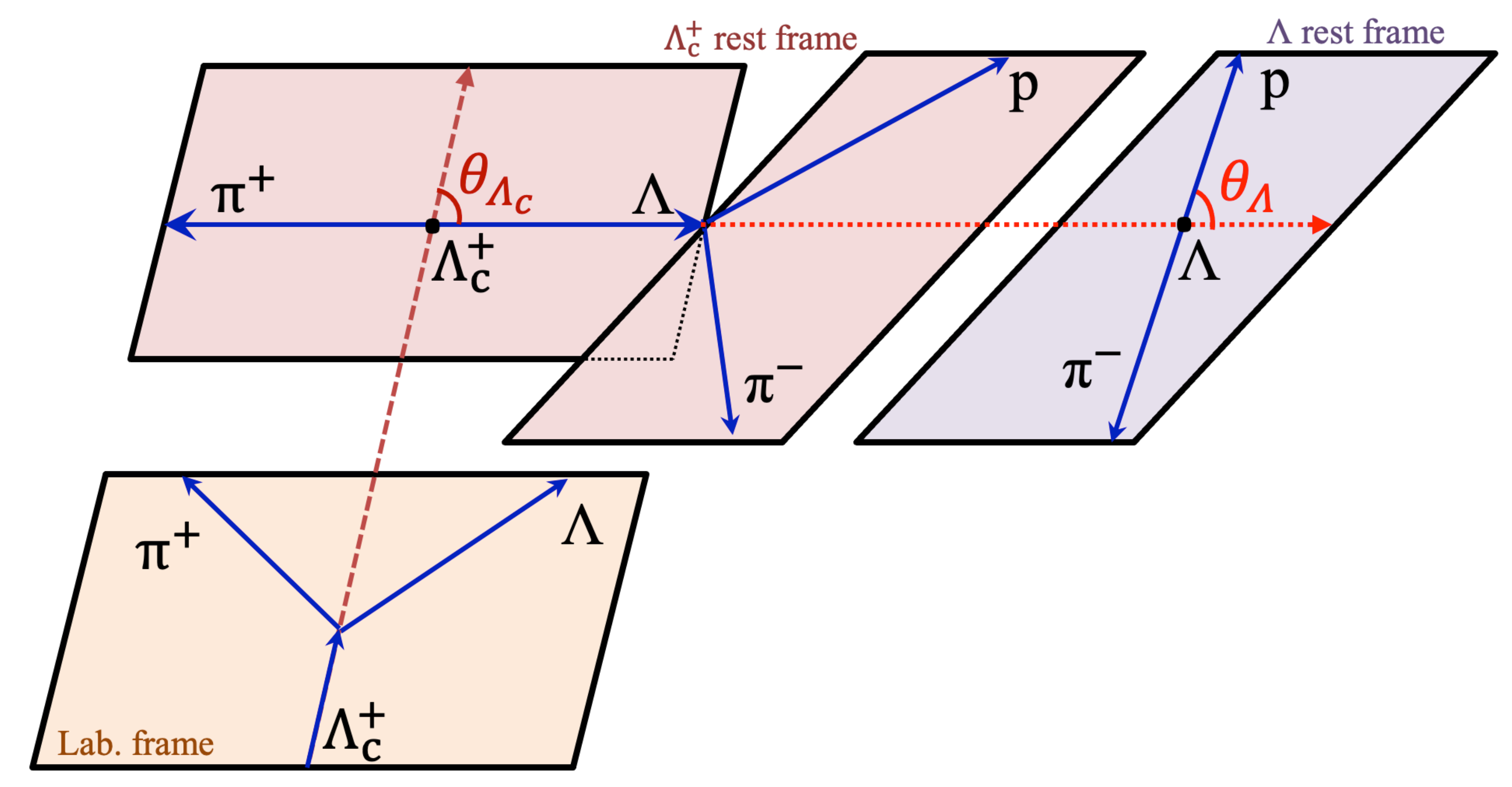}%
  \end{overpic}%
  \begin{overpic}[width=0.5\textwidth]{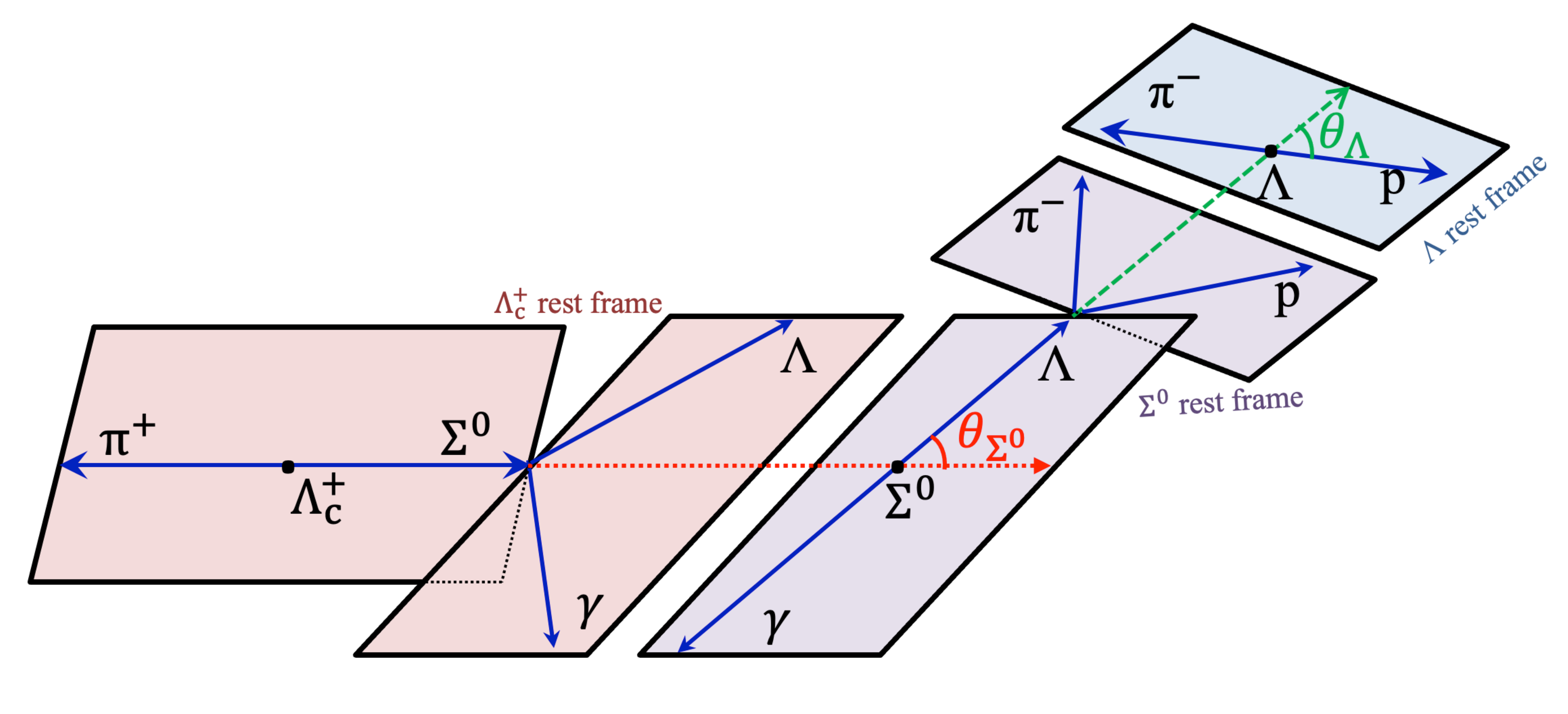}%
  \end{overpic}%
  \vskip-5pt
  \caption{\label{fig:Helicity}Schematic plot showing the helicity angles: (left) $\theta_{\Lcp}$ and $\theta_{\Lambda}$ in $\Lambda_c^+\to\Lambda\pip,\,\Lambda\to p\pim$; and (right) $\theta_{\Sigma^0}$ and $\theta_{\Lambda}$ in $\Lambda_c^+\to\Sigma^0\pip,\,\Sigma^0\to\gamma\Lambda,\,\Lambda\to p\pim$.}
  \end{centering}
\end{figure*}

\begin{figure*}[!htbp]
  \begin{centering}%
  \begin{overpic}[width=0.5\textwidth]{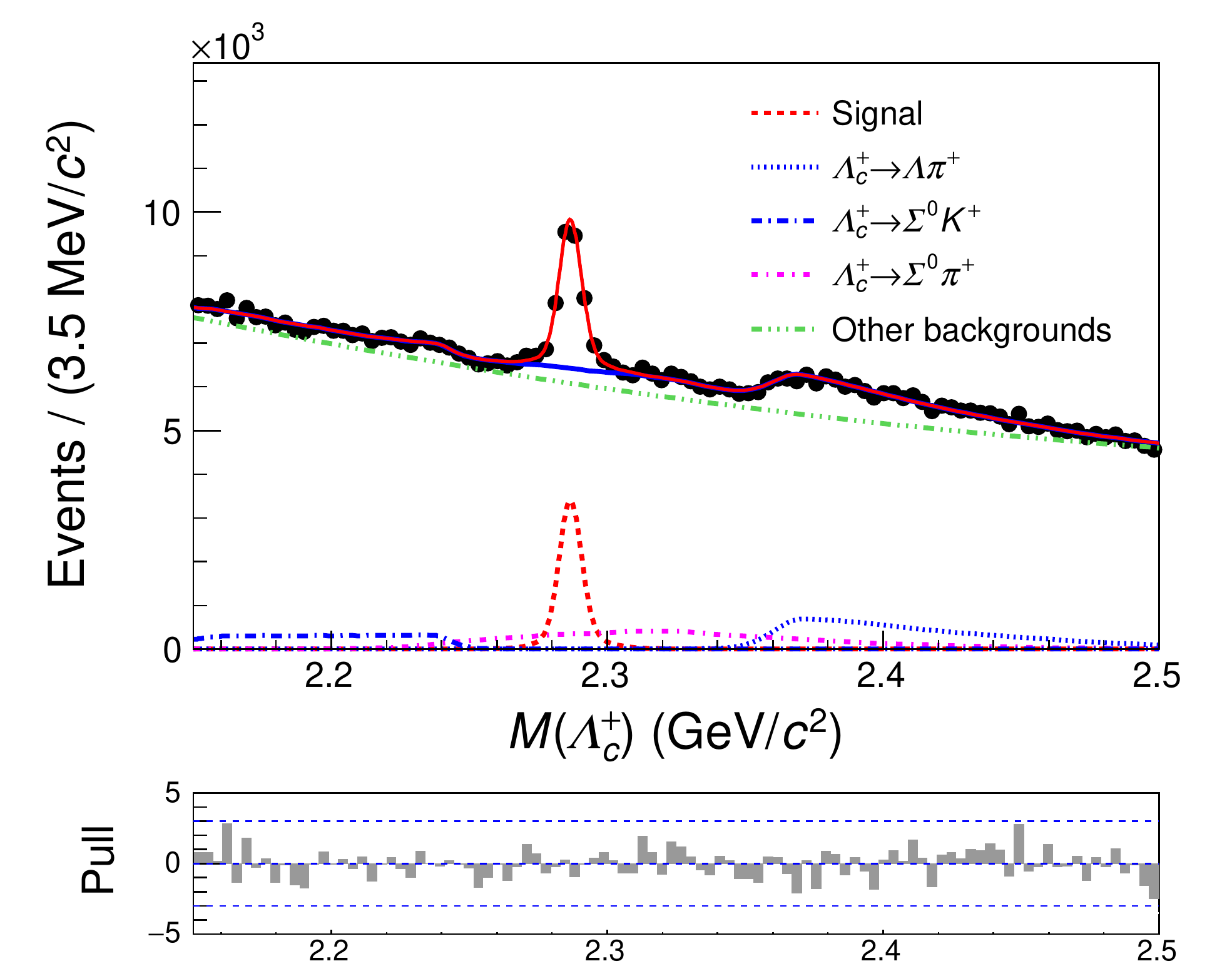}%
  \put(22,67){$\LcToLamKp$}%
  \end{overpic}%
  \begin{overpic}[width=0.5\textwidth]{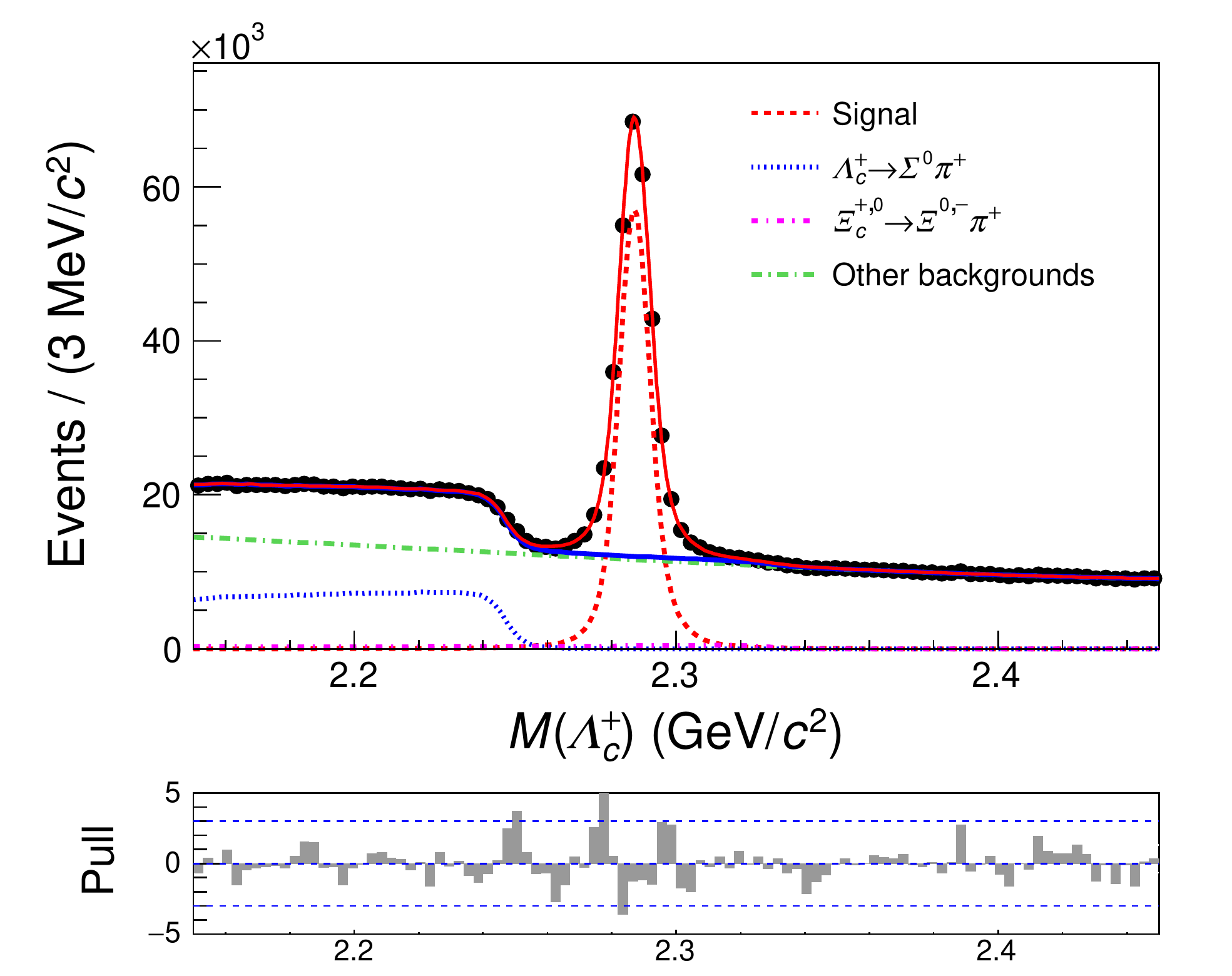}%
  \put(22,67){$\LcToLamPip$}%
  \end{overpic}
  \vskip-5pt
  \caption{\label{fig:YieldsFinal1}The fit results of $\Lcp$ invariant mass distributions for $\LcToLamKp$ and $\Lcp\to\Lambda\pip$ decays. The red curve shows the total fit result, and the blue curve the total background; the dashed curves show the components of signal and backgrounds. The fit qualities, $\chi^2$ divided by the number of degrees of freedom, are $\chi^2/91=1.12$ and $\chi^2/91=1.38$, respectively.} 
  \end{centering}
\end{figure*}

\begin{figure*}[!htbp]
  \begin{centering}%
  \begin{overpic}[width=0.5\textwidth]{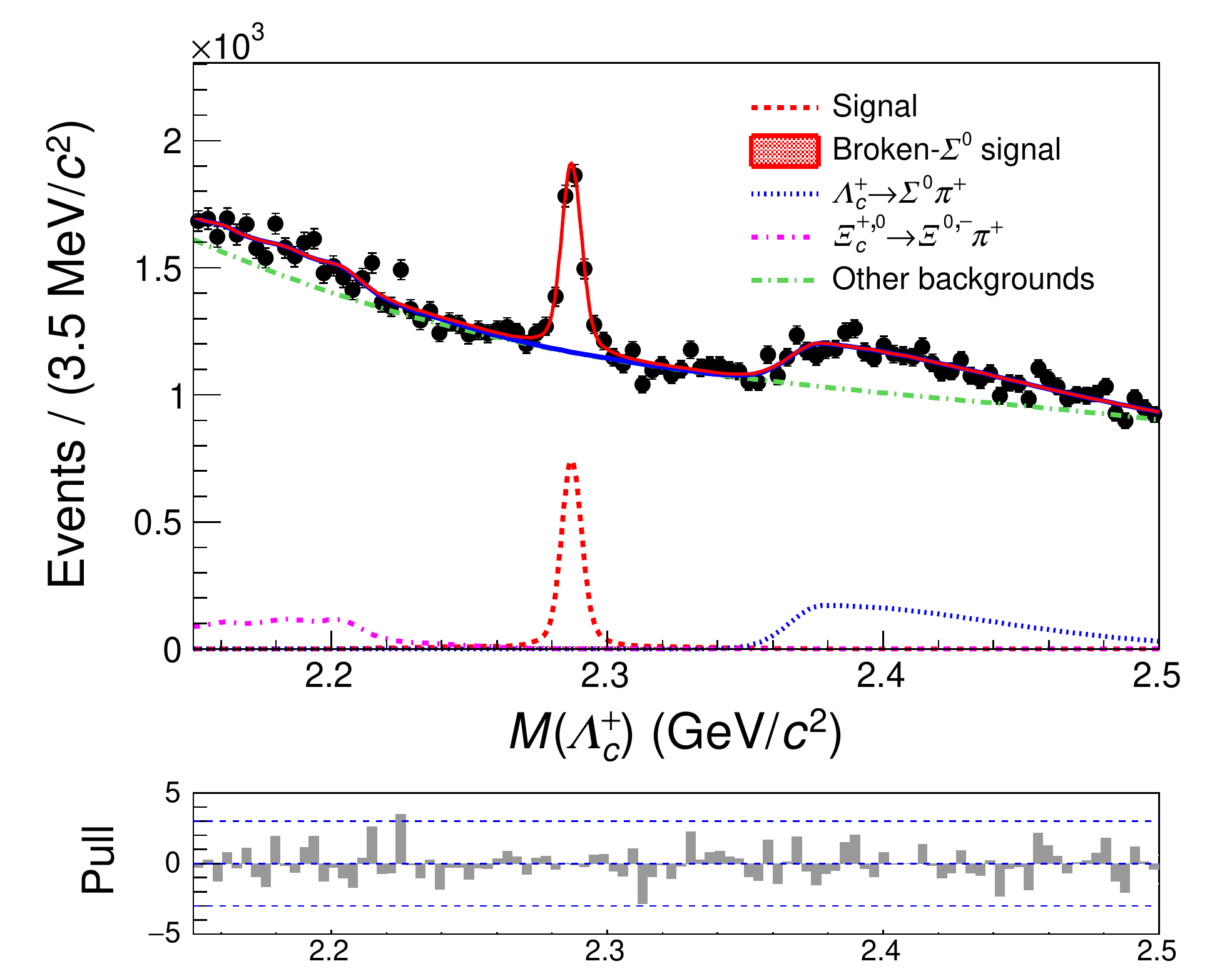}%
  \put(22,67){$\LcToSigKp$}%
  \end{overpic}%
  \begin{overpic}[width=0.5\textwidth]{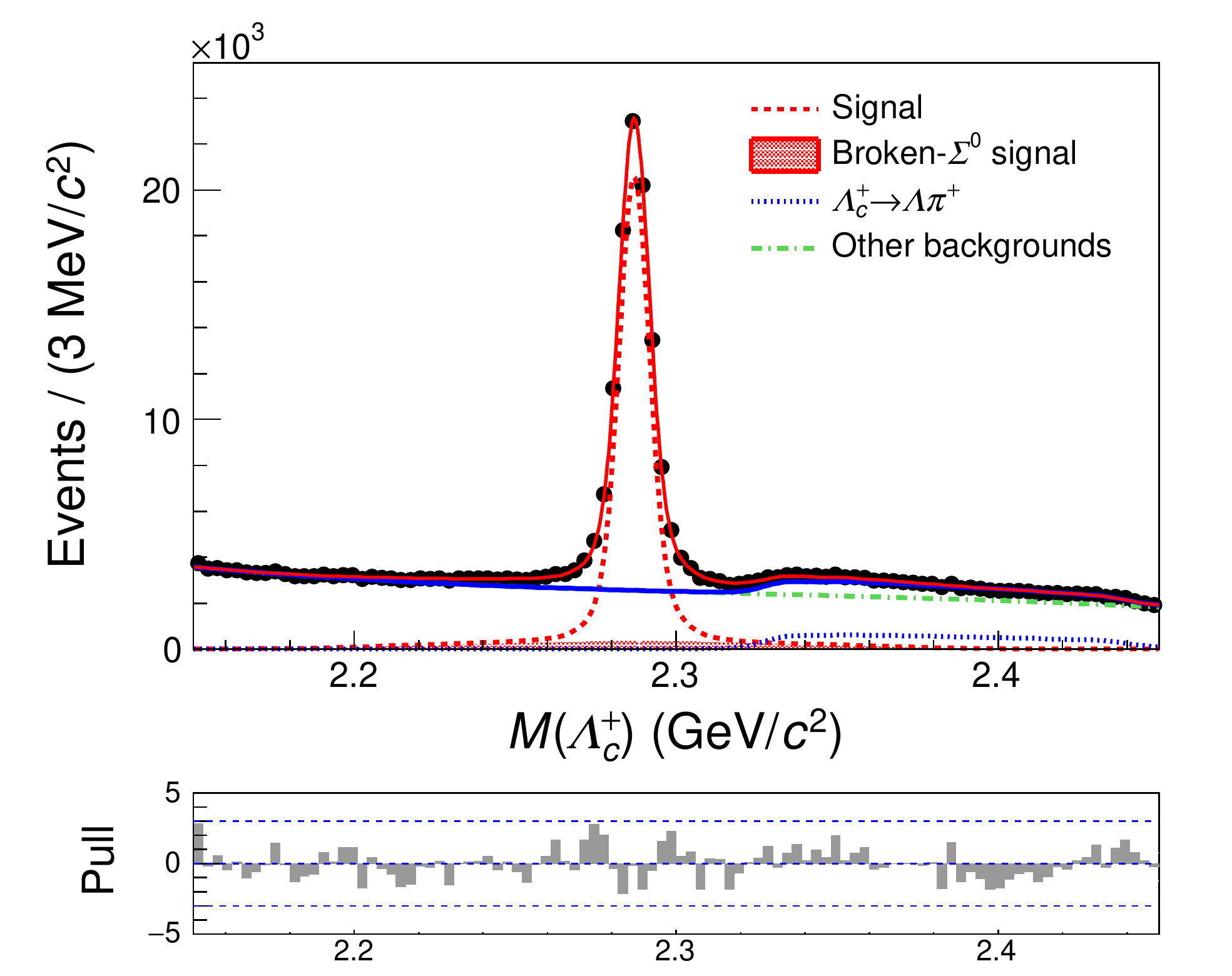}%
  \put(22,67){$\LcToSigPip$}%
  \end{overpic}\\
  \begin{overpic}[width=0.5\textwidth]{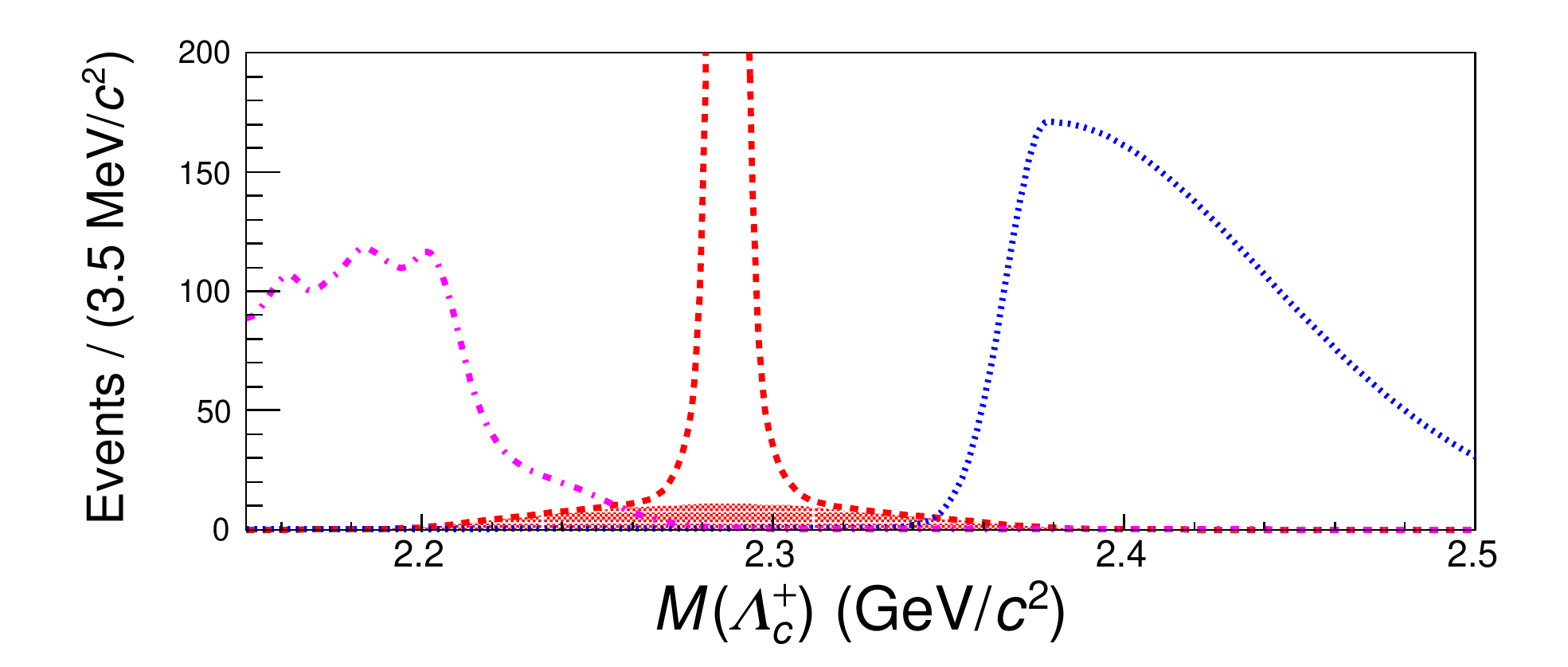}%
  \put(22,34){$\LcToSigKp$}%
  \end{overpic}%
  \begin{overpic}[width=0.5\textwidth]{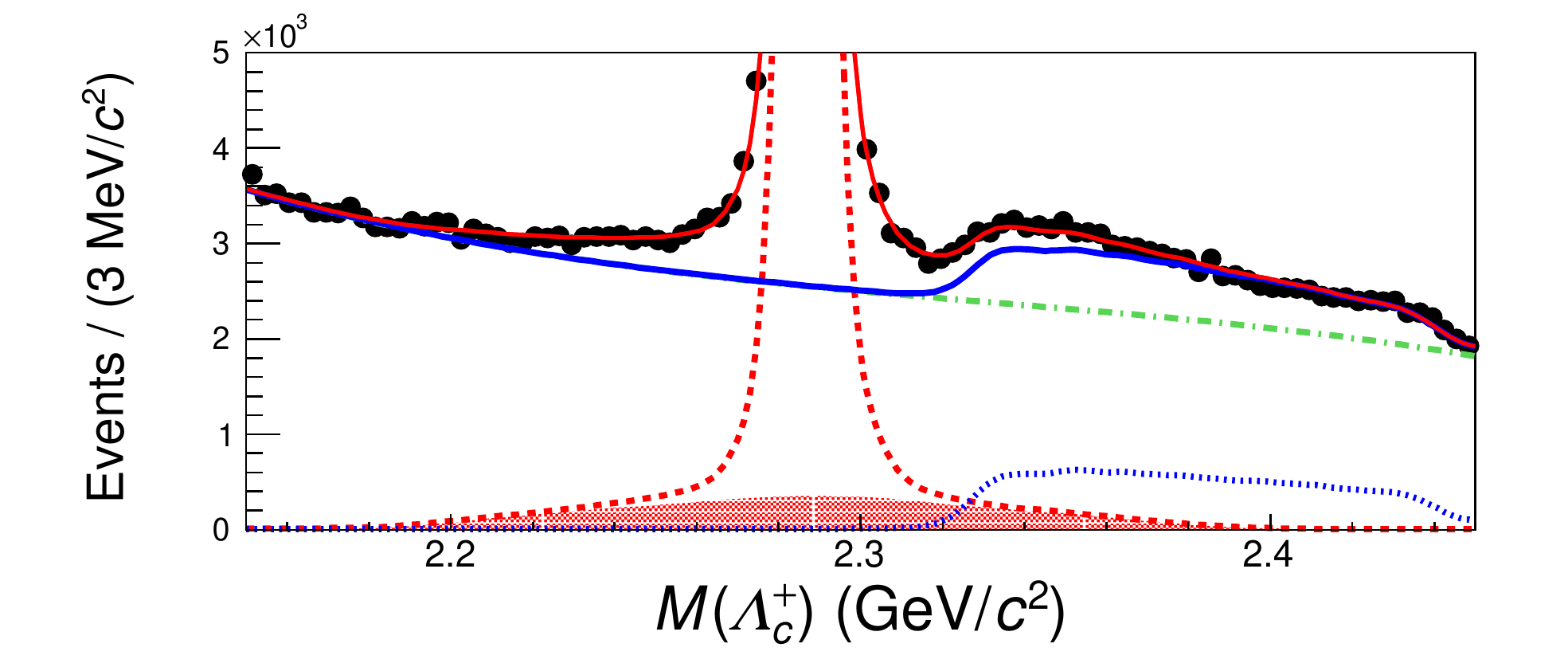}%
  \put(22,34){$\LcToSigPip$}%
  \end{overpic}
  \vskip-5pt
  \caption{\label{fig:YieldsFinal2}The fit results of $\Lcp$ invariant mass distributions for $\Lcp\to\Sigma^0\Kp$ and $\Lcp\to\Sigma^0\pip$ decays. The red curve shows the total fit result, and the blue curve the total background; the dashed curves show the components of signal (including the broken-$\Sigma^0$ signal) and backgrounds. The fit qualities, $\chi^2$ divided by the number of degrees of freedom, are $\chi^2/91=1.38$, and $\chi^2/92=1.07$, respectively. The bottoms figures are the enlarged view to show the distributions of broken-$\Sigma^0$ signal (red-filled histogram) and the peaking backgrounds more clearly.} 
  \end{centering}
\end{figure*}

\begin{figure*}[!hbtp]
  \begin{centering}
  \begin{overpic}[width=0.25\textwidth]{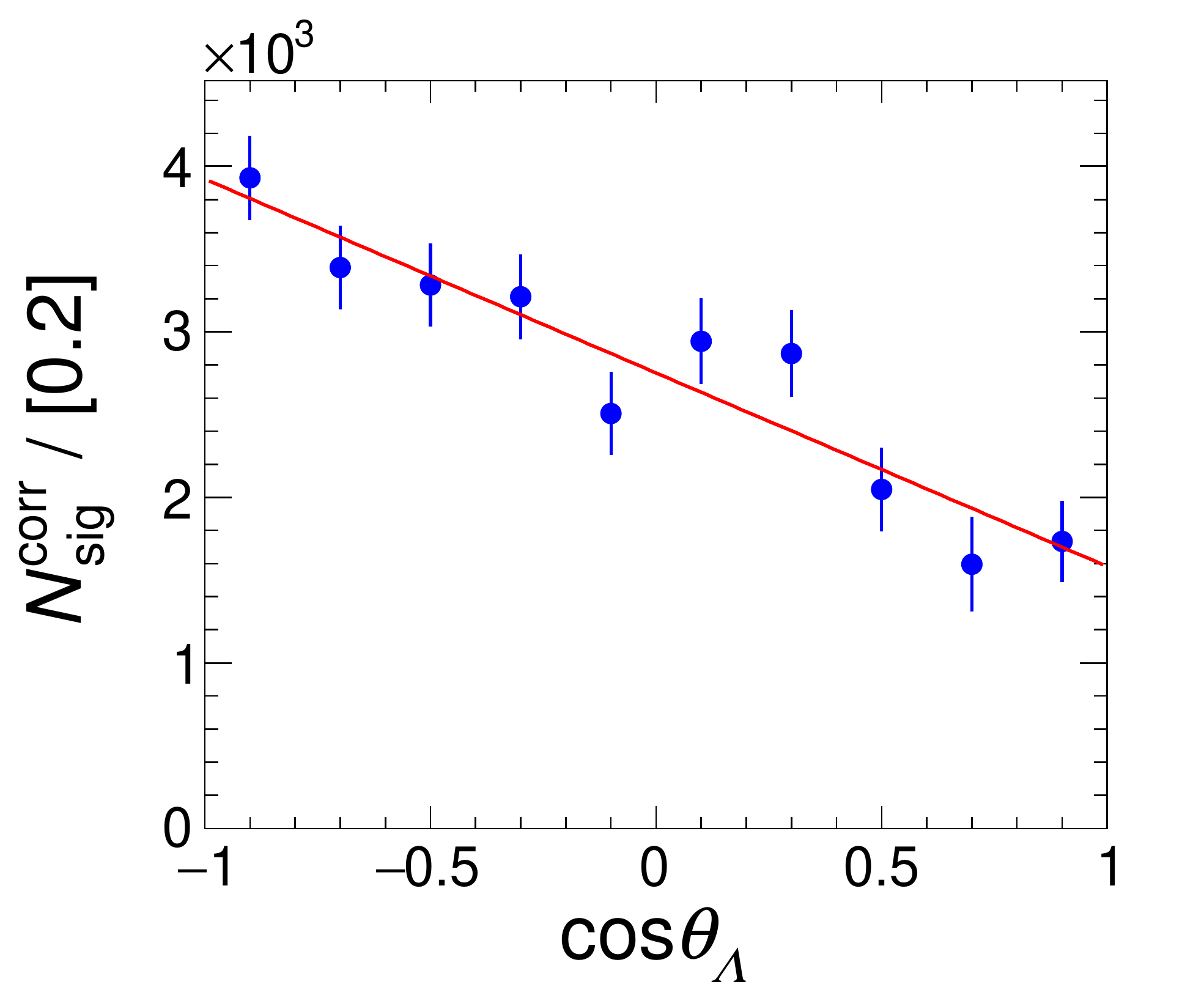}%
  \put(23,20){\small{$k=-0.425\pm 0.053$}}%
  \put(51,65){\small{$\LcToLamKp$}}%
  \end{overpic}%
  \begin{overpic}[width=0.25\textwidth]{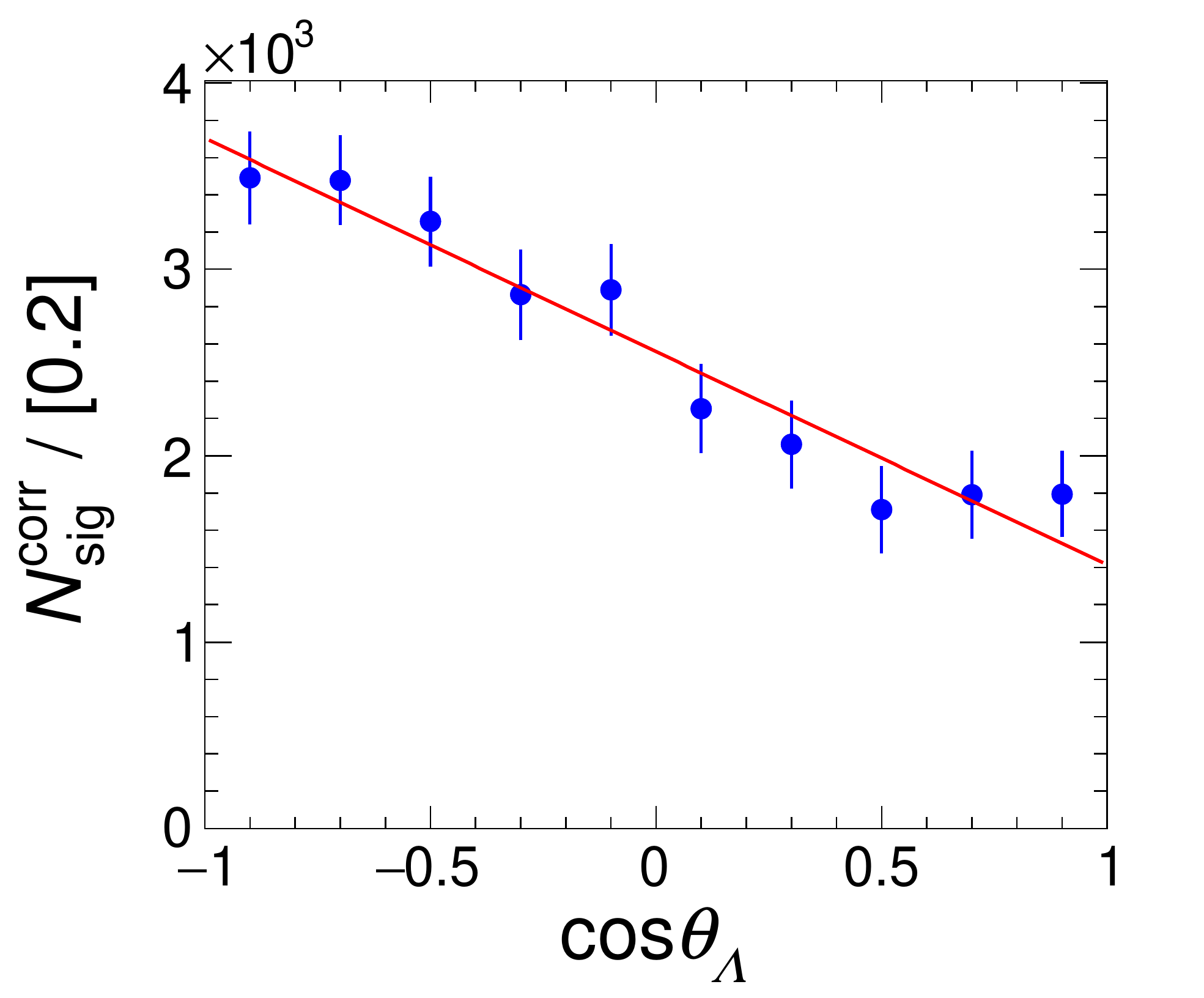}%
  \put(23,20){\small{$\overline{k}=-0.448\pm 0.053$}}%
  \put(51,65){\small{$\Lcm\to\Lbar\Km$}}%
  \end{overpic}%
  \begin{overpic}[width=0.25\textwidth]{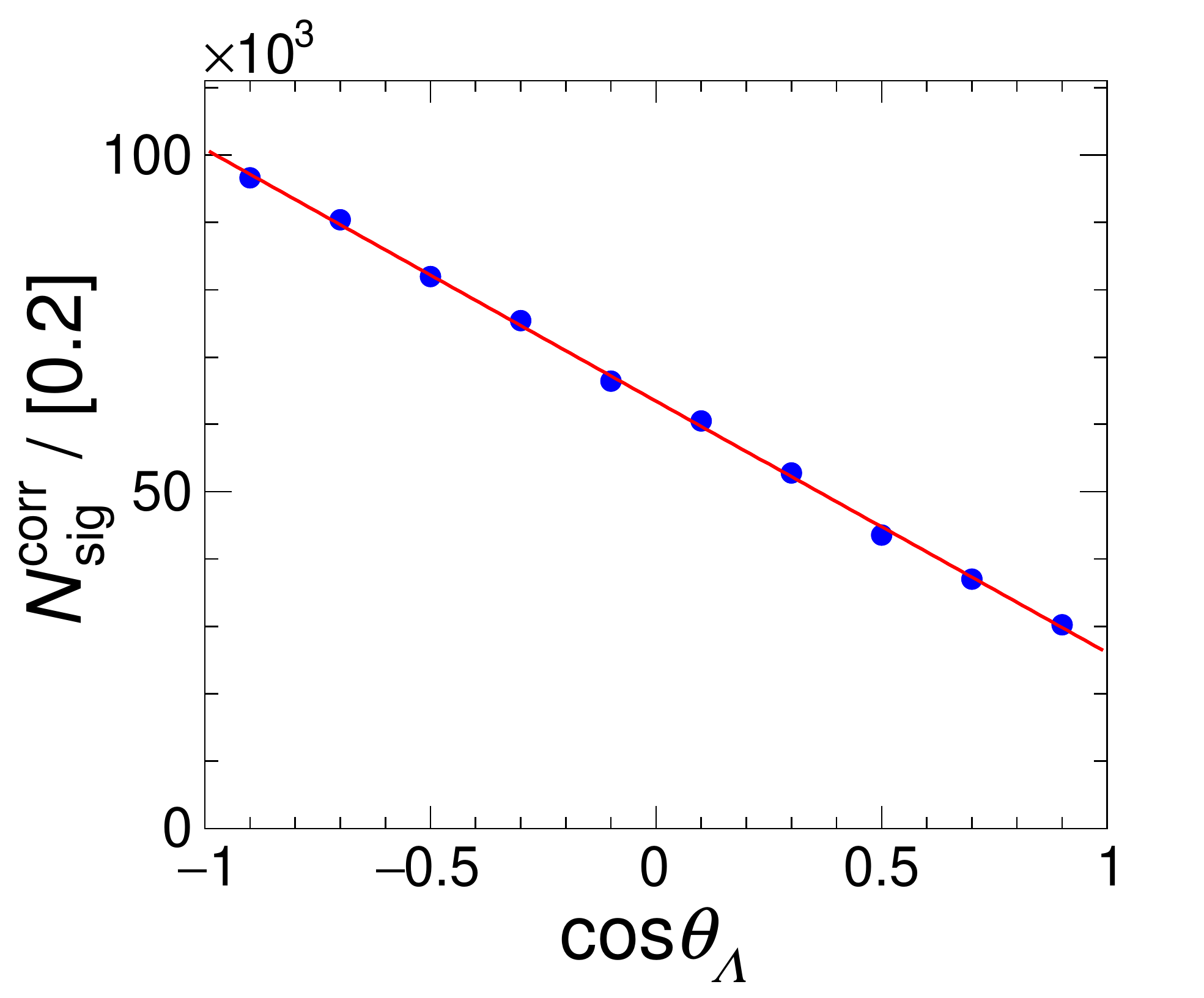}%
  \put(23,20){\small{$k=-0.590\pm 0.006$}}%
  \put(51,65){\small{$\LcToLamPip$}}%
  \end{overpic}%
  \begin{overpic}[width=0.25\textwidth]{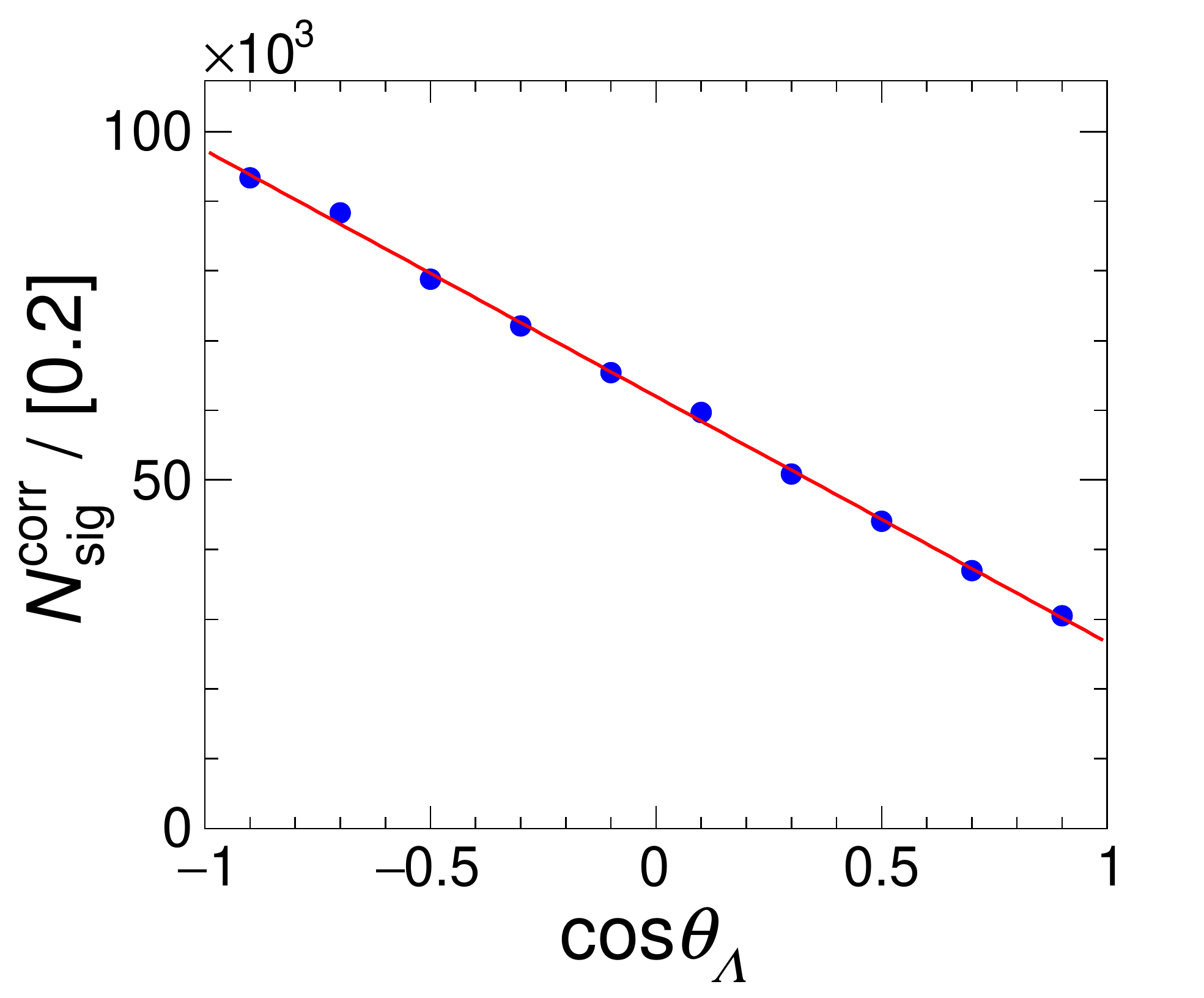}%
  \put(23,20){\small{$\overline{k}=-0.570\pm 0.006$}}%
  \put(51,65){\small{$\Lcm\to\Lbar\pim$}}%
  \end{overpic}%
  \vskip-5pt 
  \caption{\label{fig:AcpAlphaFinal_LcToLamHp}The $\cos\theta_{\Lambda}$ distributions of $\LcToLamKp$ and $\LcToLamPip$ after efficiency-correction. We fit it with a linear function of $1+\alpha_{\Lambda_c^{\pm}}\alpha_{\mp}\cos\theta_{\Lambda}$ with the $\chi^2$ divided by the number of degrees of freedom, $\chi^2/9=1.04$, 0.57, 1.25, and 0.88, respectively, and the fitted slope values ($k=\alpha_{\Lcp}\alpha_{-}$ and $\overline{k}=\alpha_{\Lcm}\alpha_{+}$) are shown.}
  \end{centering}
 \end{figure*}

\begin{figure*}[!hbtp]
  \begin{centering}
  \begin{overpic}[width=0.25\textwidth]{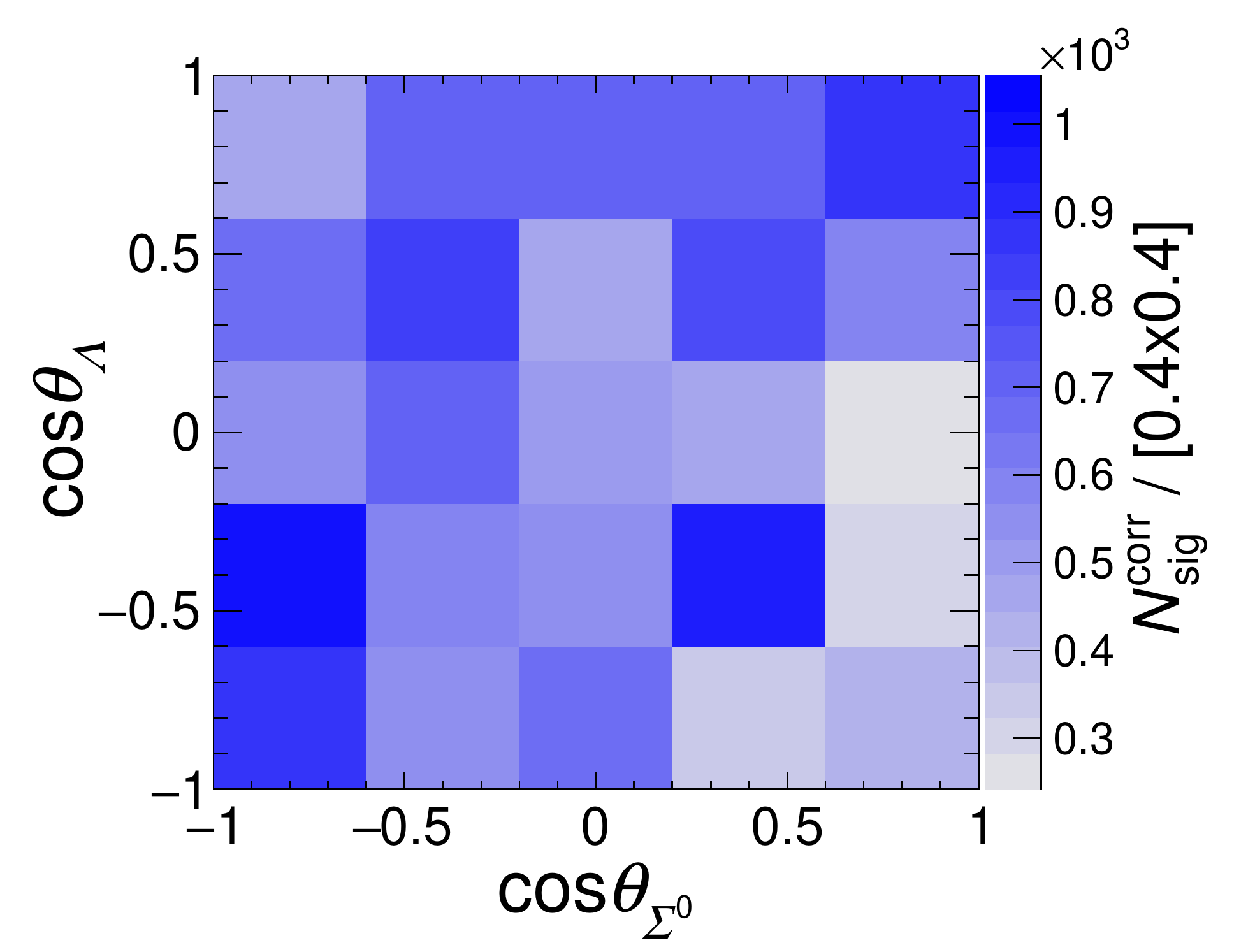}%
  \put(25,73){\small{$\LcToSigKp$}}    
  \end{overpic}%
  \begin{overpic}[width=0.25\textwidth]{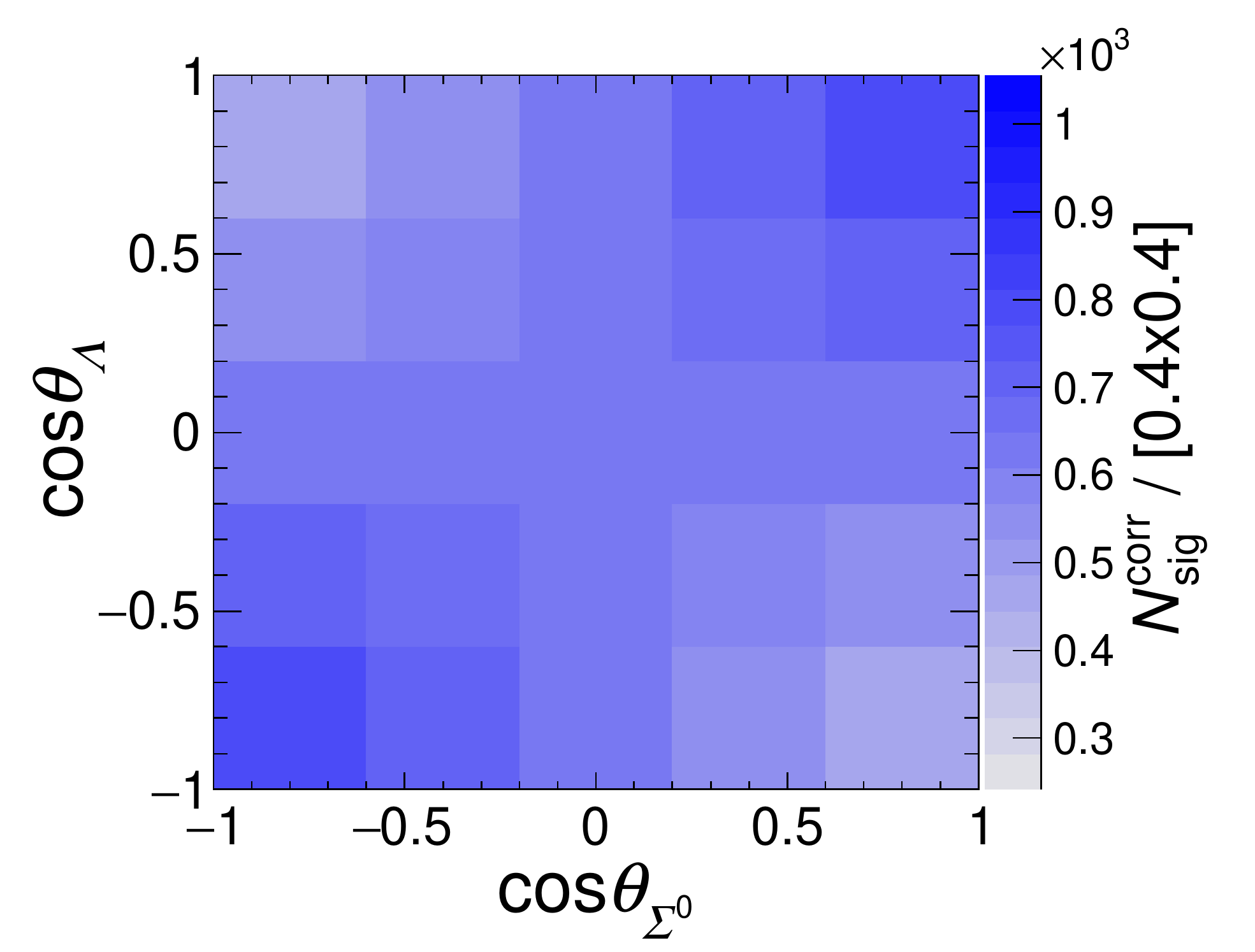}%
  \put(20,73){\small{$k=-0.434\pm 0.178$}}%
  \end{overpic}%
  \begin{overpic}[width=0.25\textwidth]{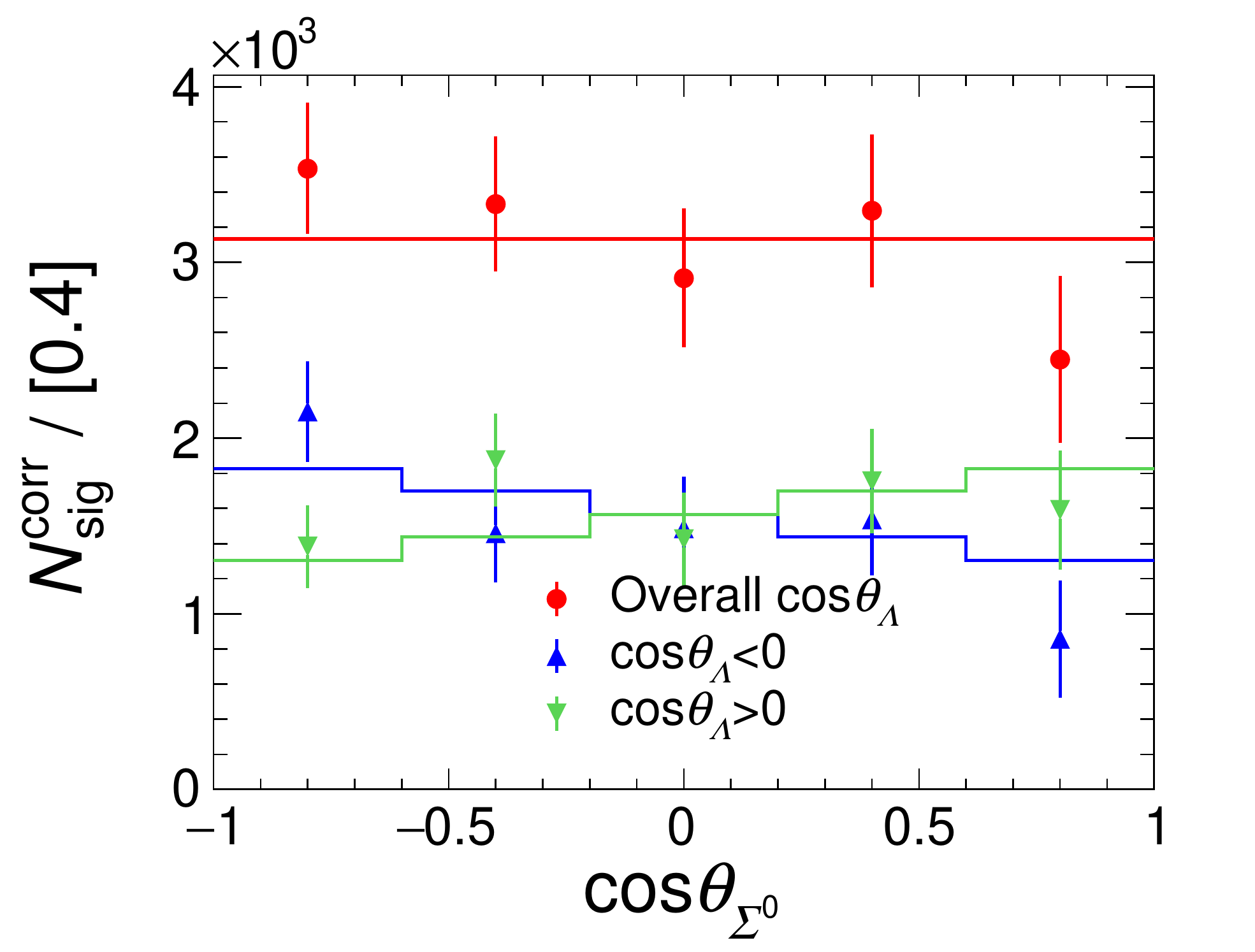}%
  \end{overpic}%
  \begin{overpic}[width=0.25\textwidth]{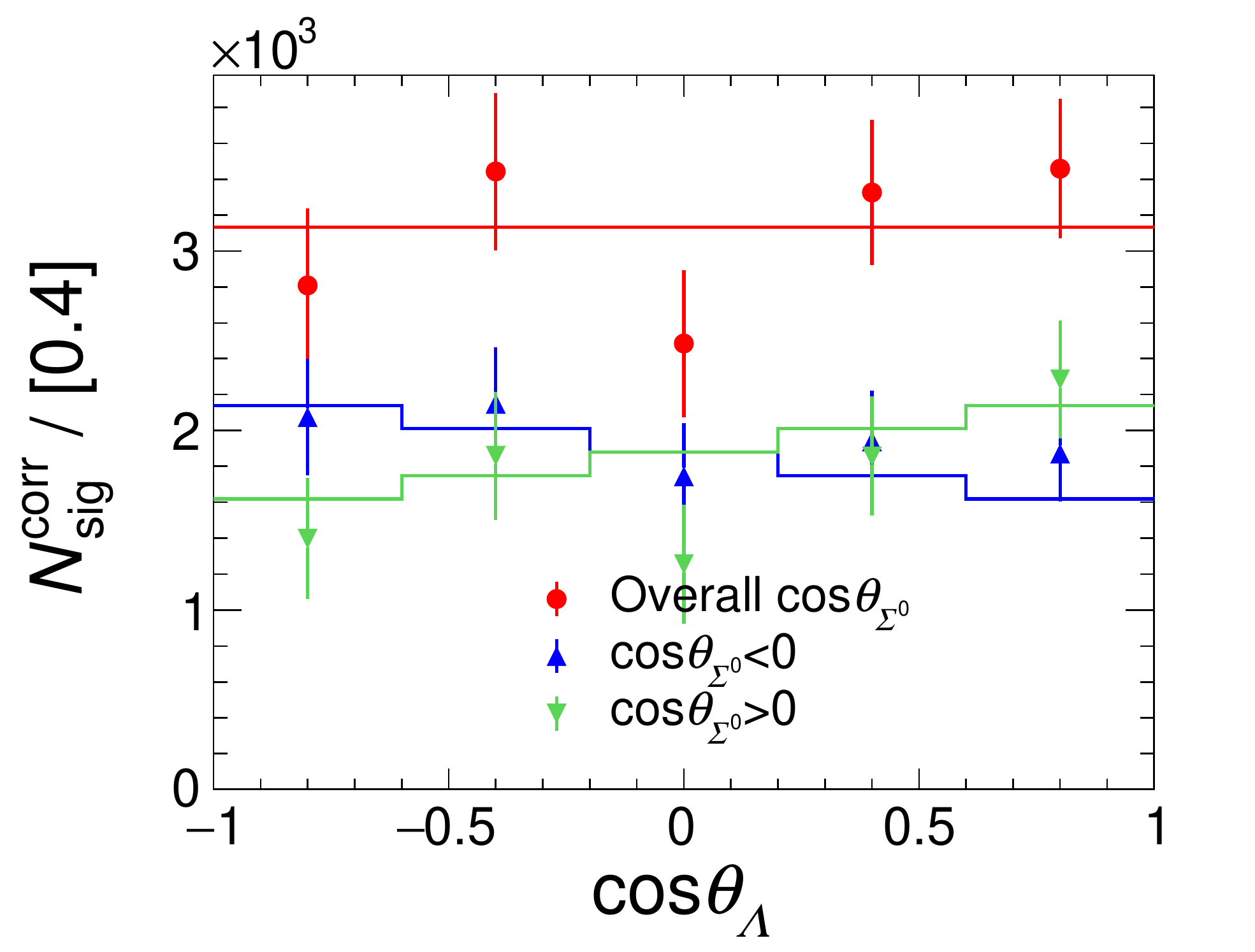}%
  \end{overpic}\\
  \vskip10pt 
  \begin{overpic}[width=0.25\textwidth]{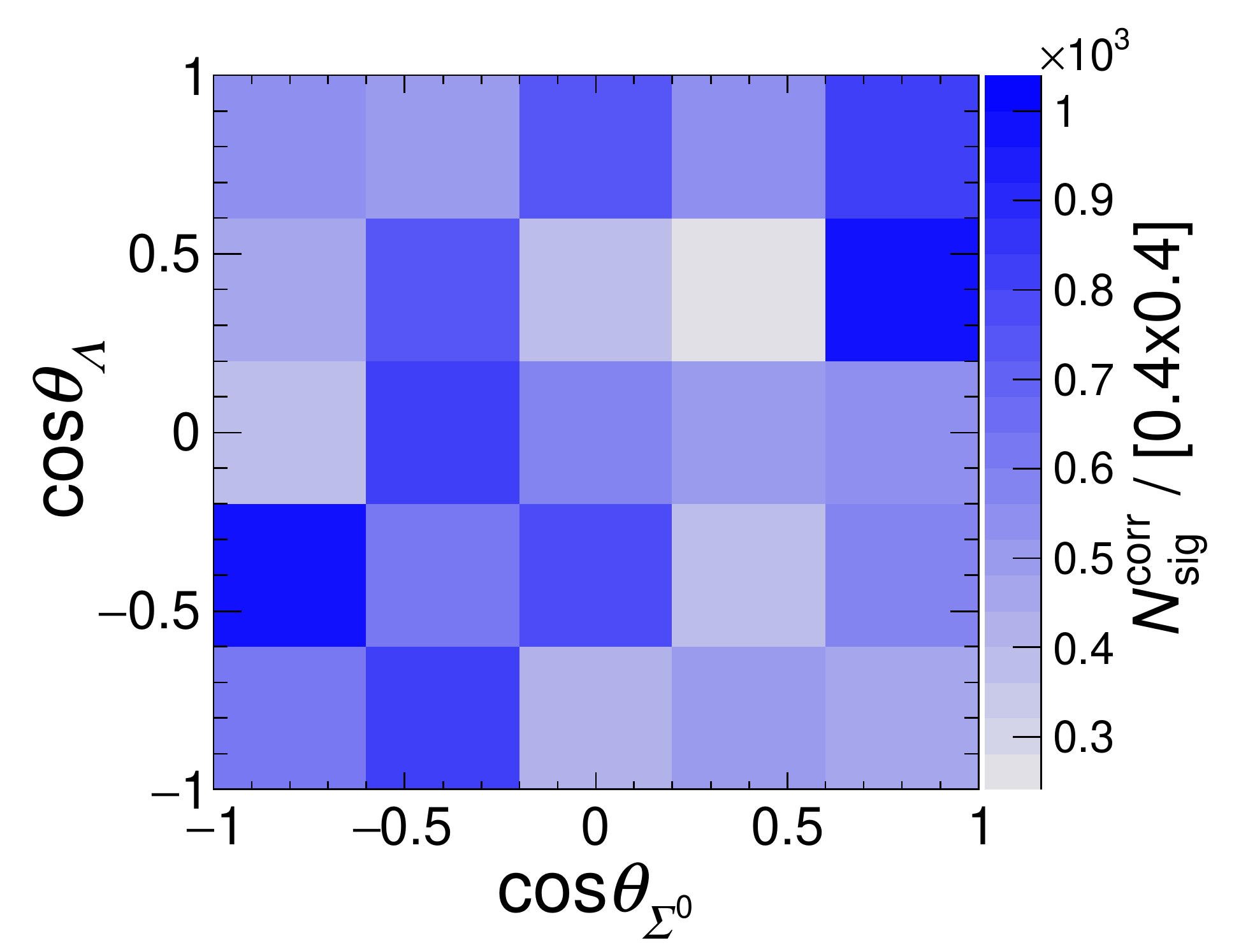}%
  \put(25,73){\small{$\Lcm\to\Sigbar{}^0\Km$}}%
  \end{overpic}%
  \begin{overpic}[width=0.25\textwidth]{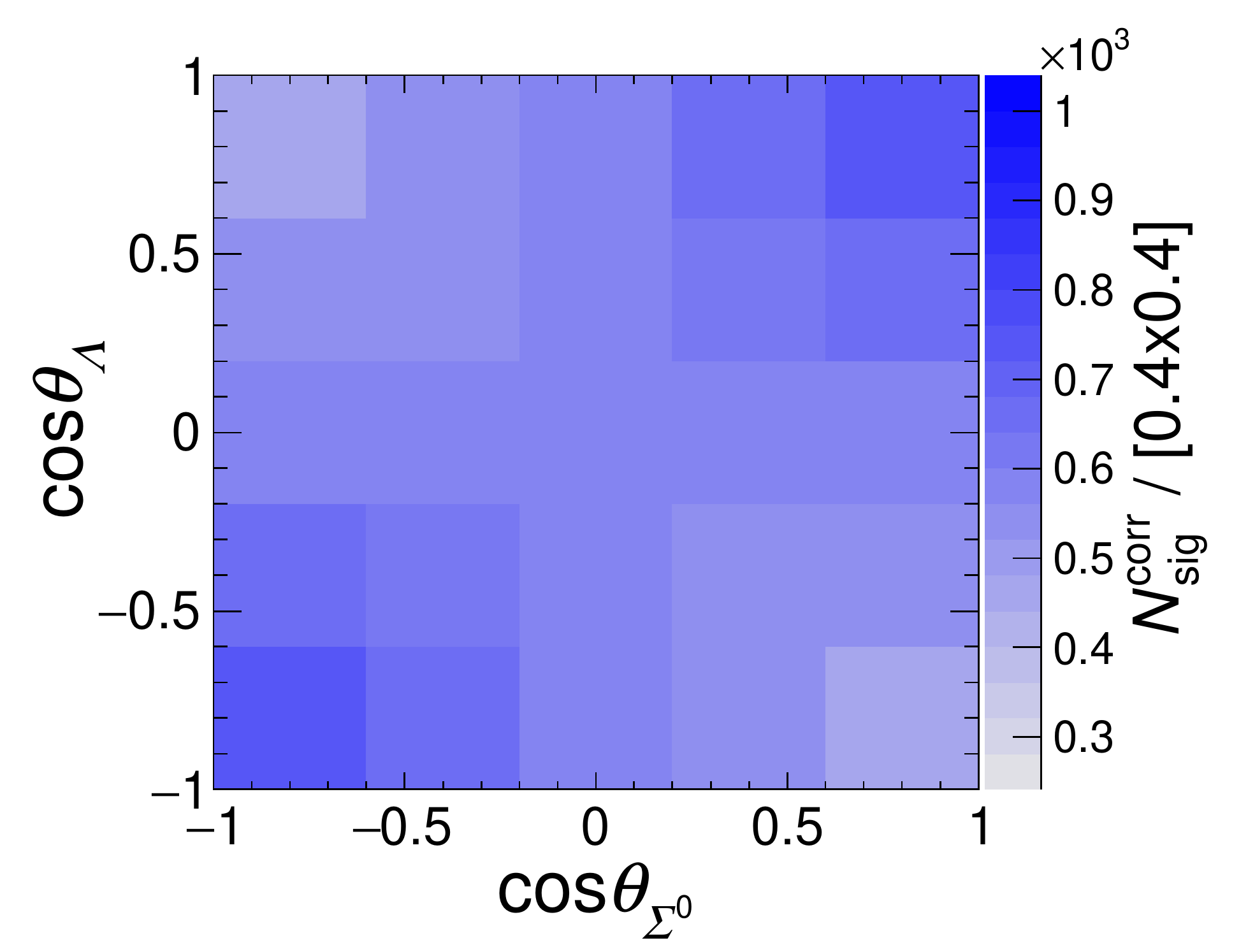}%
  \put(20,73){\small{$\overline{k}=-0.372\pm 0.209$}}%
  \end{overpic}%
  \begin{overpic}[width=0.25\textwidth]{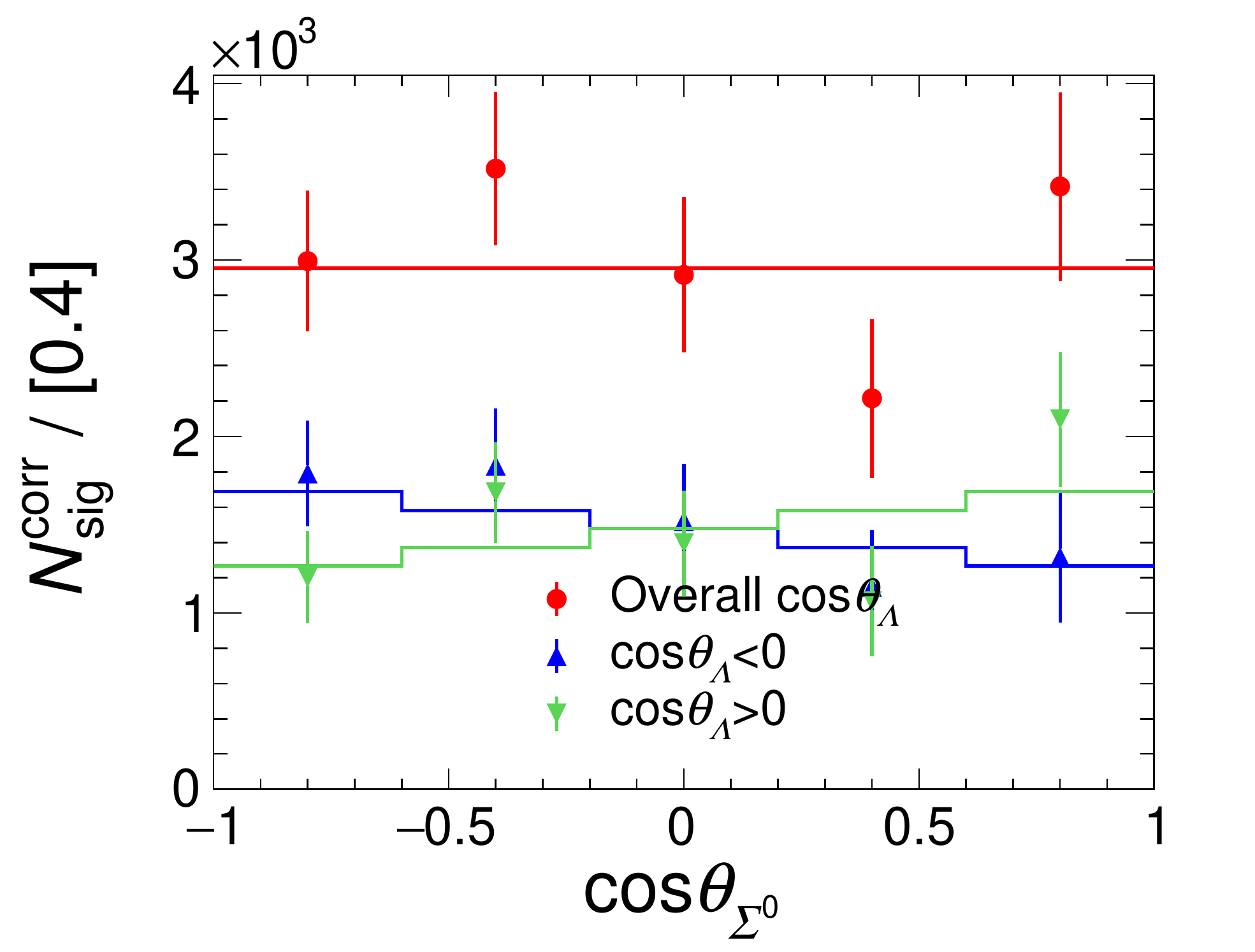}%
  \end{overpic}%
  \begin{overpic}[width=0.25\textwidth]{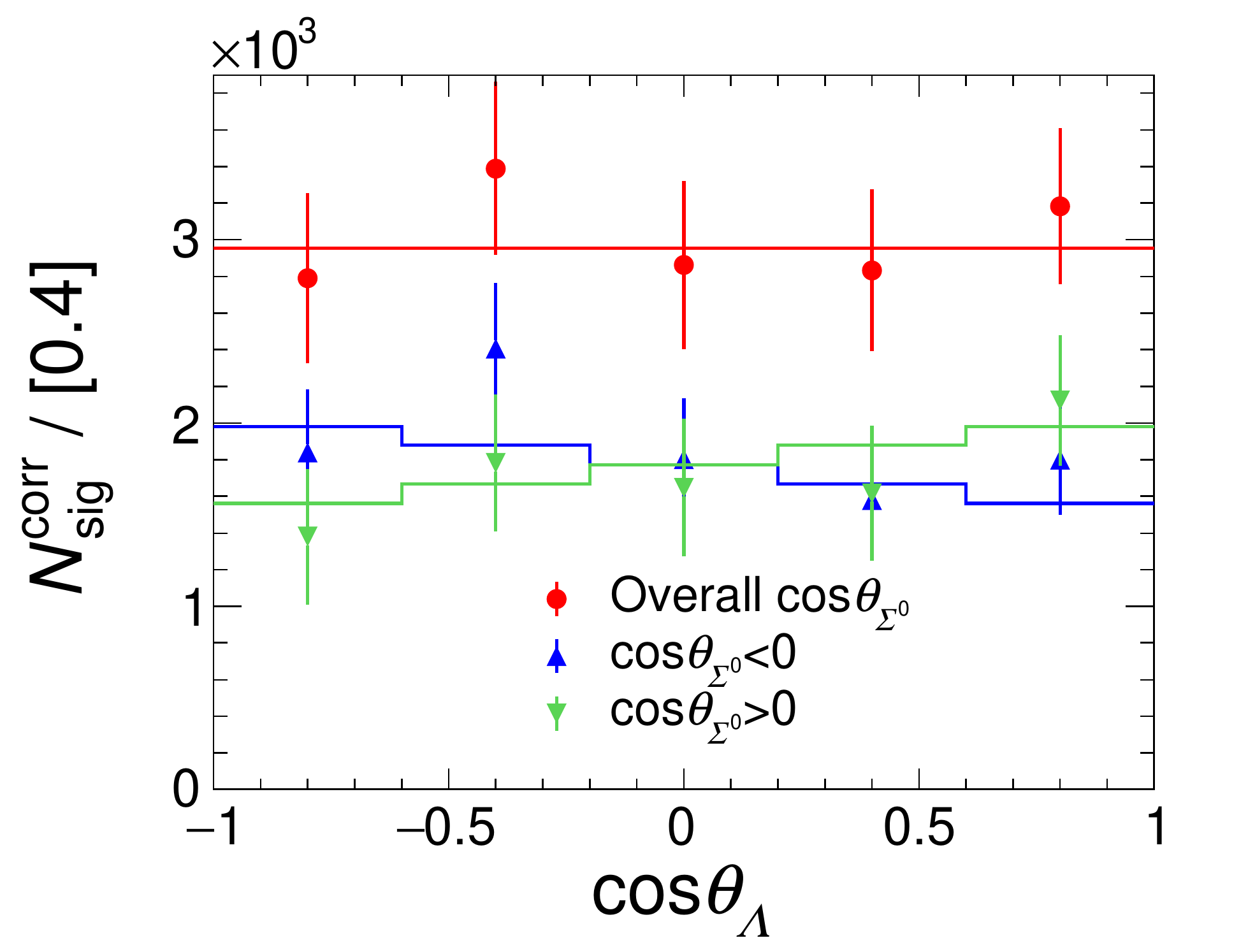}%
  \end{overpic}\\ 
  \vskip10pt   
  \begin{overpic}[width=0.25\textwidth]{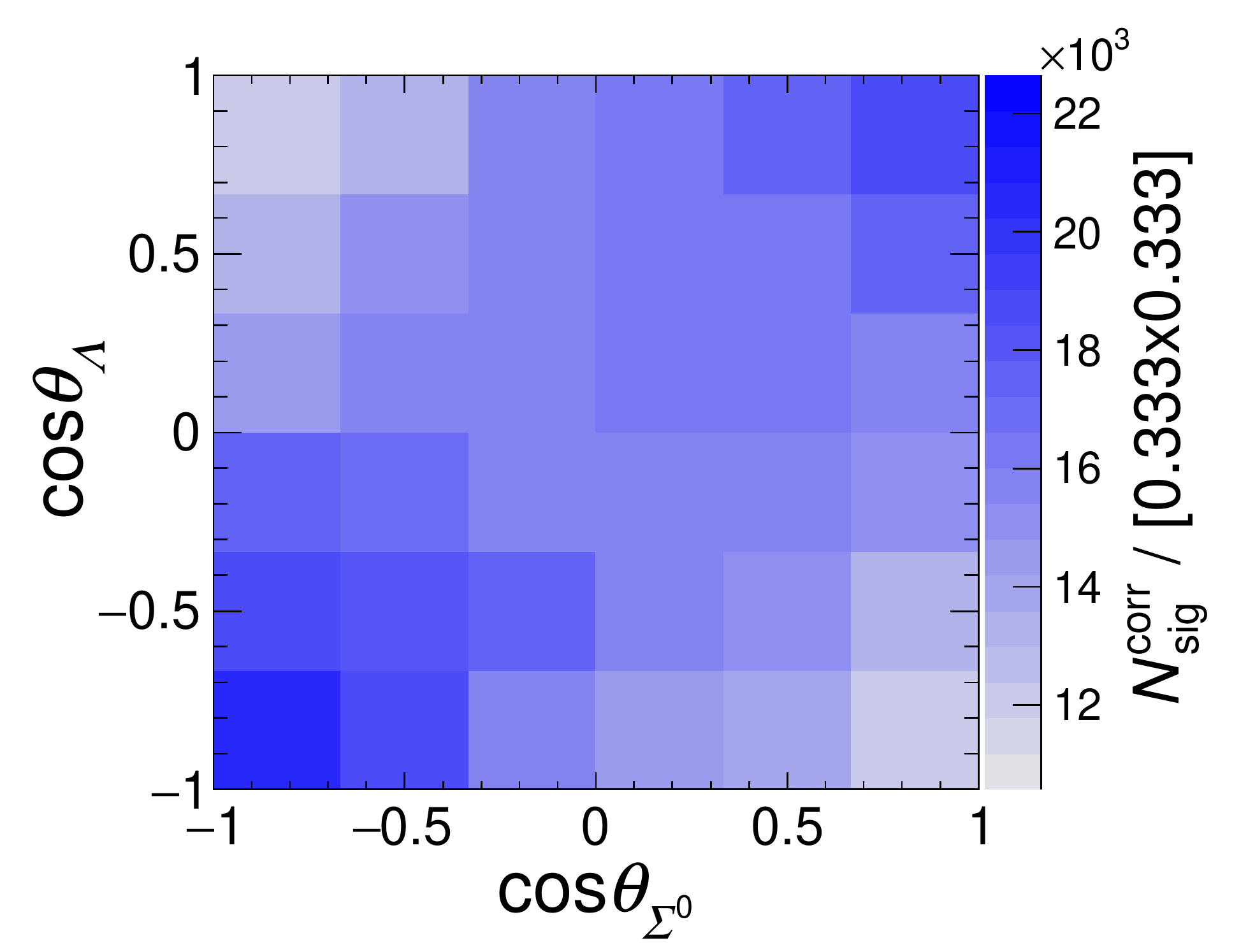}%
  \put(25,73){\small{$\LcToSigPip$}}    
  \end{overpic}%
  \begin{overpic}[width=0.25\textwidth]{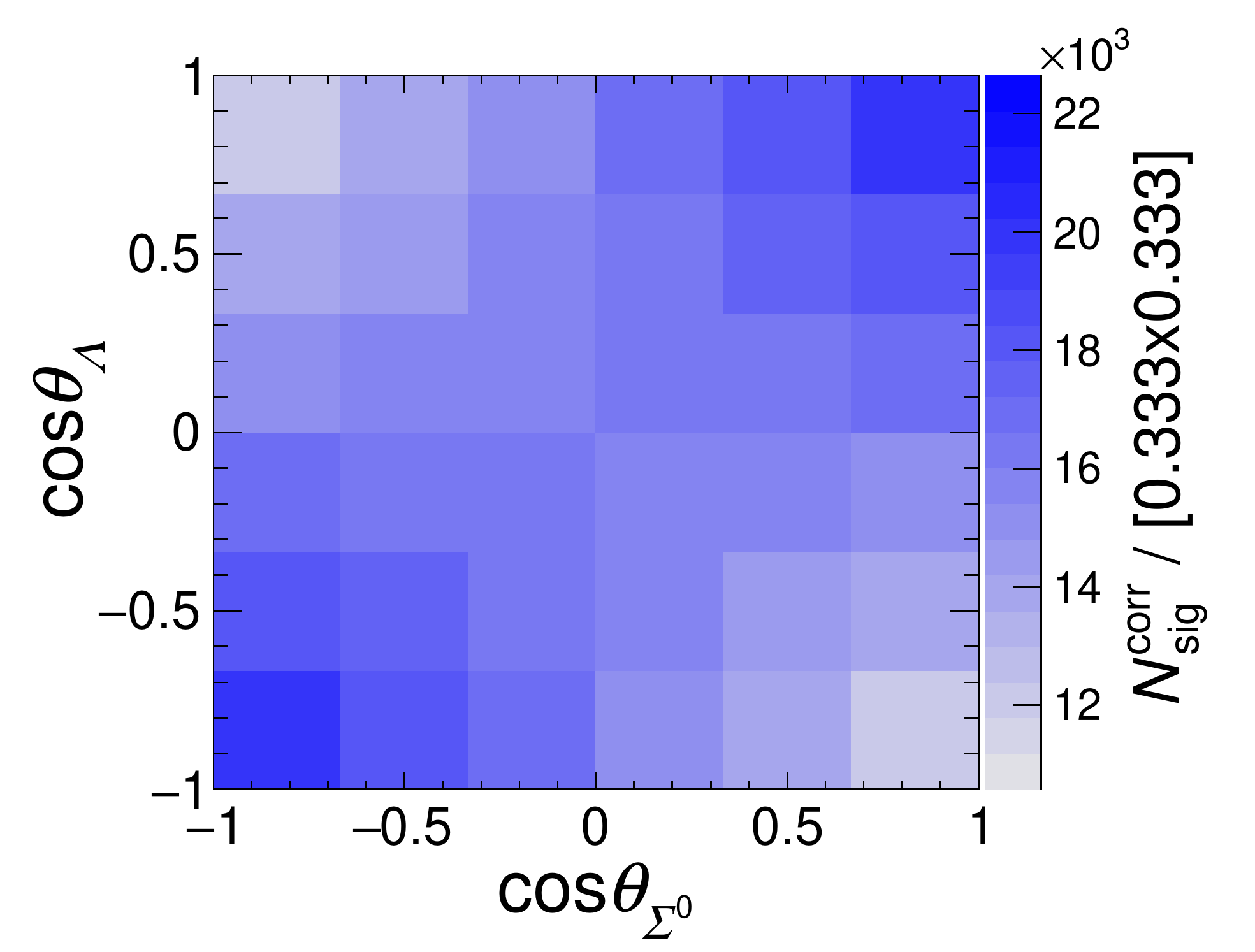}%
  \put(20,73){\small{$k=-0.340\pm 0.016$}}%
  \end{overpic}%
  \begin{overpic}[width=0.25\textwidth]{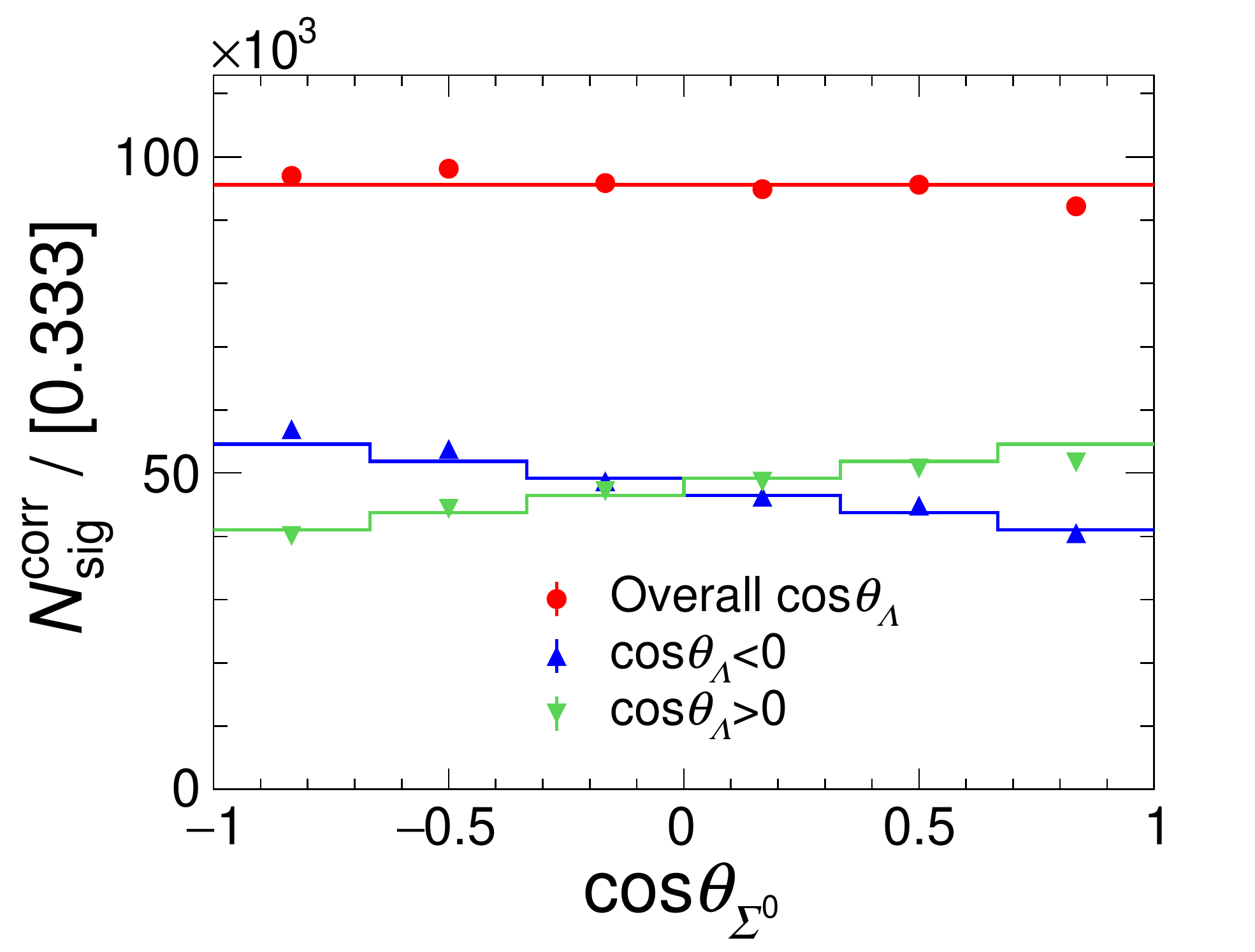}%
  \end{overpic}%
  \begin{overpic}[width=0.25\textwidth]{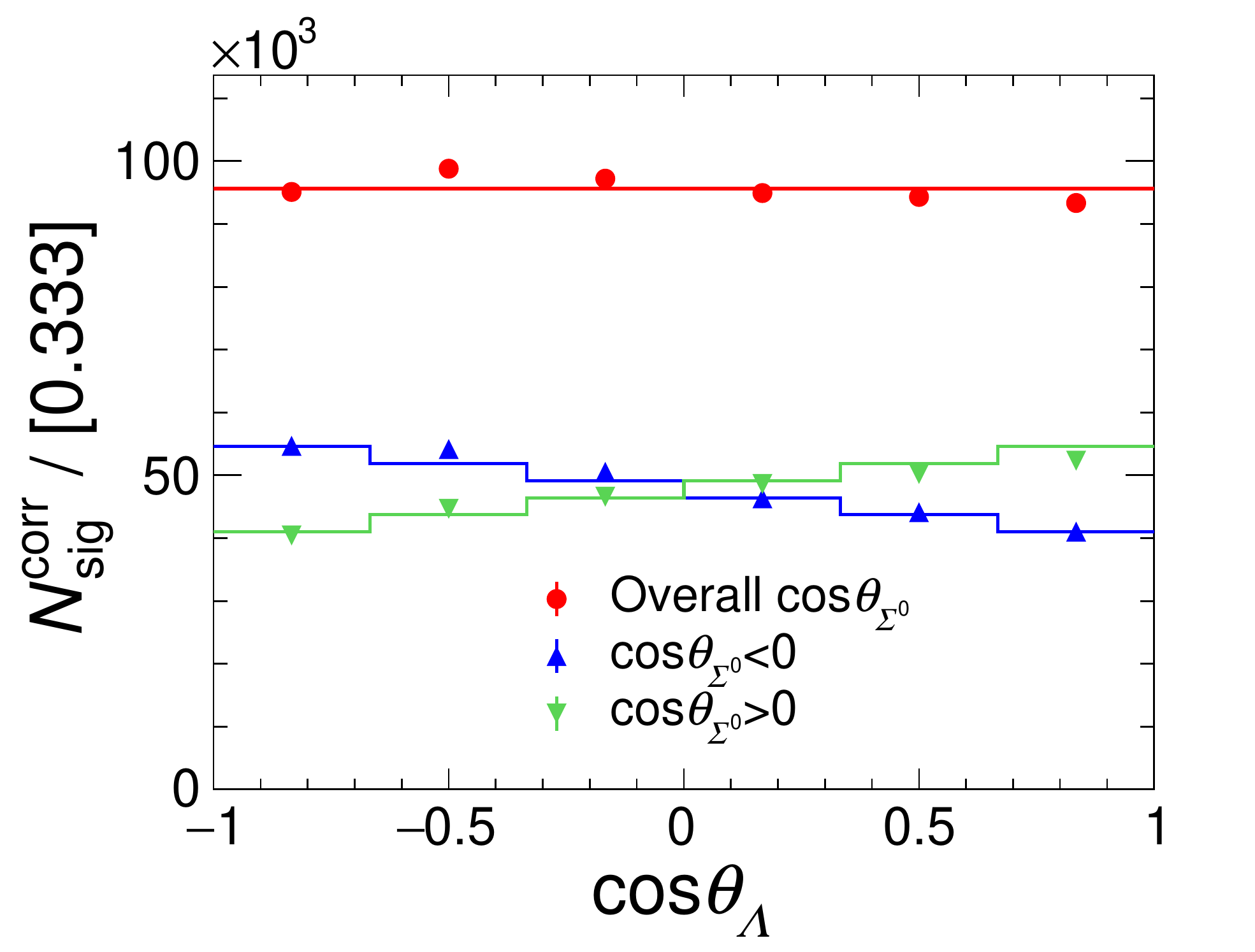}%
  \end{overpic}\\
  \vskip10pt
  \begin{overpic}[width=0.25\textwidth]{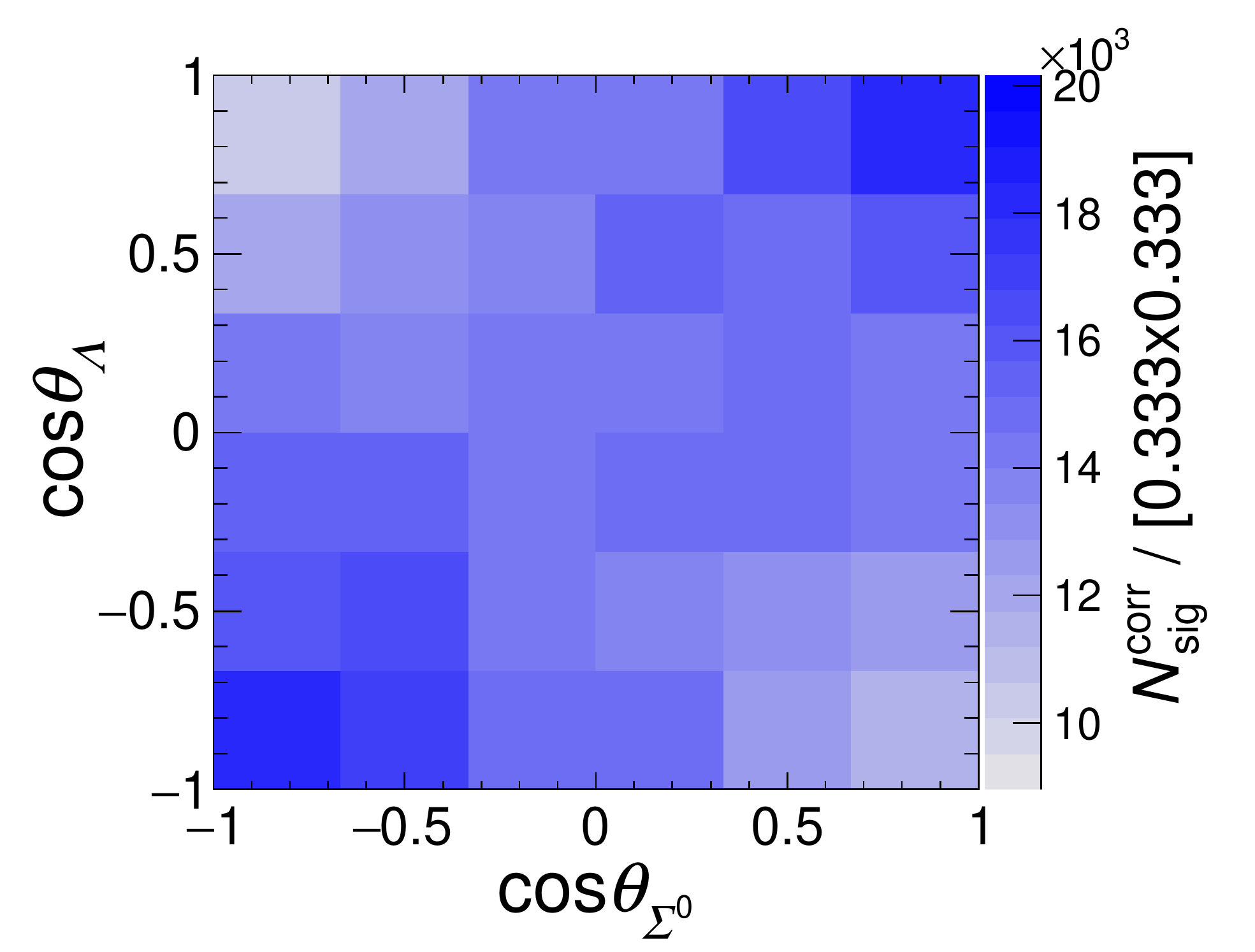}%
  \put(25,73){\small{$\Lcm\to\Sigbar{}^0\pim$}}%
  \end{overpic}%
  \begin{overpic}[width=0.25\textwidth]{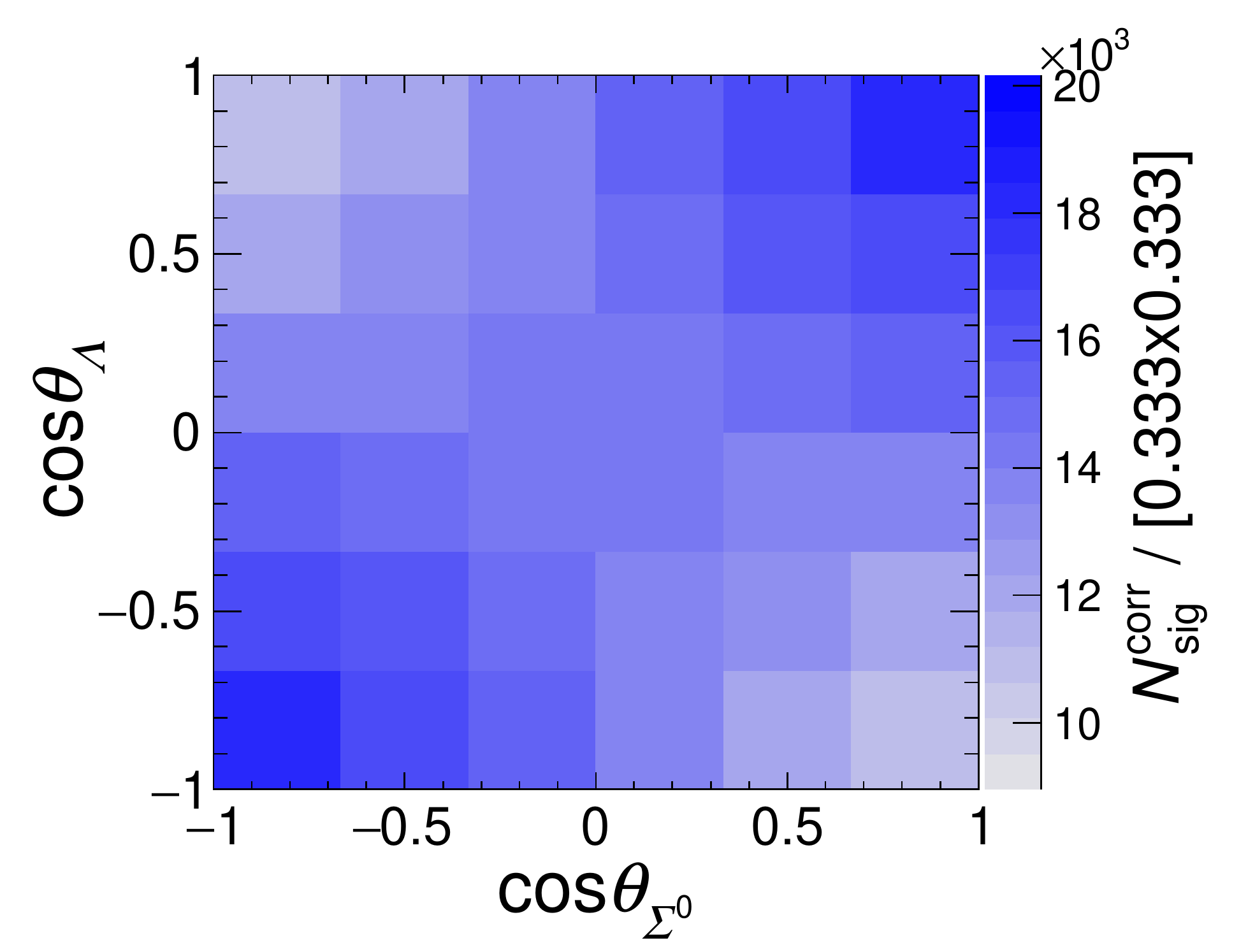}%
  \put(20,73){\small{$\overline{k}=-0.358\pm 0.017$}}%
  \end{overpic}%
  \begin{overpic}[width=0.25\textwidth]{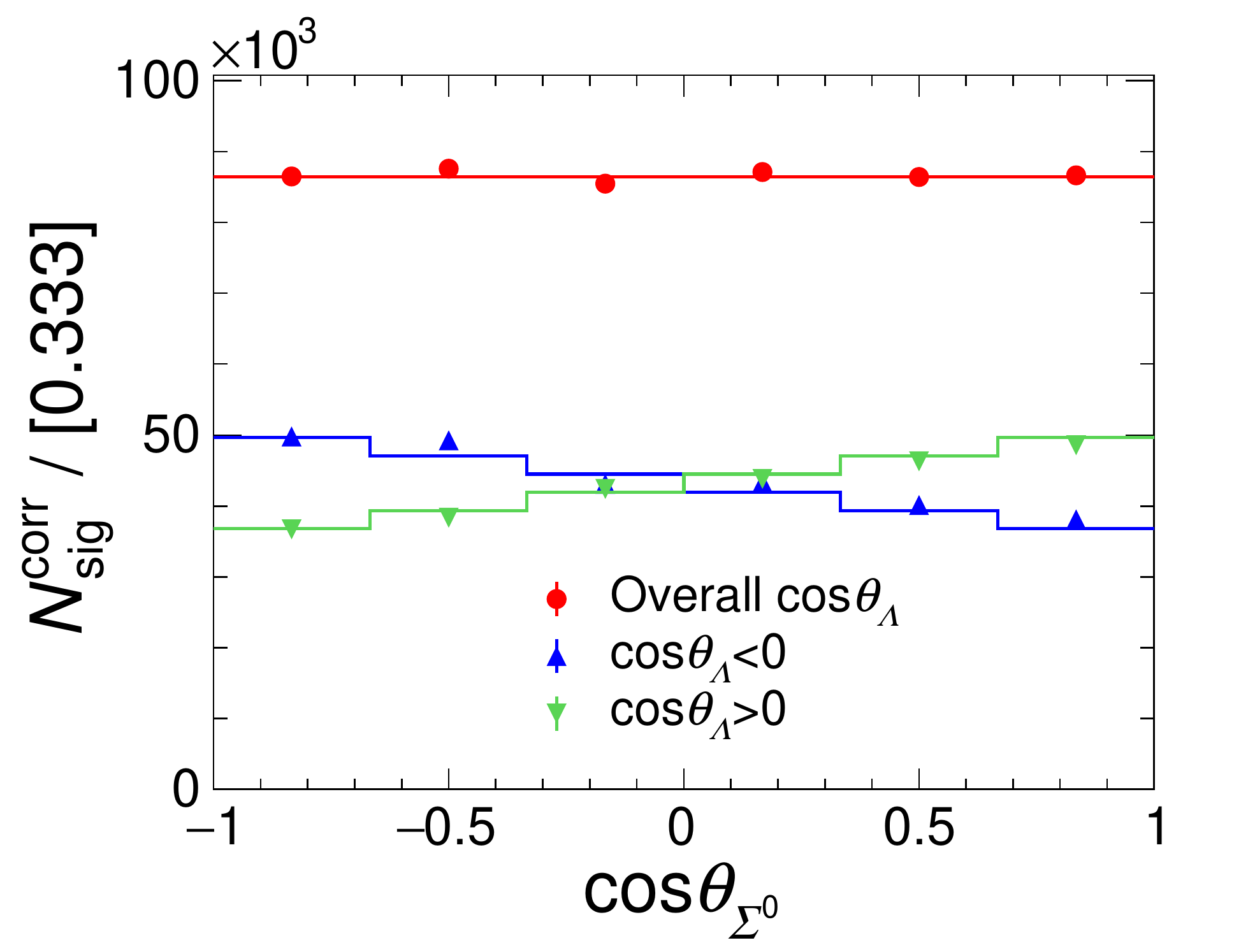}%
  \end{overpic}%
  \begin{overpic}[width=0.25\textwidth]{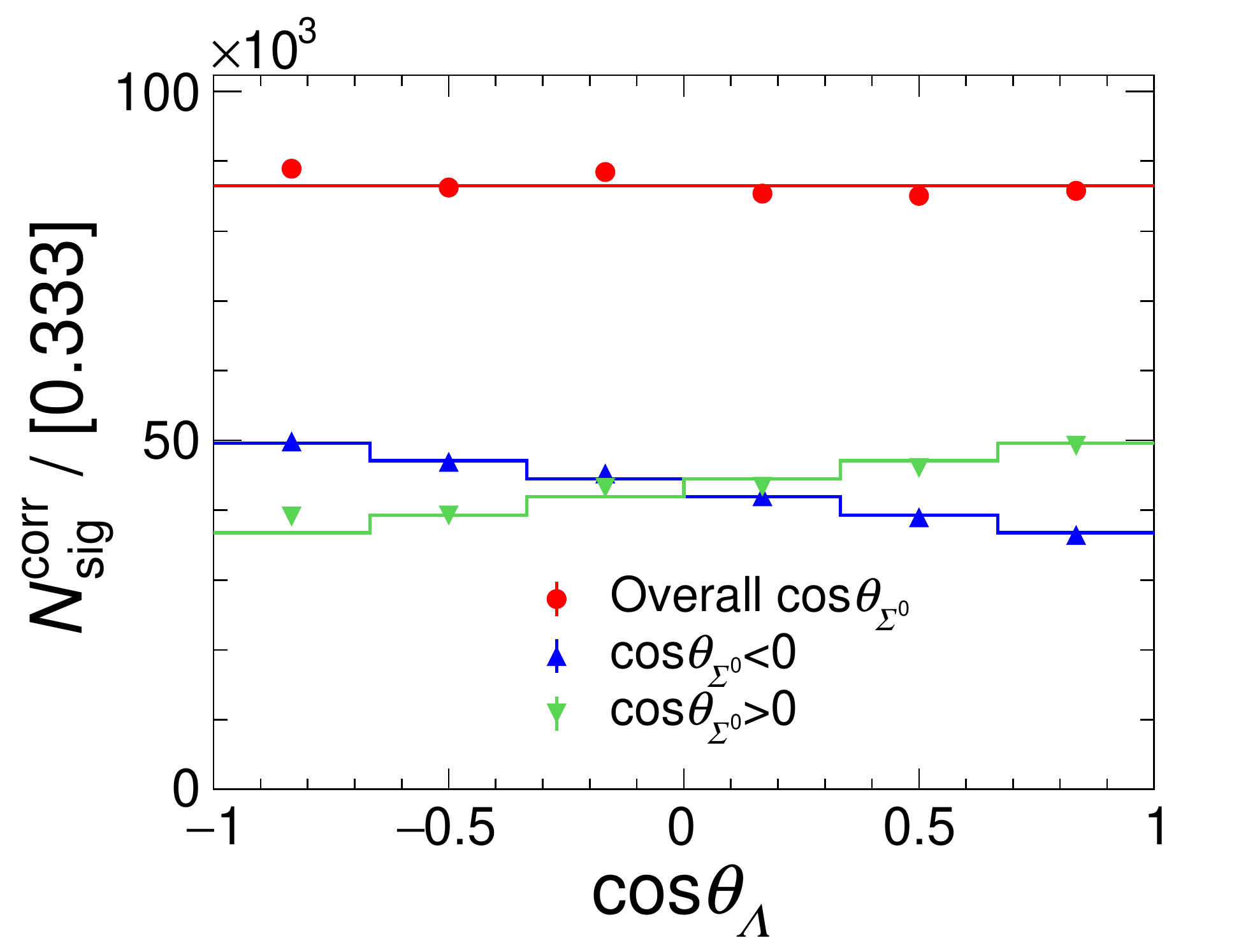}%
  \end{overpic}\\
  \vskip-5pt 
  \caption{\label{fig:AcpAlphaFinal_LcToSigHp}The $[\cos\theta_{\Sigma^0},\,\cos\theta_{\Lambda}]$ distributions of (upper plots) $\Lambda_c\to\Sigma K$ and (lower plots) $\Lambda_c\to\Sigma\pi$ decays. The first column shows the distributions after efficiency correction; the second column shows the respective fit results using a linear function $1-\alpha_{\Lambda_c^{\pm}}\alpha_{\mp}\cos\theta_{\Sigma^0}\cos\theta_{\Lambda}$. Fitted slope values ($k=\alpha_{\Lcp}\alpha_{-}$ and $\overline{k}=\alpha_{\Lcm}\alpha_{+}$) are shown. The $\chi^2$ divided by the number of degrees of freedom is $\chi^2/24=0.82$ and 0.78 for the $\Sigma K$ fits and $\chi^2/35=1.35$ and 1.05 for the $\Sigma\pi$ fits. 
  The third column shows the projections of the $\cos\theta_{\Sigma^0}$ distributions (point with error) and the fit results (histograms) in overall (red) or negative (blue) or positive (green) $\cos\theta_{\Lambda}$ region; verse visa in forth column. The absolute slopes of all projections in slices equal half of the fitted slope listed in the second column.}
  \end{centering}
\end{figure*}

\vskip-10pt
Figure~\ref{fig:Helicity} is the illustration of helicity angles definitions for $\LcToLamPip$ decays and $\LcToSigPip$ decays.
Figures~\ref{fig:YieldsFinal1} and \ref{fig:YieldsFinal2} show the $M(\Lcp)$ distribution of the combined $\Lcp$ and $\Lcm$ sample and the projection of the fit to extract the signal yields.
Figures~\ref{fig:AcpAlphaFinal_LcToLamHp} and \ref{fig:AcpAlphaFinal_LcToSigHp} show the fits to helicity angle distributions. The fitted slopes $k=\alpha_{\Lcp}\alpha_{-}$ and $\overline{k}=\alpha_{\Lcm}\alpha_{+}$ are used to determine $\alpha_{\Lcp}$ and $\alpha_{\Lcm}$ respectively, and then to calculate $\Acp^{\alpha}$. 

\end{document}